\documentclass[reprint]{revtex4-2}
\usepackage{graphicx}
\usepackage{amsmath,amsfonts}
\usepackage{bm}
\usepackage{physics}
\usepackage{enumerate}
\usepackage{color,soul}
\usepackage{hyperref}

\newcommand{\grassfint}[1]{\int\mathcal{D}\bar{#1}\mathcal{D}#1}

\begin{document}

\title{Gauge-invariant electromagnetic responses in superconductors}
\author{Sena Watanabe}\email{watanabe-sena397@g.ecc.u-tokyo.ac.jp}
\affiliation{Department of Applied Physics, The University of Tokyo, Tokyo 113-8656, Japan}
\author{Haruki Watanabe}\email{hwatanabe@g.ecc.u-tokyo.ac.jp}
\affiliation{Department of Applied Physics, The University of Tokyo, Tokyo 113-8656, Japan}
\date{\today}
\begin{abstract}
Gauge invariance is essential for making physically meaningful predictions.
In superconductors, mean-field Hamiltonians that explicitly break $U(1)$ symmetry often yield gauge-dependent results.
While this issue has been resolved for linear responses in conventional superconductors, a unified framework that also covers unconventional superconductors and nonlinear responses has yet to be established.
In this study, we present a comprehensive theoretical framework that enables gauge-invariant calculations of electromagnetic responses at arbitrary orders in external fields, applicable to both conventional and unconventional superconductors.
Our construction generalizes the consistent-fluctuation-of-the-order-parameter (CFOP) approach to full photon vertices and admits a diagrammatic representation of the response kernel in terms of Feynman diagrams.
\end{abstract}
\maketitle

\section{Introduction}
Superconductors exhibit unique electromagnetic properties such as the Meissner effect and zero electrical resistance~\cite{Schrieffer1999}. 
Microscopically, superconductivity is described by the Bardeen--Cooper--Schrieffer (BCS) theory~\cite{Bardeen1957}. 
A defining feature of this theory is that the mean-field Hamiltonian explicitly breaks $U(1)$ symmetry. 
In this framework, two electrons form a Cooper pair, resulting in the macroscopic condensation of these pairs. 
The breaking of $U(1)$ symmetry is essential to the distinctive electromagnetic properties of superconductors.

Recently, increasing attention has been paid to the electromagnetic responses of a wide variety of materials, particularly in the nonlinear regime~\cite{Sipe2000,Morimoto2016,Orenstein2021}. 
It is well established that these responses reflect the underlying symmetries and geometric structures of materials in the normal state. 
For example, second-harmonic generation occurs only in systems lacking spatial inversion symmetry, while nonreciprocal responses emerge in systems where either time-reversal or inversion symmetry is broken. 
Shift currents are nonlinear photovoltaic effects that originate from differences in the Berry phase before and after optical transitions.

Electromagnetic responses in superconductors have also been intensively studied~\cite{Wakatsuki2018,Xu2019,Daido2022,Yuan2022,Watanabe2022,Tanaka2023,Tanaka2024}, and progress has been made in understanding phenomena unique to the superconducting phase. 
Microscopic theories have been developed to describe effects such as shift currents~\cite{Xu2019} and nonreciprocal optical responses~\cite{Watanabe2022}. 
In particular, it has been shown that the superconducting Berry curvature plays a crucial role in second-order optical responses~\cite{Tanaka2023}.

While the BCS theory successfully explains many superconducting phenomena, it presents theoretical challenges concerning gauge invariance. 
The mean-field Hamiltonian lacks $U(1)$ symmetry, which is tied to charge conservation. 
As a result, electromagnetic response functions derived from such models can violate gauge invariance. 
Since gauge freedom reflects a redundancy in the mathematical formulation rather than a physical degree of freedom, gauge-dependent results are not physically trustworthy.

This issue has been recognized since the early days of BCS theory. 
There were debates about whether the Meissner effect could be derived in a gauge-invariant manner. 
Early calculations were confined to the London gauge, and potential violations of gauge-related sum rules were noted~\cite{Buckingham1957}. 
These concerns led to discussions about the role of the microscopic Hamiltonian~\cite{Schafroth1958,Rickayzen1958,Wentzel1958} and the need to include Coulomb interactions~\cite{Pines1958}. 
Through the use of the random-phase approximation, longitudinal collective modes were discovered, and the sum rule was shown to be satisfied, resulting in a gauge-invariant response kernel~\cite{Anderson1958,Bogoliubov1958,Yosida1959,Rickayzen1959a,Rickayzen1959}.

Eventually, Nambu reformulated BCS theory as a generalization of the Hartree--Fock approximation and rigorously proved gauge invariance of the electromagnetic response kernel using the Ward identity~\cite{Nambu1960}. 
This established the gauge invariance of linear electromagnetic responses in conventional superconductors. Nambu showed that gauge invariance is ensured if the microscopic Hamiltonian possesses $U(1)$ symmetry and vertex corrections are properly included through the Bethe--Salpeter equation~\cite{Nambu1960}. 
Around the same time, significant developments in the theory of many-body systems using Green functions were made~\cite{Martin1959,Kadanoff1961,Baym1961,Baym1962}. 
In particular, Baym and Kadanoff introduced the concept of ``conserving approximations'' and proved the conservation of energy, momentum, and charge from the equations of motion for Green functions~\cite{Baym1961,Baym1962}. This framework was later extended to superconductors~\cite{Ambegaokar1961}.
Building on these developments, a method known as the ``consistent fluctuations of the order parameter'' (CFOP) was introduced~\cite{Kulik1981,Zha1995,Guo2013}, providing a physically transparent interpretation of vertex corrections as fluctuations of the gap function induced by gauge fields. 
These frameworks have been employed in numerous studies of gauge-invariant linear electromagnetic responses in superconductors~\cite{Lutchyn2008,Anderson2016,Boyack2016,Dai2017,Boyack2018,Papaj2022,Oh2024}.
More recently, extensions to include impurity effects~\cite{Wang2025} and nonequilibrium dynamics~\cite{Yu2017,Yu2017a,Yang2018gi,Yang2019gi,Yang2020,Yang2020a} have also been investigated.

Although these approaches successfully describe gauge-invariant linear responses in conventional superconductors, challenges remain in formulating gauge-invariant theories of nonlinear responses and in treating more general classes of superconductors. 
Existing methods often assume continuum free-electron models for the normal state, and their validity is tied to model-specific conservation laws. 
Furthermore, proving conservation laws for nonlinear responses from the equations of motion of Green functions is not straightforward. 
An attempt to extend the Baym--Kadanoff framework to nonlinear responses was made in Ref.~\cite{Tanaka2025}, but it is limited to spatially uniform external fields. 
Huang and Wang also tried to extend Nambu's method to second-order optical responses~\cite{Huang2023}, but their formulation lacked a systematic construction of full photon vertices and did not explicitly satisfy the Ward identity, raising concerns about gauge invariance.

In our recent work~\cite{SW2024L}, we addressed this issue for spin-singlet superconductors with specific interaction forms. 
We constructed full photon vertices satisfying the Ward identities for second-order responses in $d$-wave superconductors. 
However, these results are limited to the models considered. To fully address the problem, a general theoretical framework is needed—one that enables the construction of gauge-invariant electromagnetic response functions at arbitrary order, applicable to both conventional and unconventional superconductors.

In this paper, we develop such a framework. We present a set of Feynman rules for constructing gauge-invariant electromagnetic response kernels and propose a unified approach for deriving full photon vertices, which we call the ``generalized consistent fluctuations of order parameters`` (generalized CFOP) method. This method extends the CFOP approach of Ref.~\cite{Guo2013} to higher-order responses and more general superconducting states, including spin-triplet superconductors. Our earlier results~\cite{SW2024L} are naturally recovered within this generalized framework.

The remainder of the paper is organized as follows. 
In Sec.~\ref{sec:general_theory}, we present our general method for constructing gauge-invariant response kernels in many-body systems. 
In Sec.~\ref{sec:application}, we apply this formalism to superconductors and specify the microscopic assumptions relevant to superconducting pairing. 
Section~\ref{sec:examples} provides explicit examples and numerical calculations demonstrating the effects of vertex corrections. 
We conclude in Sec.~\ref{sec:conclusion} with a summary and future perspectives.

\section{Gauge-invariant response theory of the many-body system}
\label{sec:general_theory}

In this section, we review the theoretical treatment of the electromagnetic response in interacting systems formulated in the Nambu basis to facilitate direct application to superconductors.
We summarize the Feynman rules for the electromagnetic response kernel in interacting systems and demonstrate that the gauge invariance of the response kernel is ensured by the Ward identity.

\subsection{Ward identity}

First, we introduce the Ward identity for continuum models~\cite{Ward1950,Takahashi1957,Nambu1960,Huang2023} which will guarantee the gauge invariance of the electromagnetic responses in Sec.~\ref{sec:gaugeinvariance}.
Compared to the original derivations of this identity in the quantum electrodynamics, our discussion is simpler because the electromagnetic field is not quantized.
Here, we present only the definitions and the results, and the detailed derivation will be provided in Appendix~\ref{WI}. 

Let us denote the microscopic action of an electron system with an applied gauge field as $S[A_\mu,\bar{\psi},\psi]$. Here, the Nambu spinor $\psi$ and its conjugate $\bar{\psi}$ are defined by
\begin{align}
\psi=
  \begin{pmatrix}
    \bm{c}\\
    \bar{\bm{c}}^T
  \end{pmatrix},\quad
\bar{\psi}=
  \begin{pmatrix}
   \bar{\bm{c}}^T,
    \bm{c}
  \end{pmatrix},
\end{align}
using the fermion field $\bm{c}$ and its conjugate field $\bar{\bm{c}}$ with $m$ internal degrees of freedom:
\begin{align}
&\bm{c}=\begin{pmatrix}c_1\\\vdots\\c_m\end{pmatrix},\quad
\bar{\bm{c}}=\begin{pmatrix}\bar{c}_1,\cdots,\bar{c}_m\end{pmatrix}.
\end{align}
Suppose that $S[A_\mu,\bar{\psi},\psi]$ is invariant under a local $U(1)$ gauge transformation 
\begin{align}
  A_\mu\to A_\mu-\partial_\mu\theta, \bar{\psi}\to\bar{\psi}e^{-i\theta\tau_3},\psi\to e^{i\theta\tau_3}\psi,\label{GaugeTransformation}
\end{align}
where $\tau_i$ ($i=0,1,2,3$) are the Pauli matrices act on the Nambu basis.
Namely, we have
\begin{align}
  S[A_\mu-\partial_\mu\theta,\bar{\psi}e^{-i\theta\tau_3},e^{i\theta\tau_3}\psi]=S[A_\mu,\bar{\psi},\psi].\label{ActionInvariance}
\end{align}
The Green function $G_A(x,y)$ in the presence of the gauge field is defined as
\begin{align}
  &G_A(x,y)=-\langle T_\tau\hat{\psi}(x)\hat{\psi}^\dagger(y)\rangle\notag\\
  &=-\frac{1}{Z_A}\grassfint{\psi}\psi(x)\bar{\psi}(y)\exp[-S[A_\mu,\bar{\psi},\psi]],
\end{align}
where $T_\tau$ is the imaginary-time ordered product and $Z_A=\grassfint{\psi}\exp[-S[A_\mu,\bar{\psi},\psi]]$ is the partition function.

The full $n$-photon vertex is defined as the functional derivative of the inverse of the Green function with respect to the gauge field:
\begin{align}
  \Gamma&^{\alpha_1\cdots\alpha_n}(x,y,w_1,\cdots,w_n)\nonumber\\
  &=(-1)^{n-1}\left.\left(\prod_{i=1}^n\fdv{A_{\alpha_i}(w_i)}\right)G_A^{-1}(x,y)\right|_{A=0}.\label{eq:def_fullvertex}
\end{align}

The Ward identity originates from the local $U(1)$ symmetry of the microscopic action $S[A_\mu,\bar{\psi},\psi]$, and it asserts a relationship of the form
\begin{align}
  \partial_{z^\mu}\fdv{G_A^{-1}(x,y)}{A_\mu(z)}=&i\delta(x-z)\tau_3G_A^{-1}(x,y)\nonumber\\
    &-iG_A^{-1}(x,y)\tau_3\delta(y-z)
    \label{general_wi}
\end{align}
among the Green functions in the presence of the gauge field.
By performing successive functional derivatives with respect to the gauge field, one obtains relations among the full $n$-photon vertice and full $(n-1)$-photon vertice:
\begin{align}
  &\partial_{w_n^{\alpha_n}}\Gamma^{\alpha_1\cdots\alpha_n}(x,y,w_1,\cdots,w_n)\nonumber\\
  =&i\Gamma^{\alpha_1\cdots\alpha_{n-1}}(x,y,w_1,\cdots,w_{n-1})\tau_3\delta(y-w_n)\nonumber\\
    &-i\delta(x-w_n)\tau_3\Gamma^{\alpha_1\cdots\alpha_{n-1}}(x,y,w_1,\cdots,w_{n-1}).\label{WIRS}
\end{align}
This relation is known as the Ward identity. 
To discuss the current response in interacting systems in a gauge-invariant manner, it is necessary to use the full photon vertex that manifestly satisfies this relation. 
In the translational invariant case, the Fourier transformation yields the Ward identity in the momentum space
\begin{align}
  (q_n)_{\alpha_n}&\Gamma^{\alpha_1\cdots\alpha_n}(k,q_1,\cdots,q_n)\nonumber\\
  =&\tau_3\Gamma^{\alpha_1\cdots\alpha_{n-1}}(k,q_1,\cdots,q_{n-1})\nonumber\\
  &-\Gamma^{\alpha_1\cdots\alpha_{n-1}}(k+q_n,q_1,\cdots,q_n)\tau_3.\label{WIMS}
\end{align}
\begin{figure}[t]
  \centering
  \includegraphics[width=0.9\columnwidth]{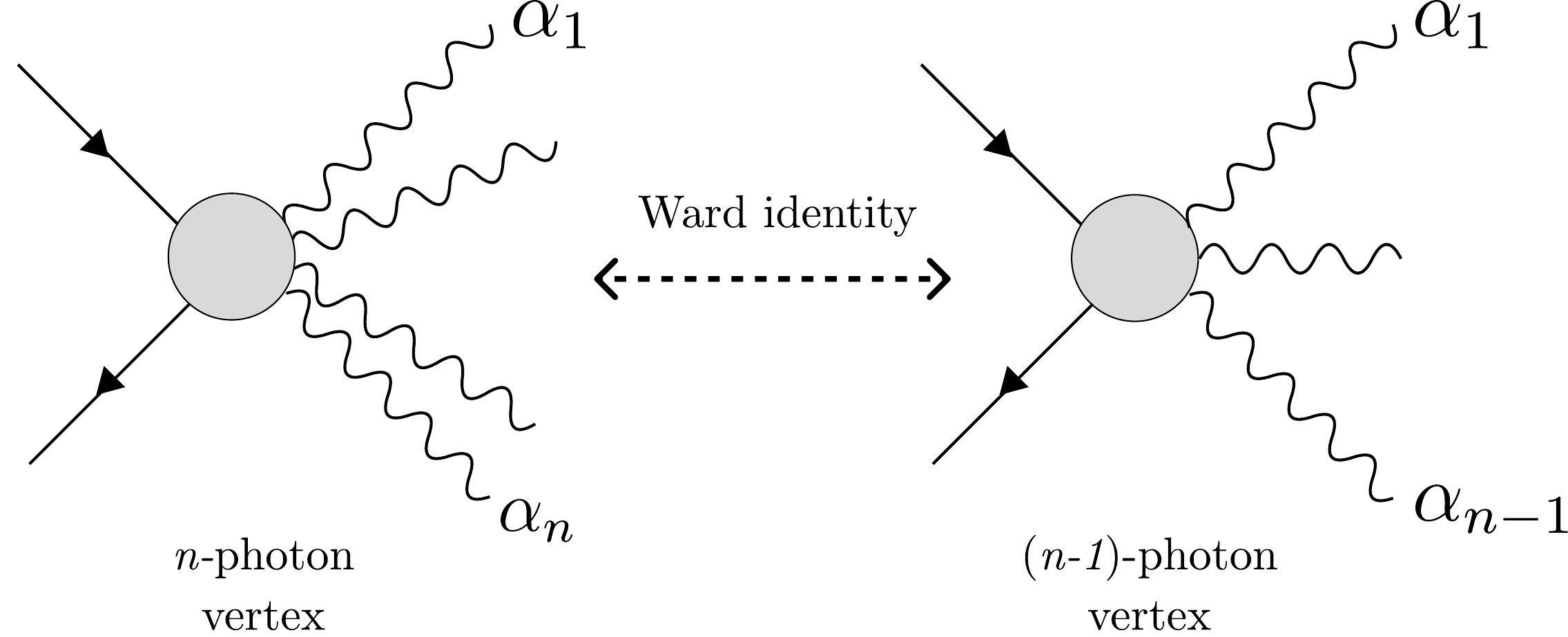}
  \caption{Ward identities relating photon vertices with $n$ and $n-1$ external photon lines.}
  \label{fig:WI}
\end{figure}
One can decompose the full photon vertex $\Gamma^{\alpha_1\cdots\alpha_n}$ into the bare vertex $\gamma^{\alpha_1\cdots\alpha_n}$ and the correction part $\Lambda^{\alpha_1\cdots\alpha_n}$ originating from the many-body effects: 
\begin{align}
\Gamma^{\alpha_1\cdots\alpha_n}=\gamma^{\alpha_1\cdots\alpha_n}+\Lambda^{\alpha_1\cdots\alpha_n}.
\label{eq:decompose_fullvertex}
\end{align}
To give definitions of these quantities, let us introduce the action $S_0[A_\mu,\bar{\psi},\psi]$ for the noninteracting limit that is also invariant under the local $U(1)$ gauge transformation in Eq.~\eqref{GaugeTransformation}.
The free Green function $G_{0,A}(x,y)$ in the presence of gauge field is defined by
\begin{align}
  &G_{0,A}(x,y)\notag\\
  &=-\frac{1}{Z_{0,A}}\grassfint{\psi}\psi(x)\bar{\psi}(y)\exp[-S_0[A_\mu,\bar{\psi},\psi]],
\end{align}
where $Z_{0,A}=\grassfint{\psi}\exp[-S_0[A_\mu,\bar{\psi},\psi]]$.
The bare vertices are defined by
\begin{align}
  \gamma&^{\alpha_1\cdots\alpha_n}(x,y,w_1,\cdots,w_n)\nonumber\\
  &=(-1)^{n-1}\left.\left(\prod_{i=1}^n\fdv{A_{\alpha_i}(w_i)}\right)G_{0,A}^{-1}(x,y)\right|_{A=0},
  \label{eq:def_barevertex}
\end{align}
which by themselves satisfy the Ward identity for bare one-photon vertices
\begin{align}
  \gamma^\mu(k,q)q_\mu=G_0^{-1}(k+q)\tau_3-\tau_3G_0^{-1}(k)
  \label{eq:wi_bare}
\end{align}
and for bare multi-photon vertices:
\begin{align}
  (q_n)_{\alpha_n}&\gamma^{\alpha_1\cdots\alpha_n}(k,q_1,\cdots,q_n)\nonumber\\
  =&\tau_3\gamma^{\alpha_1\cdots\alpha_{n-1}}(k,q_1,\cdots,q_{n-1})\nonumber\\
  &-\gamma^{\alpha_1\cdots\alpha_{n-1}}(k+q_n,q_1,\cdots,q_{n-1})\tau_3.
\end{align}
The Dyson equation
\begin{align}
  G_A^{-1}(x,y)=G_{0,A}^{-1}(x,y)-\Sigma_A(x,y)\label{eq:dyson_real}
\end{align}
relates the full Green function $G_A(x,y)$ and the bare Green function $G_{0,A}(x,y)$, which in turn defines the self-energy $\Sigma_A(x,y)$.
We define the correction part for the photon vertex  by
\begin{align}
  \Lambda&^{\alpha_1\cdots\alpha_n}(x,y,w_1,\cdots,w_n)\nonumber\\
  &=(-1)^n\left.\left(\prod_{i=1}^n\fdv{A_{\alpha_i}(w_i)}\right)\Sigma_A(x,y)\right|_{A=0},
  \label{eq:def_correction}
\end{align}
which satisfies the Ward identity for the correction part:
\begin{align}
 (q_n)_{\alpha_n}&\Lambda^{\alpha_1\cdots\alpha_n}(k,q_1,\cdots,q_n)\nonumber\\
 =&\tau_3\Lambda^{\alpha_1\cdots\alpha_{n-1}}(k,q_1,\cdots,q_{n-1})\nonumber\\
 &-\Lambda^{\alpha_1\cdots\alpha_{n-1}}(k+q_n,q_1,\cdots,q_{n-1})\tau_3.
 \label{eq:wi_correction}
\end{align}

\subsection{Diagrammatic representation of the electromagnetic response kernel}

Here, we examine how the full photon vertex and the bare photon vertex, defined in the previous section, appear in the expression of the response kernel.
We summarize the Feynman rules for the electromagnetic response kernel in many-body systems.

\begin{figure}[t]
  \centering
  \includegraphics[width=\columnwidth]{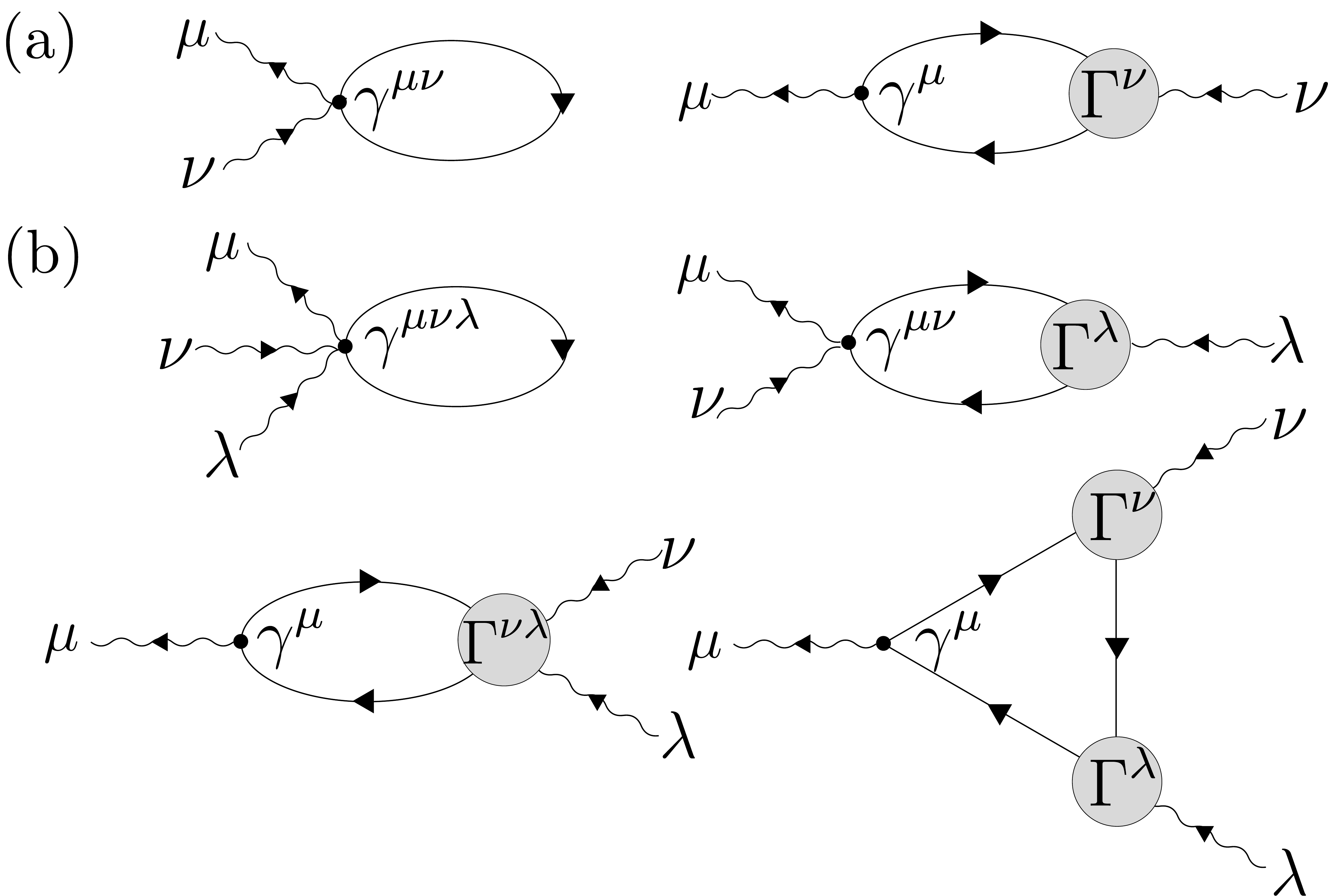}
  \caption{Diagrammatic representations of electromagnetic response kernels. 
(a) Linear response kernel $K^{\mu\nu}$. 
(b) Second-order response kernel $K^{\mu\nu\lambda}$.}
  \label{fig:kernel}
\end{figure}

Let us assume that the microscopic action of a system with a background gauge field can be decomposed as
\begin{align}
  S[A_\mu,\bar{\psi},\psi]=S_0[A_\mu,\bar{\psi},\psi]+S_{\mathrm{int}}[\bar{\psi},\psi].
  \label{eq:action_int}
\end{align}
Here, the first term represents the noninteracting limit of the action and the second term represents the interaction between electrons.
Crucially, we assume that the interactions do not depend on the gauge field.
Under this assumption, the current is defined as
\begin{align}
  J_A^\mu(x)=-\fdv{S_0[A_\mu,\bar{\psi},\psi]}{A_\mu(x)}.
\end{align}
The noninteracting action can be expressed in terms of the bare Green function $G_{0,A}(x,y)$ as
\begin{align}
  S_0[A_\mu,\bar{\psi},\psi]=-\frac{1}{2}\bar{\psi}(\overline{w}_1)G_{0,A}^{-1}(\overline{w}_1,\overline{w}_2)\psi(\overline{w}_2).
\end{align}
The factor $1/2$ represents the particle-hole doubling due to the Nambu basis. 
Here and hereafter, we use the abbreviation for integral
\begin{align}
  f(\overline{x})g(\overline{x})=\int d^4x f(x)g(x).
\end{align}
Namely, a pair of overlined variables implies an integration over those variables.

The current can be expressed as
\begin{align}
  J_A^\mu(x)=\frac{1}{2}\bar{\psi}(\overline{w}_1)\gamma_A^\mu(\overline{w}_1,\overline{w}_2,x)\psi(\overline{w}_2)
\end{align}
using the generalized bare photon vertex $\gamma_A^\mu$ defined by
\begin{align}
  \gamma_A^\mu(w_1,w_2,x)=\fdv{G_{0,A}^{-1}(w_1,w_2)}{A_\mu(x)}.
\end{align}
Therefore, the expectation value of the current can be expressed in terms of the Green function as
\begin{align}
  \expval{J_A^\mu(x)}=\frac{1}{2}\mathrm{Tr}\left[\gamma_A^\mu(\overline{w}_1,\overline{w}_2,x)G_A(\overline{w}_2,\overline{w}_1)\right].\label{JA}
\end{align}
Expanding this into the series of the gauge field $A$, we find
\begin{align}
 \expval{J_A^\mu(x)}&=\sum_{n=0}^\infty\frac{1}{n!}K^{\mu\alpha_1\cdots\alpha_n}(x,\overline{w}_1,\cdots,\overline{w}_n)\nonumber\\
  &\qquad\qquad\times A_{\alpha_1}(\overline{w}_1)\cdots A_{\alpha_n}(\overline{w}_n).\label{eq:expval_real}
\end{align}
Thus the $n$-th order electromagnetic response kernel is obtained by $n$-th functional derivative of $\expval{J_A^\mu(x)}$ with respect to the gauge field $A$:
\begin{align}
  K^{\mu\alpha_1\cdots\alpha_n}(x,y_1,\cdots,y_n)=\left.\frac{\delta^n\langle J_A^\mu(x)\rangle}{\delta A_{\alpha_1}(y_1)\cdots\delta A_{\alpha_n}(y_n)}\right|_{A=0}.
\end{align}
For example, the linear electromagnetic response kernel is defined as
\begin{align}
  K^{\mu\nu}(x,y)=\left.\fdv{\expval{J_A^\mu(x)}}{A_\nu(y)}\right|_{A=0},\label{linearK}
\end{align}
and the second-order response kernel is given by
\begin{align}
  K^{\mu\nu\lambda}(x,y,z)=\left.\frac{\delta^2\expval{J_A^\mu(x)}}{\delta A_\nu(y)\delta A_\lambda(z)}\right|_{A=0}.\label{secondK}
\end{align}

\begin{widetext}
Plugging Eq.~\eqref{JA} into Eq.~\eqref{linearK}, we obtain the explicit form
\begin{align}
  K^{\mu\nu}(x,y)&=\frac{1}{2}\left.\mathrm{Tr}\big[\gamma_A^\mu(\overline{w}_1,\overline{w}_2,x)\fdv{G_A(\overline{w}_2,\overline{w}_1)}{A_\nu(y)}+\fdv{\gamma_A^\mu(\overline{w}_1,\overline{w}_2,x)}{A_\nu(y)}G_A(\overline{w}_2,\overline{w}_1)\big]\right|_{A=0}\nonumber\\
  &=\frac{1}{2}\bigg(-\mathrm{Tr}\big[\gamma^\mu(\overline{w}_1,\overline{w}_2,x)G(\overline{w}_2,\overline{w}_3)\Gamma^\nu(\overline{w}_3,\overline{w}_4,y)G(\overline{w}_4,\overline{w}_1)\big]-\mathrm{Tr}\big[\gamma^{\mu\nu}(\overline{w}_1,\overline{w}_2,x,y)G(\overline{w}_2,\overline{w}_1)\big]\bigg).\label{eq:lin_kernel_real}
\end{align}
In the derivation, we used
\begin{align}
  \left.\fdv{G_A(w_2,w_1)}{A_\nu(y)}\right|_{A=0}&=-\left.G_A(w_2,\overline{w}_3)\fdv{G_A^{-1}(\overline{w}_3,\overline{w}_4)}{A_\nu(y)}G_A(\overline{w}_4,w_1)\right|_{A=0}=-G(w_2,\overline{w}_3)\Gamma^\nu(\overline{w}_3,\overline{w}_4,y)G(\overline{w}_4,w_1)
\end{align}
and
\begin{align}
  \left.\fdv{\gamma_A^\mu(w_1,w_2,x)}{A_\nu(y)}\right|_{A=0}&=\left.\frac{\delta^2G_{0,A}^{-1}(w_1,w_2)}{\delta A_\mu(x)\delta A_\nu(y)}\right|_{A=0}=-\gamma^{\mu\nu}(w_1,w_2,x,y),
\end{align}
and we defined the Green function without a gauge field as
\begin{align}
  G(x,y)=\left.G_A(x,y)\right|_{A=0}.
\end{align}
The Feynman diagrams for Eq.~\eqref{eq:lin_kernel_real} can be illustrated as in Fig.~\ref{fig:kernel}(a).

Similarly, for the second-order response kernel, we obtain
\begin{align}
  K^{\mu\nu\lambda}(x,y,z)=\frac{1}{2}\bigg(&\frac{1}{2}\mathrm{Tr}\big[\gamma^{\mu\nu\lambda}(\overline{w}_1,\overline{w}_2,x,y,z)G(\overline{w}_2,\overline{w}_1)\big]+\mathrm{Tr}\big[\gamma^{\mu\nu}(\overline{w}_1,\overline{w}_2,x,y)G(\overline{w}_2,\overline{w}_3)\Gamma^\lambda(\overline{w}_3,\overline{w}_4,z)G(\overline{w}_4,\overline{w_1})\big]\nonumber\\
    &\qquad+\frac{1}{2}\mathrm{Tr}\big[\gamma^\mu(\overline{w}_1,\overline{w_2},x)G(\overline{w}_2,\overline{w}_3)\Gamma^{\nu\lambda}(\overline{w}_3,\overline{w}_4,y,z)G(\overline{w}_4,\overline{w}_1)\big]\nonumber\\
    &\qquad+\mathrm{Tr}\big[\gamma^\mu(\overline{w}_1,\overline{w}_2,x)G(\overline{w}_2,\overline{w}_3)\Gamma^\lambda(\overline{w}_3,\overline{w}_4,z)G(\overline{w}_4,\overline{w}_5)\Gamma^\nu(\overline{w}_5,\overline{w}_6,y)G(\overline{w}_6,\overline{w}_1)\big]\nonumber\\
    &\qquad+[(\nu,y)\leftrightarrow(\lambda,z)]\bigg).\label{eq:sec_kernel_real}
\end{align}
\end{widetext}

The corresponding Feynman diagrams are presented in Fig.~\ref{fig:kernel}(b).
A higher-order response kernel can also be derived systematically.

Assuming translational symmetry in the system, one can switch to momentum space by Fourier transformation.
The Fourier transformation of Eq.~\eqref{eq:expval_real} is given by
\begin{align}
  \expval{\hat{J}_A^\mu(q)}=\sum_{n=0}^\infty\frac{1}{n!}&\delta(q-\sum_{i=1}^n q_i)K^{\mu\alpha_1\cdots\alpha_n}(q_1,\cdots,q_n)\nonumber\\
  &\times A_{\alpha_1}(q_1)\cdots A_{\alpha_n}(q_n)\label{eq:expval_current}
\end{align}
and the $n$-th order response kernel in momentum space is defined.
Note that the response kernel has an intrinsic permutation symmetry~\cite{Rostami2021}:
\begin{align}
  K^{\mu\alpha_{p(1)}\cdots\alpha_{p(n)}}(q_{p(1)},\cdots,q_{p(n)})=K^{\mu\alpha_1\cdots\alpha_n}(q_1,\cdots,q_n)
\end{align}
for any permutation $p$.

\begin{table}[t]
  \caption{Feynman rules for the electromagnetic response kernel in interacting systems.}
  \label{tab:feynman_rule}
  \begin{tabular}{cc|c}\hline
    \multicolumn{2}{c|}{Diagrams} & value\\ \hline\hline
    photon line &
    \begin{minipage}{0.48\columnwidth}
      \centering
      \includegraphics[width=0.5\columnwidth]{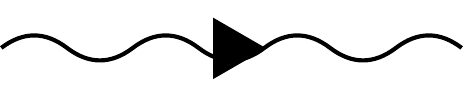}
    \end{minipage}& $1$\\
    electron line &
    \begin{minipage}{0.48\columnwidth}
      \centering
      \includegraphics[width=0.5\columnwidth]{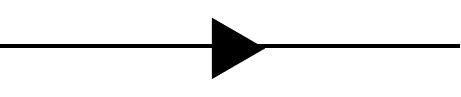}
    \end{minipage}& $G$ \\
    \begin{tabular}{c}
      $(m+1)$-photon \\output vertex
    \end{tabular} &
    \begin{minipage}{0.48\columnwidth}
      \centering
      \includegraphics[width=0.8\columnwidth]{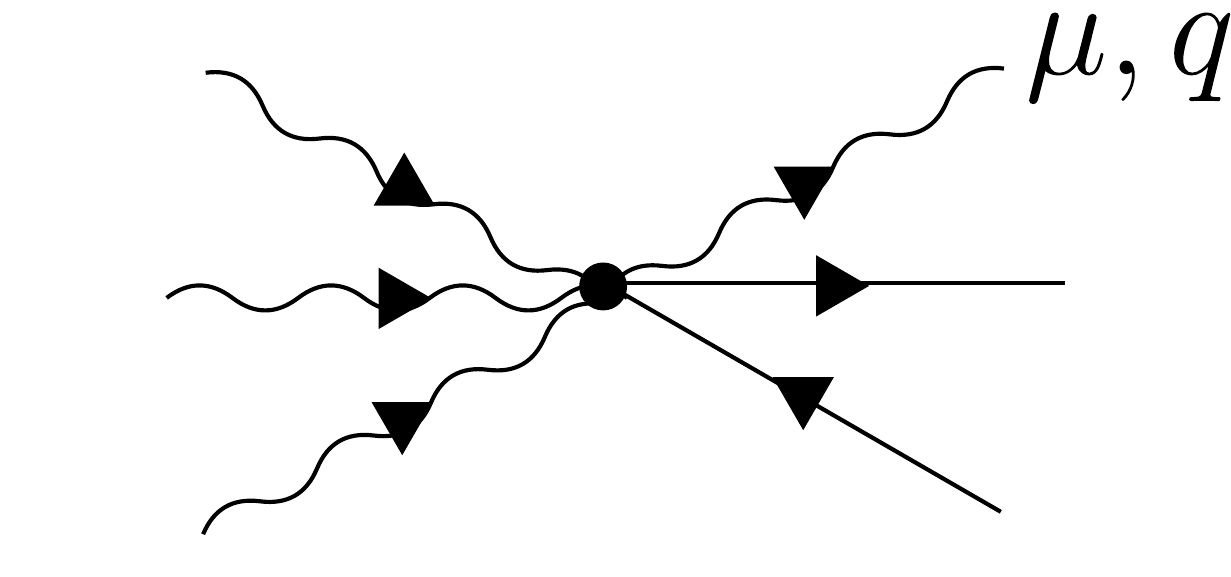}
    \end{minipage} & $\frac{1}{m!}\gamma^{\mu\alpha_{i_1}\cdots\alpha_{i_m}}$\\
    \begin{tabular}{c}
      $m$-photon \\input vertex
    \end{tabular} &
    \begin{minipage}{0.48\columnwidth}
      \centering
      \includegraphics[width=0.9\columnwidth]{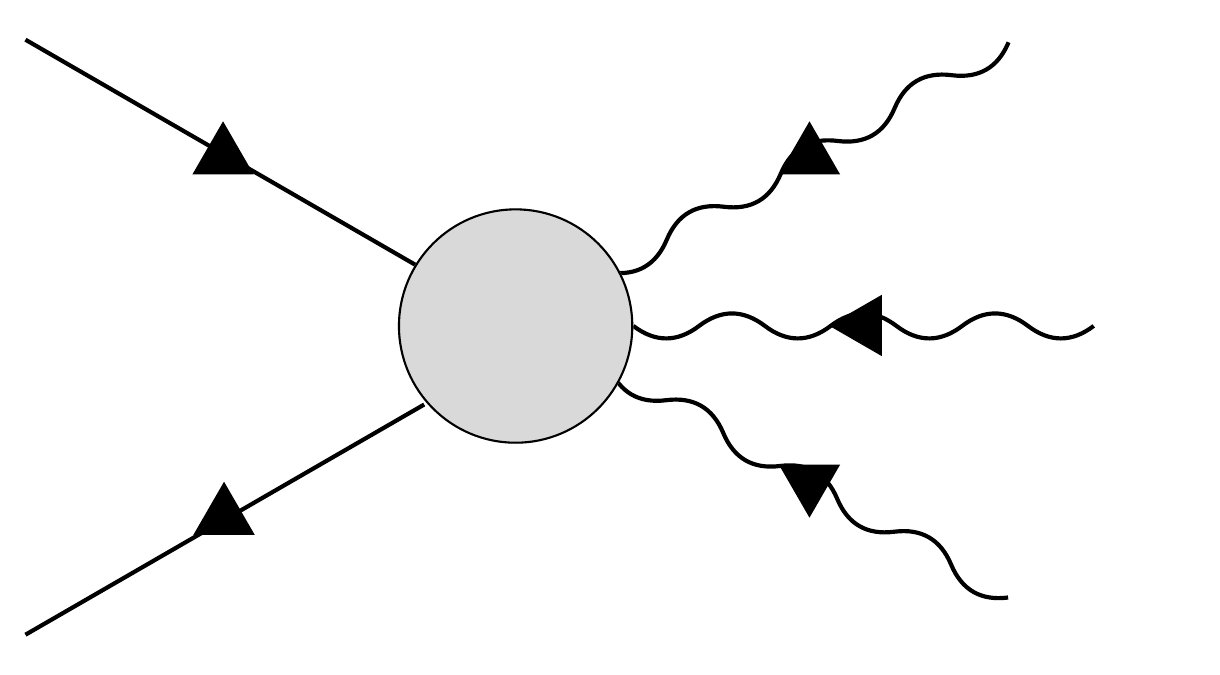}
    \end{minipage} & $\frac{1}{m!}\Gamma^{\alpha_{i_1}\cdots\alpha_{i_m}}$\\\hline\hline
  \end{tabular}
\end{table}

Continuing the above calculations, we obtain the following Feynman rules for $K^{\mu\alpha_1\cdots\alpha_n}(q_1,\cdots,q_n)$:
\begin{enumerate}[(1)]
  \setlength{\parskip}{0cm}
  \setlength{\itemsep}{0cm}
  \item Each diagram contains $n+1$ external photon lines.
  \begin{enumerate}
    \setlength{\parskip}{0cm}
    \setlength{\itemsep}{0cm}
    \item [(1a)] Among them, one is the outgoing line $(\mu,q_1+\cdots+q_n)$ and the others are incoming lines $(\alpha_i,q_i)$. Whole diagrams are symmetric about the $n$ input lines.
    \item [(1b)] A vertex containing the outgoing photon line is called an output vertex. All the others are called input vertices.
  \end{enumerate}
  \item Internal electron lines form a loop.
  \item The entire diagram comes with a factor of $(-1)^n/2$.
\end{enumerate}
For example, the linear response kernel is given by 
\begin{widetext}
\begin{align}
  K^{\mu\nu}(q)&=\frac{1}{2}\bigg(-\mathrm{Tr}\big[\gamma^\mu(\overline{k}+q,-q)G(\overline{k}+q)\Gamma^\nu(\overline{k},q)G(\overline{k})\big]-\mathrm{Tr}\big[\gamma^{\mu\nu}(\overline{k},-q,q)G(\overline{k})\big]\bigg),\label{eq:lin_kernel_momentum}
  \end{align}
 and the second-order response kernel is 
\begin{align}
  K^{\mu\nu\lambda}(q_1,q_2)&=\frac{1}{2}\bigg(\frac{1}{2}\mathrm{Tr}\big[\gamma^{\mu\nu\lambda}(\overline{k},-q,q_1,q_2)G(\overline{k})\big]+\mathrm{Tr}\big[\gamma^{\mu\nu}(\overline{k}+q_2,-q,q_1)G(\overline{k}+q_2)\Gamma^\lambda(\overline{k},q_2)G(\overline{k})\big]\nonumber\\
  &\qquad+\frac{1}{2}\mathrm{Tr}\big[\gamma^\mu(\overline{k}+q,-q)G(\overline{k}+q)\Gamma^{\nu\lambda}(\overline{k},q_1,q_2)G(\overline{k})\big]\nonumber\\
  &\qquad+\mathrm{Tr}\big[\gamma^\mu(\overline{k}+q,-q)G(\overline{k}+q)\Gamma^\lambda(\overline{k}+q_1,q_2)G(\overline{k}+q_1)\Gamma^\nu(\overline{k},q_1)G(\overline{k})\big]+[(\nu,q_1)\leftrightarrow(\lambda,q_2)]\bigg).\label{eq:sec_kernel_momentum}
\end{align}
\end{widetext}
These results agree with the Fourier transform of Eqs.~\eqref{eq:lin_kernel_real} and \eqref{eq:sec_kernel_real}.
The explicit formula for the $n$-th order response kernel can also be obtained by drawing all the diagrams that obey these Feynman rules and assigning the values corresponding to each diagram summarized in Table~\ref{tab:feynman_rule}.

\subsection{Gauge invariance of response kernel}
\label{sec:gaugeinvariance}
In this section, we demonstrate the gauge invariance of the electromagnetic response kernel obtained above based on the Ward identity.

The gauge transformation
\begin{align}
A_{\alpha_i}(q_i)\to A_{\alpha_i}(q_i)-i(q_i)_{\alpha_i}\theta(q_i)
\end{align}
generates various additional terms in the expression of the current response in Eq.~\eqref{eq:expval_current}.
For the gauge invariance, all these terms must vanish.
This requires
\begin{align}
  K^{\mu\alpha_1\cdots\alpha_n}(q_1,\cdots,q_n)(q_n)_{\alpha_n}=0.
\end{align}
Let us show that the electromagnetic response kernel constructed above satisfies this condition. 
For example, the linear  response kernel in Eq.~\eqref{eq:lin_kernel_momentum} satisfies
\begin{widetext}
\begin{align}
  2K^{\mu\nu}(q)q_\nu&=-\mathrm{Tr}\big[\gamma^\mu(\overline{k}+q,-q)G(\overline{k}+q)\Gamma^\nu(\overline{k},q)q_\nu G(\overline{k})+\gamma^{\mu\nu}(\overline{k},-q,q)q_\nu G(\overline{k})\big]\nonumber\\
    &=-\mathrm{Tr}\big[\gamma^\mu(\overline{k}+q,-q)G(\overline{k}+q)\left(G^{-1}(\overline{k}+q)\tau_3-\tau_3G^{-1}(\overline{k})\right)G(\overline{k})\big]\nonumber\\
    &\qquad-\mathrm{Tr}\big[\left(\tau_3\gamma^\mu(\overline{k},-q)-\gamma^\mu(\overline{k}+q,-q)\tau_3\right)G(\overline{k})\big]\nonumber\\
    &=0.
\end{align}
In going to the second line, we used the Ward identity for the bare two-photon vertices
\begin{align}
  \gamma^{\mu\nu}(k,q_1,q_2)(q_2)_{\nu}=\tau_3\gamma^\mu(k,q_1)-\gamma^\mu(k+q_2,q_1)\tau_3
\end{align}
and the Ward identity for the full one-photon vertices
\begin{align}
  \Gamma^\nu(k,q)q_\nu=G^{-1}(k+q)\tau_3-\tau_3G^{-1}(k).
\end{align}
Also, for the second-order response kernel,
\begin{align}
  2K^{\mu\nu\lambda}(q_1,q_2)(q_2)_\lambda&=\mathrm{Tr}\big[\gamma^{\mu\nu\lambda}(\overline{k},-q,q_1,q_2)(q_2)_\lambda G(\overline{k})\big]+\mathrm{Tr}\big[\gamma^{\mu\nu}(\overline{k}+q_2,-q,q_1)G(\overline{k}+q_2)\Gamma^\lambda(\overline{k},q_2)(q_2)_\lambda G(\overline{k})\big]\nonumber\\
  &\quad+\mathrm{Tr}\big[\gamma^{\mu\lambda}(\overline{k}+q_1,-q,q_2)(q_2)_\lambda G(\overline{k}+q_1)\Gamma^\nu(\overline{k},q_1)G(\overline{k})\big]\nonumber\\
  &\quad+\mathrm{Tr}\big[\gamma^\mu(\overline{k}+q,-q)G(\overline{k}+q)\Gamma^{\nu\lambda}(\overline{k},q_1,q_2)(q_2)_\lambda G(\overline{k})\big]\nonumber\\
  &\quad+\mathrm{Tr}\big[\gamma^\mu(\overline{k}+q,-q)G(\overline{k}+q)\Gamma^\lambda(\overline{k}+q_1,q_2)(q_2)_\lambda G(\overline{k}+q_1)\Gamma^\nu(\overline{k},q_1)G(\overline{k})\big]\nonumber\\
  &\quad+\mathrm{Tr}\big[\gamma^\mu(\overline{k}+q,-q)G(\overline{k}+q)\Gamma^\nu(\overline{k}+q_2,q_1)G(\overline{k}+q_2)\Gamma^\lambda(\overline{k},q_2)(q_2)_\lambda G(\overline{k})\big]\nonumber\\
  &=\mathrm{Tr}\big[\big(\tau_3\gamma^{\mu\nu}(\overline{k},-q,q_1)-\gamma^{\mu\nu}(\overline{k}+q_2,-q,q_1)\tau_3\big) G(\overline{k})\big]\nonumber\\
  &\quad+\mathrm{Tr}\big[\gamma^{\mu\nu}(\overline{k}+q_2,-q,q_1)G(\overline{k}+q_2)(G^{-1}(\overline{k}+q_2)\tau_3-\tau_3G^{-1}(\overline{k}))G(\overline{k})\big]\nonumber\\
  &\quad+\mathrm{Tr}\big[\big(\tau_3\gamma^\mu(\overline{k}+q_1,-q)-\gamma^\mu(\overline{k}+q,-q)\tau_3\big)G(\overline{k}+q_1)\Gamma^\nu(\overline{k},q_1)G(\overline{k})\big]\nonumber\\
  &\quad+\mathrm{Tr}\big[\gamma^\mu(\overline{k}+q,-q)G(\overline{k}+q)\big(\tau_3\Gamma^\nu(\overline{k},q_1)-\Gamma^\nu(\overline{k}+q_2,q_1)\tau_3\big)G(\overline{k})\big]\nonumber\\
  &\quad+\mathrm{Tr}\big[\gamma^\mu(\overline{k}+q,-q)G(\overline{k}+q)\big(G^{-1}(\overline{k}+q)\tau_3-\tau_3G^{-1}(\overline{k}+q_1)\big)G(\overline{k}+q_1)\Gamma^\nu(\overline{k},q_1)G(\overline{k})\big]\nonumber\\
  &\quad+\mathrm{Tr}\big[\gamma^\mu(\overline{k}+q,-q)G(\overline{k}+q)\Gamma^\nu(\overline{k}+q_2,q_1)G(\overline{k}+q_2)\big(G^{-1}(\overline{k}+q_2)\tau_3-\tau_3G^{-1}(\overline{k})\big)G(\overline{k})\big]\nonumber\\
  &=0.
\end{align}
\end{widetext}

Even for higher-order responses, one can prove their gauge invariance in the same way regardless of the specific details of the model.

\subsection{Optical responses in many-body systems}
Next, we discuss the optical response as the $\bm{q}\to \bm{0}$ limit of the electromagnetic response.
For this purpose, we assume a spatially uniform and time-dependent electric field $\bm{E}$.
Since the gauge invariance of electromagnetic responses has been confirmed, we can fix the gauge to the one convenient for our purpose
\begin{align}
  A_0(q)=0, ~\bm{A}(q)=\frac{\bm{E}(i\Omega)}{i(i\Omega)},
\end{align}
which is called the velocity gauge~\cite{Sipe2000,Parker2019}.

In this case, the $n$-th order current in Eq.~\eqref{eq:expval_current} can be rewritten as 
\begin{align}
  &\langle J_A^\mu(i\Omega)\rangle=\sum_{n=0}^\infty\frac{1}{n!}\sigma^{\mu\alpha_1\cdots\alpha_n}(i\Omega_1,\cdots,i\Omega_n)\nonumber\\
  &\qquad\times \delta(i\Omega-\sum_{j=1}^ni\Omega_j)E_{\alpha_1}(i\Omega_1)\cdots E_{\alpha_n}(i\Omega_n), 
\end{align}
where the $n$-th order optical conductivity is defined as
\begin{align}
  &\sigma^{\mu\alpha_1\cdots\alpha_n}(i\Omega_1,\cdots,i\Omega_n)\nonumber\\
  &=\bigg(\prod_{j=1}^n\frac{1}{i(i\Omega_j)}\bigg)\left.K^{\mu\alpha_1\cdots\alpha_n}(q_1,\cdots,q_n)\right|_{\bm{q_i=0}}.
\end{align}
The frequency response of the system is obtained via analytic continuation $i\Omega_i\to\omega_i+i\eta$, where $\eta$ is an infinitesimal positive parameter.

\section{Applications to superconductors}
\label{sec:application}

In this section, we apply the electromagnetic response theory developed in Sec.~\ref{sec:general_theory} to superconductors.

\subsection{The mean-field theory of superconductors}
First, we review the mean-field theory of superconductors.
We assume the microscopic action $S[A_\mu,\bar{\psi},\psi]$ possesses local $U(1)$ symmetry and the interaction part $S_{\mathrm{int}}[\bar{\psi},\psi]$ does not depend on the gauge field as shown in Eqs.~\eqref{ActionInvariance} and~\eqref{eq:action_int}.
Under these assumptions, the microscopic interaction between electrons should take the form
\begin{align}
  \hat{H}_{\mathrm{int}}=-\frac{1}{2}\sum_{\bm{x,y}}\sum_{\alpha\beta\gamma\delta}V^{\alpha\beta\gamma\delta}(\bm{x-y})\hat{c}_{\bm{x}\alpha}^\dagger\hat{c}_{\bm{y}\beta}^\dagger\hat{c}_{\bm{y}\gamma}\hat{c}_{\bm{x}\delta},
  \label{eq:gi_interaction}
\end{align}
where indices $\alpha$,$\beta$,$\gamma$, and $\delta$ represent internal degrees of freedom.
Due to the anticommutation relation of the electrons, $V^{\beta\alpha\delta\gamma}(\bm{y-x})=V^{\alpha\beta\gamma\delta}(\bm{x-y})$ should hold.
In momentum space, this interaction can be expressed as
\begin{align}
  \hat{H}_{\mathrm{int}}=-\frac{1}{2N}&\sum_{\bm{k,p,q}}\sum_{\alpha\beta\gamma\delta}V^{\alpha\beta\gamma\delta}(\bm{k-p})\nonumber\\
    &\times\hat{c}_{\bm{k+q}\alpha}^\dagger\hat{c}_{\bm{-k+q}\beta}^\dagger\hat{c}_{\bm{-p+q}\gamma}\hat{c}_{\bm{p+q}\delta},
\end{align}
where $V^{\alpha\beta\gamma\delta}(\bm{k-p})$ is the Fourier transform of $V^{\alpha\beta\gamma\delta}(\bm{x-y})$.
Applying the mean-field approximation in the Cooper channel, we define the gap function as
\begin{align}
  [\Delta(\bm{k})]_{\alpha\beta}=-\frac{1}{N}&\sum_{\bm{p}}\sum_{\gamma\delta}V^{\alpha\beta\gamma\delta}(\bm{k-p})\langle\hat{c}_{\bm{-p}\gamma}\hat{c}_{\bm{p}\delta}\rangle.
\end{align}
With this approximation, the interaction term is simplified to
\begin{align}
  \hat{H}_{\mathrm{int}}\to\frac{1}{2}\sum_{k}\sum_{\alpha\beta}\bigg(\hat{c}_{\bm{k}\alpha}^\dagger[\Delta(\bm{k})]_{\alpha\beta}\hat{c}_{\bm{-k}\beta}^\dagger+h.c.\bigg).
\end{align}
Adding this to the normal-state Hamiltonian
\begin{align}
  \hat{H}_{N}=\sum_{\bm{k}}\sum_{\alpha\beta}\hat{c}_{\bm{k}\alpha}^\dagger [H_N(\bm{k})]_{\alpha\beta}\hat{c}_{\bm{k}\beta},
\end{align}
we obtain the mean-field Hamiltonian in the Bogoliubov-de Gennes (BdG) form:
\begin{align}
  \hat{H}_{\mathrm{MF}}=\frac{1}{2}\sum_{\bm{k}}\hat{\psi}_{\bm{k}}^\dagger H_{\mathrm{BdG}}(\bm{k})\hat{\psi}_{\bm{k}},
\end{align}
where
\begin{align}
  H_{\mathrm{BdG}}(\bm{k})=
  \begin{pmatrix}
    H_N(\bm{k}) & \Delta(\bm{k})\\
    \Delta^\dagger(\bm{k}) & -[H_N(-\bm{k})]^T
  \end{pmatrix},
\end{align}
is the BdG Hamiltonian and
\begin{align}
  \hat{\psi}_{\bm{k}}=\begin{pmatrix}
    \hat{\bm{c}}_{\bm{k}}\\
    (\hat{\bm{c}}_{\bm{-k}}^\dagger)^T
  \end{pmatrix}
\end{align} 
is the Nambu spinor.
This mean-field Hamiltonian no longer has $U(1)$ symmetry.

Next, we introduce the gauge field $A_\mu$ into the system to discuss electromagnetic responses later.
In general, the gauge field breaks translational symmetry, so it is necessary to discuss the mean-field theory of superconductors in real space.
Since we assume the gauge field does not affect the interaction term, the interaction in the presence of the gauge field remains identical to Eq.~\eqref{eq:gi_interaction}.
Under the mean-field approximation in the Cooper channel, the real space representation of the gap function is defined as
\begin{align}
  [\Delta_{A}(\bm{x,y})]_{\alpha\beta}=-\sum_{\gamma\delta}V^{\alpha\beta\gamma\delta}(\bm{x-y})\langle\hat{c}_{\bm{y}\gamma}\hat{c}_{\bm{x}\delta}\rangle.
\end{align}
This gap function depends on $A_\mu$, as the gauge field modifies the normal-state Hamiltonian $\hat{H}_N\to\hat{H}_{N,A}$, and consequently alters the expectation value $\langle\hat{c}_{\bm{y}\gamma}\hat{c}_{\bm{x}\delta}\rangle$.
The expectation value can be rewritten using the Green function as
\begin{align}
  \langle\hat{c}_{\bm{y}\gamma}\hat{c}_{\bm{x}\delta}\rangle&=-\langle T_\tau\hat{c}_{\bm{x}\delta}(0^-)\hat{c}_{\bm{y}\gamma}(0)\rangle\nonumber\\
  &=\mathrm{Tr}\big[P_{\gamma\delta}G_A(\bm{x,y};0^-)\big].
\end{align}
where we defined the matrices
\begin{align}
  P_{\gamma\delta}=\begin{pmatrix}
    O&O\\
    E_{\gamma\delta}&O
  \end{pmatrix}
\end{align}
and $[E_{\gamma\delta}]_{ab}=\delta_{a\gamma}\delta_{b\delta}$.
The self-energy is given by
\begin{align}
  \Sigma_A(x,y)=\begin{pmatrix}
    O&\Delta_A(x,y)\\
    \Delta_A^\dagger(y,x)&O
  \end{pmatrix}
  \label{eq:se_sc}
\end{align}
where
\begin{align}
  [\Delta_A(x,y)]_{\alpha\beta}=-\sum_{\gamma\delta}V^{\alpha\beta\gamma\delta}(x-y)\mathrm{Tr}\big[P_{\gamma\delta}G_A(x,y)\big]
  \label{eq:def_mf}
\end{align}
and $V^{\alpha\beta\gamma\delta}(x-y)=V^{\alpha\beta\gamma\delta}(\bm{x-y})\delta(x_0-y_0-0^-)$.

\subsection{The Ward identities under the mean-field approximation}
We now turn to the discussion of electromagnetic responses.
While it may seem straightforward to apply the gauge-invariant theory for many-body systems summarized earlier, it is necessary to ensure one crucial condition is met.
Namely, we must examine whether the Ward identities are consistent with the approximation employed.

The whole Ward identities for photon vertices are derived by the functional derivatives of Eq.~\eqref{general_wi}.
Plugging the Dyson equation in Eq.\eqref{eq:dyson_real} into Eq.~\eqref{general_wi}, we obtain the relation that should hold between the self-energies:
\begin{align}
  &\partial_{z^\mu}\fdv{\Sigma_A(x,y)}{A_\mu(z)}\nonumber\\
  &=i\delta(x-z)\tau_3\Sigma_A(x,y)-i\Sigma_A(x,y)\tau_3\delta(y-z).
  \label{wi_se}
\end{align}
Since the self-energy of superconductors is given by Eq.~\eqref{eq:se_sc}, the gap function should satisfy
\begin{align}
  &\partial_{z^\mu}\fdv{\Delta_A(x,y)}{A_\mu(z)}\nonumber\\
  &=i\Delta_A(x,y)\delta(y-z)+i\Delta_A(x,y)\delta(x-z).
\end{align}
We can prove that this relation is compatible with the adopted approximations as follows, using the definition of the gap function in Eq.~\eqref{eq:def_mf}:
\begin{widetext}
\begin{align}
  &\partial_{z^\mu}\fdv{[\Delta_A(x,y)]_{\alpha\beta}}{A_\mu(z)}=\sum_{\gamma\delta}V^{\alpha\beta\gamma\delta}(x-y)\partial_{z^\mu}\mathrm{Tr}\big[P_{\gamma\delta}G_A(x,\overline{x}')\fdv{G_A^{-1}(\overline{x}',\overline{y}')}{A_\mu(z)}G_A(\overline{y}',y)\big]\nonumber\\
  &=\sum_{\delta\gamma}V^{\alpha\beta\gamma\delta}(x-y)\partial_{z^\mu}\mathrm{Tr}\big[P_{\gamma\delta}G_A(x,\overline{x}')\big(i\delta(\overline{x}'-z)\tau_3G_A^{-1}(\overline{x}',\overline{y}')-iG_A^{-1}(\overline{x}',\overline{y}')\tau_3\delta(\overline{y}'-z)\big)G_A(\overline{y}',y)\big]\nonumber\\
  &=i[\Delta_A(x,y)]_{\alpha\beta}\delta(y-z)+i[\Delta_A(x,y)]_{\alpha\beta}\delta(x-z).\label{eq:wi_verify}
\end{align}
\end{widetext}
Therefore, it is possible to apply the electromagnetic theory above to superconductors which are originating from the microscopic interactions given by Eq.~\eqref{eq:gi_interaction}.

The calculation of electromagnetic responses can be done given the Green function, bare photon vertices, and full photon vertices.
The Green function is obtained by
\begin{align}
  G^{-1}(k)=k_0\tau_0-H_{\mathrm{BdG}}(\bm{k}).
\end{align}
The bare and full photon vertices are obtained by definitions in Eqs.~\eqref{eq:def_barevertex} and~\eqref{eq:def_fullvertex} and they automatically satisfy the Ward identities.

For instance, let us consider the photon vertices in the velocity gauge.
Since the gauge field is introduced by the minimal coupling $\bm{k}\to\bm{k+A}$ for the electron part $H_N(\bm{k})$ and $\bm{-k}\to\bm{-k+A}$ for the hole part  $H_N(-\bm{k})^T$, the bare $n$-photon vertices are given by
\begin{align}
  \gamma^{\alpha_1\cdots\alpha_n}(k)=\prod_{i=1}^n(-\tau_3\partial_{k_i})
  \begin{pmatrix}
    H_N(\bm{k})&\\
    &-[H_N(-\bm{k})]^T
  \end{pmatrix}.
\end{align}

The full $n$-photon vertex $\Gamma^{\alpha_1\cdots\alpha_n}$ can be decomposed into the sum of the bare $n$-photon vertex $\gamma^{\alpha_1\dots\alpha_n}$ and the correction part $\Lambda^{\alpha_1\cdots\alpha_n}$ as shown in Eq.~\eqref{eq:decompose_fullvertex}.
The latter is obtained by the functional derivatives of the self-energy as in Eq.~\eqref{eq:def_correction}.
Since the self-energy $\Sigma_A[G_A]$ is a functional of the Green function, the functional derivatives of the self-energy $\delta\Sigma/\delta A$ are related to the functional derivatives of the Green function $\delta G/\delta A$, which is again related to the full photon vertex.
Therefore, the full photon vertices are given by the solutions to some integral equations.

The concrete form of the integral equation for the full photon vertex is determined by specifying the form of the self-energy.
The Fock approximation leads to the Bethe-Salpeter equation for the correction part $\Lambda^\mu$, as we review in Sec.~\ref{BCS}.
However, this method has a subtlety associated to the diagonal components of the self-energy.
Instead, we use the more precise expression of the self-energy in Eq.~\eqref{eq:def_mf}.
The correction part of the full one-photon vertex $\Lambda^\mu=-\delta\Sigma_A/\delta A_\mu|_{A=0}$ is then directly calculated by the functional derivatives of the gap equation.
This approach at the linear response level for BCS superconductors was called ``consistent fluctuations of order parameters`` (CFOP) method in Ref.~\cite{Guo2013}, whose name reflects the fact that $\delta\Delta_A/\delta A_\mu$ represents the fluctuations of the gap function induced by the gauge field. 
Similar discussions of linear responses of BCS superconductors can be found in Refs.~\cite{Kulik1981,Zha1995}.
Our derivation in this work is more general and is capable of handling responses of arbitrary order and more general superconductors as long as the microscopic interaction leading to the superconductivity takes the form of Eq.~\eqref{eq:gi_interaction}.
Hence, we call our method generalized CFOP method and study the optical responses in superconductors in this framework.
The concrete form of the integral equations for the full photon vertices are presented in Sec.~\ref{sec:examples}.
In Sec.~\ref{subsec:multi_band}, we will see that our method at the linear order level is equivalent to the random phase approximations addressed in Refs.~\cite{Kamatani2022,Nagashima2024,Nagashima2024a} in the different context.

\subsection{Finite momentum Cooper pairing}
\label{sec:finite_momentum}
Our formulation can also be applied to the case where the Cooper pairs have a finite momentum $2\bm{Q}$.
In this situation, the supreconducting gap is assumed to be
\begin{align}
  [\Delta(\bm{k})]_{\alpha\beta}=-\frac{1}{N}&\sum_{\bm{p}}\sum_{\gamma\delta}V^{\alpha\beta\gamma\delta}(\bm{k-p})\langle\hat{c}_{\bm{-p+Q}\gamma}\hat{c}_{\bm{p+Q}\delta}\rangle,
\end{align}
and the interaction term is reduced to
\begin{align}
  \hat{H}_{\mathrm{int}}\to\sum_{\bm{k}}\sum_{\alpha\beta}\bigg(\hat{c}_{\bm{k+Q}\alpha}^\dagger[\Delta(\bm{k})]_{\alpha\beta}\hat{c}_{\bm{-k+Q}\beta}^\dagger+h.c.\bigg).
\end{align}
The mean-field Hamiltonian is given by
\begin{align}
  \hat{H}_{\mathrm{MF}}=\frac{1}{2}\sum_{\bm{k}}(\hat{\psi}_{\bm{k}}^{\bm{Q}})^\dagger H_{\mathrm{BdG}}^{\bm{Q}}(\bm{k})\hat{\psi}_{\bm{k}}^{\bm{Q}},
\end{align}
where
\begin{align}
  H_{\mathrm{BdG}}^{\bm{Q}}(\bm{k})=
  \begin{pmatrix}
    H_N(\bm{k+Q})&\Delta(\bm{k})\\
    \Delta^\dagger(\bm{k})&-H_N^T(\bm{-k+Q})
  \end{pmatrix}
\end{align}
and $\hat{\psi}_{\bm{k}}^{\bm{Q}}=(\hat{\bm{c}}_{\bm{k+Q}}^T,\hat{\bm{c}}_{\bm{-k+Q}}^\dagger)^T$.

To consider electromagnetic responses, we also present the case where the gauge field is present and translational symmetry is broken.
The corresponding mean-field approximation in real space can be formulated by defining
\begin{align}
  \hat{\bm{c}}_{\bm{x}}^{\bm{Q}}=\hat{\bm{c}}_{\bm{x}}e^{-i\bm{Q}\cdot\bm{x}}, (\hat{\bm{c}}_{\bm{x}}^{\bm{Q}})^\dagger=\hat{\bm{c}}_{\bm{x}}^\dagger e^{i\bm{Q}\cdot\bm{x}}.
\end{align}
The interaction term in Eq.~\eqref{eq:gi_interaction} can be rewritten as
\begin{align}
  \hat{H}_{\mathrm{int}}=\sum_{\bm{x,y}}\sum_{\alpha\beta\gamma\delta}V^{\alpha\beta\gamma\delta}(\bm{x-y})(\hat{c}_{\bm{x}\alpha}^{\bm{Q}})^\dagger(\hat{c}_{\bm{y}\beta}^{\bm{Q}})^\dagger\hat{c}_{\bm{y}\gamma}^{\bm{Q}}\hat{c}_{\bm{x}\delta}^{\bm{Q}}.
\end{align}
The gap function in real space is defined by
\begin{align}
  [\Delta_A(\bm{x,y})]_{\alpha\beta}=\sum_{\gamma\delta}V^{\alpha\beta\gamma\delta}(\bm{x-y})\langle\hat{c}_{\bm{y}\gamma}^{\bm{Q}}\hat{c}_{\bm{x}\delta}^{\bm{Q}}\rangle.
\end{align}
Redefining the Nambu spinor in real space as
\begin{align}
  \hat{\psi}_{\bm{x}}^{\bm{Q}}=
  \begin{pmatrix}
    \hat{\bm{c}}_{\bm{x}}^{\bm{Q}}\\
    ((\hat{\bm{c}}_{\bm{x}}^{\bm{Q}})^\dagger)^T
  \end{pmatrix},
\end{align}
the self-energy is expressed in exactly the same form as Eqs.~\eqref{eq:se_sc} and~\eqref{eq:def_mf}.
Therefore, one can show the Ward identities are consistent with this approximation and discuss the gauge-invariant electromagnetic responses even in the situation where Cooper pairs have the finite momentum.

\subsection{Spin rotational symmetry}
If we further assume spin-singlet pairing and spin rotational symmetry,
the size of the BdG Hamiltonian and the Nambu spinor can be halved.
Since the introduction of the gauge field is independent of the spin operations, this dimensional reduction can be done even in the presence of the gauge field.
For simplicity, we do not introduce the gauge field here.

We consider the degrees of freedom defined by sublattices $l=1,\cdots,n$ and spin $s={\uparrow},{\downarrow}$.
The spin operator $\hat{S}^i_{\bm{x}l}$ on site $\bm{x}$ and sublattice $l$ is defined by
\begin{align}
  \hat{S}^i_{\bm{x}l}=
  \begin{pmatrix}
    \hat{c}_{\bm{x}l\uparrow}^\dagger&\hat{c}_{\bm{x}l\downarrow}^\dagger
    \end{pmatrix}\frac{\sigma_i}{2}
    \begin{pmatrix}
      \hat{c}_{\bm{x}l\uparrow}\\
      \hat{c}_{\bm{x}l\downarrow}
    \end{pmatrix},
\end{align}
where $\sigma_i$ are the Pauli matrices that act on the spin space.
The total spin operator is given by $\hat{S}^i=\sum_{\bm{x},l}\hat{S}^i_{\bm{x}l}$.
The spin rotations are generated by $\hat{S}^i$, and a spin rotation about axis $\bm{n}$ by an angle $\phi$ is represented by
\begin{align}
  \hat{U}(\bm{n},\phi)=\exp(i\phi\bm{n}\cdot\hat{\bm{S}}).
\end{align}
This operation does not change the sublattice degrees of freedom.

Assuming spin rotational symmetry, the normal-state Hamiltonian takes the form
\begin{align}
  \hat{H}_N=\sum_{\bm{x,y}}\sum_{l,l'}\sum_{s}\hat{c}_{\bm{x}ls}^\dagger[\tilde{H}_{N}(\bm{x,y})]_{ll'}\hat{c}_{\bm{y}l's}.
\end{align}
If the interaction term is also invariant under spin rotation, it consists of the interactions such as spin-spin interaction
\begin{align}
  V_S^{l_1l_2l_3l_4}(\bm{x-y})(\bm{\sigma})_{s_1s_4}\cdot(\bm{\sigma})_{s_2s_3}
\end{align}
and density-density interaction
\begin{align}
  V_n^{l_1l_2l_3l_4}(\bm{x-y})(\sigma_0)_{s_1s_4}(\sigma_0)_{s_2s_3}.
\end{align}
For spin-singlet pairing, we assume the interaction term $\hat{H}_{\mathrm{int}}$ is represented by
\begin{align}
  V^{\alpha\beta\gamma\delta}(\bm{x-y})=V^{l_1l_2l_3l_4}(\bm{x-y})(i\sigma_2)_{s_1s_2}(i\sigma_2)^\dagger_{s_3s_4},
\end{align}
where $V^{l_1l_2l_3l_4}(\bm{x-y})=V^{l_2l_1l_4l_3}(\bm{y-x})$.
This pairing interaction is invariant under spin rotation due to the equality
\begin{align}
  &(i\sigma_2)_{s_1s_2}(i\sigma_2)^\dagger_{s_3s_4}\nonumber\\
  &=-\frac{1}{2}\big[(\bm{\sigma})_{s_1s_4}\cdot(\bm{\sigma})_{s_2s_3}-(\sigma_0)_{s_1s_4}(\sigma_0)_{s_2s_3}\big].
\end{align}
Therefore, this microscopic Hamiltonian for spin-singlet pairing possesses spin rotational symmetry:
\begin{align}
  \hat{U}(\bm{n},\phi)(\hat{H}_N+\hat{H}_{\mathrm{int}})\hat{U}^\dagger(\bm{n},\phi)=\hat{H}_N+\hat{H}_{\mathrm{int}}.
\end{align}

Performing mean-field approximation, we define the gap function as
\begin{align}
  [\Delta(\bm{x,y})]_{l_1s_1,l_2s_2}&=[\tilde{\Delta}(\bm{x,y})]_{l_1l_2}(i\sigma_2)_{s_1s_2},
\end{align}
where
\begin{align}
  [\tilde{\Delta}(\bm{x,y})]_{l_1l_2}&=-\sum_{l_3s_3l_4s_4}V^{l_1l_2l_3l_4}(\bm{x-y})\nonumber\\
  &\qquad\times(i\sigma_2)^\dagger_{s_3s_4}\langle\hat{c}_{\bm{y}l_3s_3}\hat{c}_{\bm{x}l_4s_4}\rangle,
\end{align}
and the mean-field Hamiltonian is written as
\begin{align}
  \hat{H}_{\mathrm{MF}}=\frac{1}{2}\sum_{\bm{x,y}}\hat{\psi}_{\bm{x}}^\dagger H_{\mathrm{BdG}}(\bm{x,y})\hat{\psi}_{\bm{y}},
\end{align}
where the BdG Hamiltonian is given by
\begin{align}
&H_{\mathrm{BdG}}(\bm{x,y})\nonumber\\
  &=\begin{pmatrix}
    \tilde{H}_{N}(\bm{x,y})\otimes\sigma_0&\tilde{\Delta}(\bm{x,y})\otimes(i\sigma_2)\\
    \tilde{\Delta}^\dagger(\bm{y,x})\otimes(i\sigma_2)^\dagger& -\tilde{H}_{N}^T(\bm{y,x})\otimes\sigma_0
  \end{pmatrix}.
\end{align}
Redefining the Nambu spinor as
\begin{align}
  \hat{\tilde{\psi}}_{\bm{x}}=
  \begin{pmatrix}
    \hat{\bm{c}}_{\bm{x}\uparrow}\\
    (\hat{\bm{c}}_{\bm{x}\downarrow}^\dagger)^T
  \end{pmatrix}
\end{align}
where $\hat{\bm{c}}_{\bm{x}s}=(\hat{c}_{\bm{x}1s},\cdots,\hat{c}_{\bm{x}ns})^T$, the mean-field Hamiltonian reduces to
\begin{align}
  \hat{H}_{\mathrm{MF}}=\sum_{\bm{x,y}}\hat{\tilde{\psi}}_{\bm{x}}^\dagger\tilde{H}_{\mathrm{BdG}}(\bm{x,y})\hat{\tilde{\psi}}_{\bm{y}}
\end{align}
with
\begin{align}
  \tilde{H}_{\mathrm{BdG}}(\bm{x,y})=
  \begin{pmatrix}
    \tilde{H}_{N}(\bm{x,y})&\tilde{\Delta}(\bm{x,y})\\
    \tilde{\Delta}^\dagger(\bm{y,x})&-\tilde{H}_{N}^T(\bm{y,x})
  \end{pmatrix}.
\end{align}
This redefinition avoids the particle-hole doubling.
The Green function and the self-energy are then defined as
\begin{align}
  &\tilde{G}(x,y)=-\langle T_\tau\hat{\tilde{\psi}}(x)\hat{\tilde{\psi}}^\dagger(y)\rangle\\
  &\tilde{\Sigma}(x,y)=
  \begin{pmatrix}
    &\tilde{\Delta}(x,y)\\
    \tilde{\Delta}^\dagger(y,x)&
  \end{pmatrix},
  \label{eq:su2_se}
\end{align}
where
\begin{align}
  [\tilde{\Delta}(x,y)]_{l_1l_2}=&-2\sum_{l_3l_4}V^{l_1l_2l_3l_4}(x-y)\mathrm{Tr}\big[P_{l_3l_4}\tilde{G}(x,y)\big]
  \label{eq:su2_gapeq}
\end{align}
and $V^{l_1l_2l_3l_4}(x-y)=V^{l_1l_2l_3l_4}(\bm{x-y})\delta(x_0-y_0-0^-)$.
This approach removes the $1/2$ factors often introduced in definitions of current and response kernels due to the particle-hole doubling.

\section{Examples}
\label{sec:examples}
In this section, we discuss electromagnetic responses of superconductors in several models.
The first example clarifies the relation between the generalized CFOP method and previous studies~\cite{Nambu1960,Kulik1981,Zha1995,Guo2013}.
The second and third examples, which are multi-band $s$-wave superconductors and single-band $d$-wave superconductors, are used to demonstrate the significance of vertex corrections in optical response.
Since all of these examples possess spin rotational symmetry and spin-singlet pairing is assumed,
we omit all tildes, for example, in the Green function $\tilde{G}$ and the Nambu spinor $\tilde{\psi}$.

\subsection{BCS single-band superconductors}\label{BCS}
In the BCS theory, the internal degree of freedom is given by spin $s={\uparrow},{\downarrow}$.
The interaction between electrons can be written as
\begin{align}
  V^{s_1s_2s_3s_4}(\bm{x-y})=\frac{1}{2}V(\bm{x-y})(i\sigma_2)_{s_1,s_2}(i\sigma_2)^\dagger_{s_3,s_4},
\end{align}
where $V(\bm{x-y})$ is real function.
Then, the self-energy in Eq.~\eqref{eq:def_mf} of this superconductor is given by
\begin{align}
  \Sigma_A(x,y)=\begin{pmatrix}
    &\Delta_A(x,y)\\
    \Delta_A^\dagger(y,x)&
  \end{pmatrix},\\
  \Delta_A(x,y)=-V(x-y)\mathrm{Tr}\big[PG_A(x,y)\big].\label{DVPG}
\end{align}

\begin{figure}[t]
  \centering
  \includegraphics[width=1\linewidth]{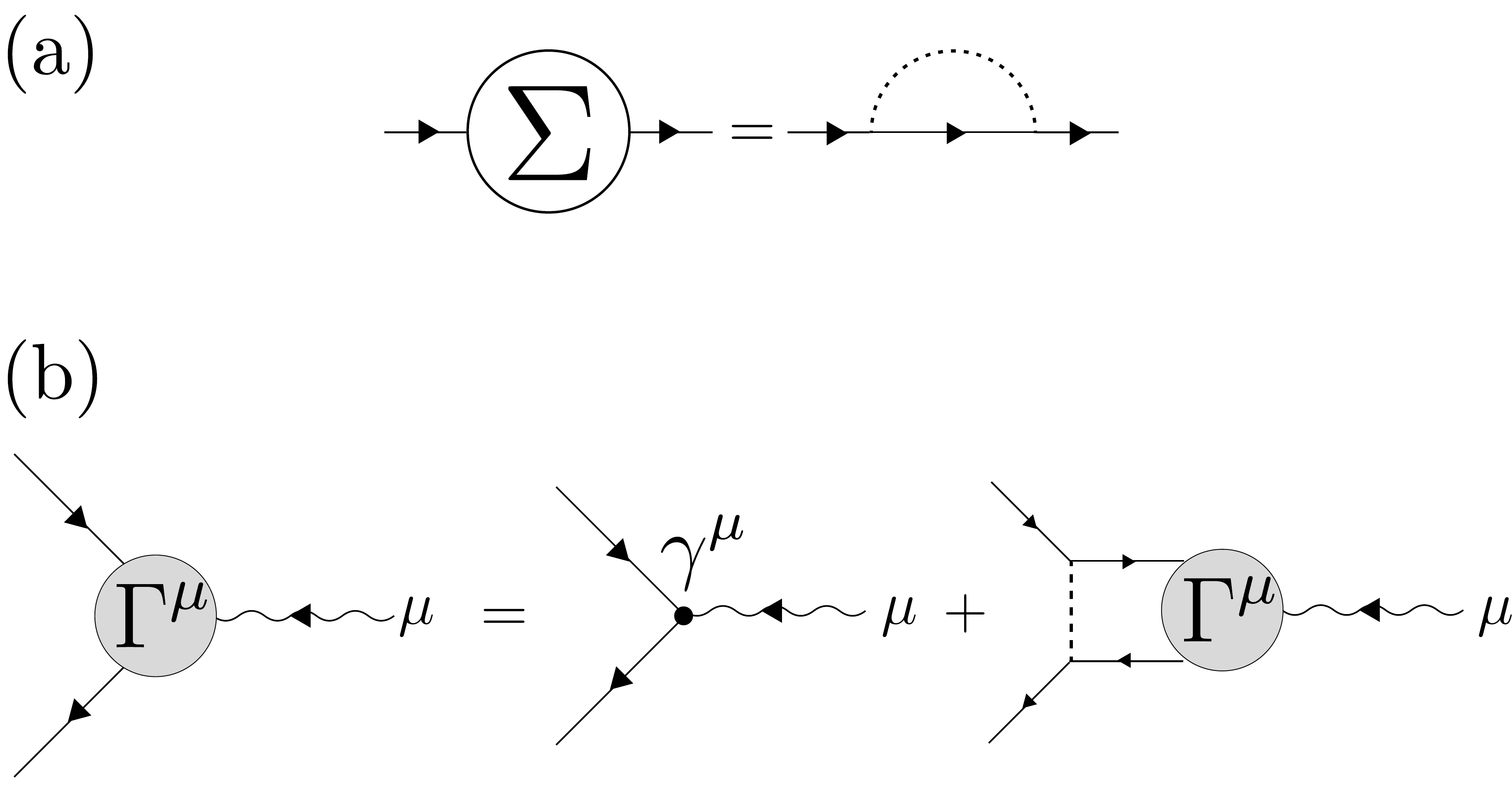}
  \caption{Green function approach to BCS superconductors. 
(a) Fock approximation to the self-energy. The dotted line represents the microscopic electron-electron interaction $V$. 
(b) Bethe--Salpeter equation, representing the ladder approximation to the bare one-photon vertex.}
  \label{fig:fock_bse}
\end{figure}

Instead of Eq.~\eqref{DVPG}, one often uses the Fock approximation to the self-energy illustrated in Fig.~\ref{fig:fock_bse}(a):
\begin{align}
  \Sigma_A(x,y)=V(x-y)\tau_3G_A(x,y)\tau_3\label{se_gf},
\end{align}
which, combined with Eqs.~\eqref{eq:def_fullvertex} and \eqref{eq:dyson_real}, gives the full one-photon vertex:
\begin{align}
  &\Gamma^\mu(x,y,z)=\gamma^\mu(x,y,z)-\left.\fdv{\Sigma_A(x,y)}{A_\mu(z)}\right|_{A=0}\nonumber\\
  &=\gamma^\mu(x,y,z)\nonumber\\
  &\quad+V(x-y)\tau_3G(x,\overline{x}')\Gamma^\mu(\overline{x}',\overline{y}',z)G(\overline{y}',y)\tau_3.
\end{align}
See Fig.~\ref{fig:fock_bse}(b) for the diagrammatic expression. 
Performing the Fourier transform, we obtain the Bethe-Salpeter equation~\cite{Nambu1960}:
\begin{align}
  \Gamma^\mu(k,q)=\gamma^\mu(k,q)+V(k-\overline{p})\tau_3G(\overline{p}+q)\Gamma^\mu(\overline{p},q)G(\overline{p})\tau_3.\label{eq:BSE_main}
\end{align}
While Nambu~\cite{Nambu1960} argued only the linear electromagnetic responses, our method can also derive its nonlinear extension, which will be discussed in Appendix~\ref{ap:BSE}.

It should be noted that Eq.~\eqref{se_gf} differs from Eq.~\eqref{DVPG} in its diagonal component.
This mismatch is often justified from the fact that the diagonal components of the self-energy only affect the band dispersion of the normal state and do not change the superconducting gap. However, the diagonal components of the self-energy may affect the calculation of the full photon vertices and electromagnetic responses.
For these reasons, we will not use this approach in this work.

The other examples are presented based on the generalized CFOP method.
We note that the generalized CFOP method yields the full photon vertex which is the same as the solution to the Bethe-Salpeter equation in the case of BCS single-band superconductors.

\subsection{Optical responses in multi-band superconductors}
\label{subsec:multi_band}
\subsubsection{Model}
\begin{figure}[t]
  \centering
  \includegraphics[width=0.5\linewidth]{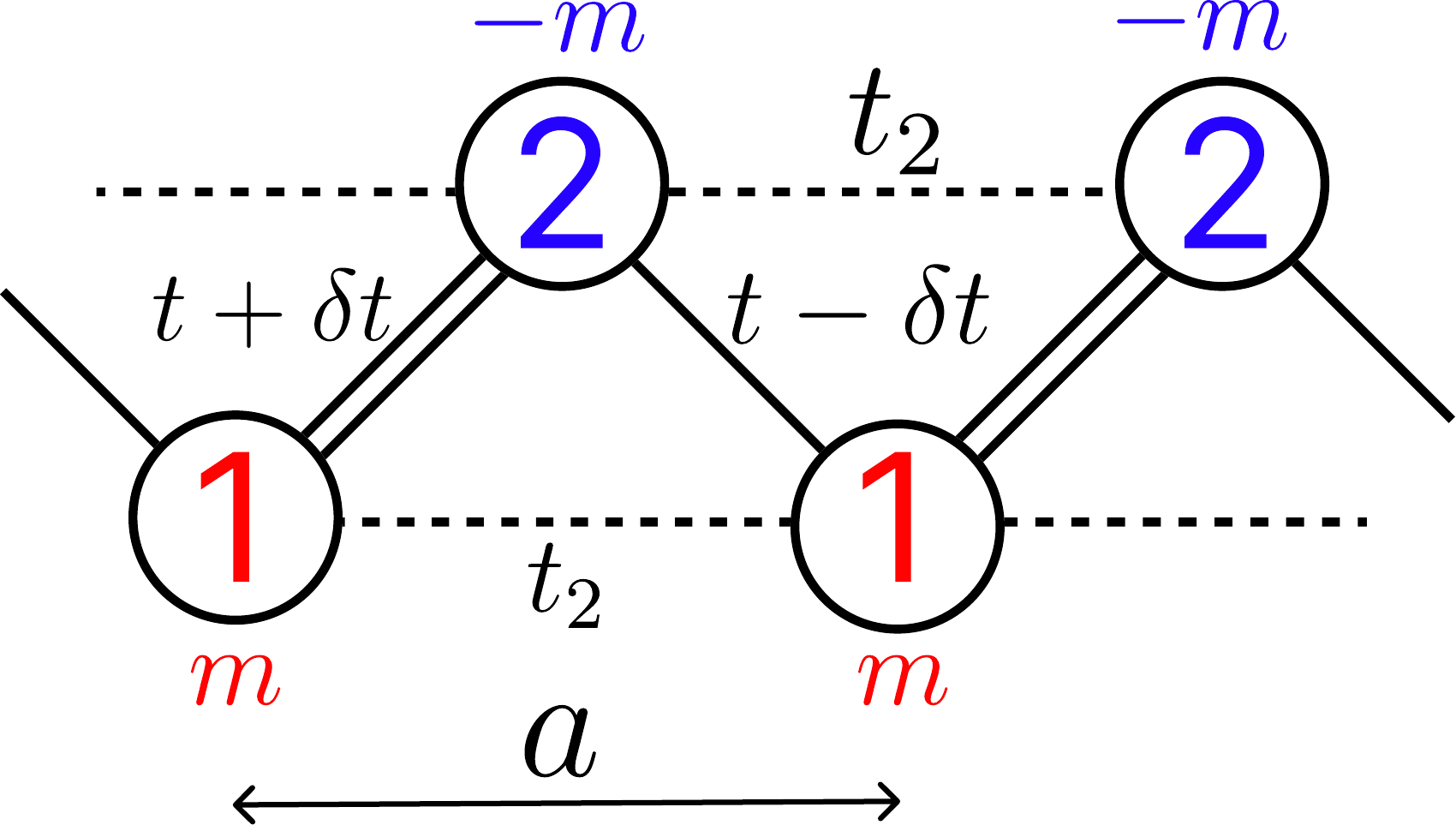}
\caption{Illustration of the Hamiltonian defined in Eq.~\eqref{eq:model}.}
  \label{fig:model}
\end{figure}

Let us consider a chain with the sublattice degrees of freedom $l=1,2$ as well as the spin degrees of freedom $s={\uparrow},{\downarrow}$.
We use the Rice-Mele model~\cite{Rice1982} with the next-nearest-neighbor hopping $t_2$ as the Hamiltonian in normal state~[Fig.\ref{fig:model}]:
\begin{align}
  \hat{H}_N&=\sum_{x,s}\bigg(\frac{1}{2}(t+\delta t)\hat{c}_{x,1s}^\dagger \hat{c}_{x,2s}+\frac{1}{2}(t-\delta t)\hat{c}_{x,2s}^\dagger \hat{c}_{x+1,1s}\nonumber\\
  &\qquad+\frac{t_2}{2}\hat{c}_{x,1s}^\dagger\hat{c}_{x+1,1s}+\frac{t_2}{2}\hat{c}_{x,2s}^\dagger\hat{c}_{x+1,2s}+h.c.\bigg)\nonumber\\
  &\qquad+\sum_{x,s}\bigg((m-\mu)\hat{n}_{x,1s}-(m+\mu)\hat{n}_{x,2s}\bigg).\label{eq:model}
\end{align}
We assume spin-singlet microscopic interactions
\begin{align}
  \hat{H}_{\mathrm{int}}=-\sum_{i,l}g_l\hat{c}_{x,l\uparrow}^\dagger\hat{c}_{x,l\downarrow}^\dagger\hat{c}_{x,l\downarrow}\hat{c}_{x,l\uparrow},
\end{align}
where $g_l>0$ is the coupling constant, for which we have
\begin{align}
  V^{l_1l_2l_3l_4}(x-y)=\frac{1}{2}\delta_{l_1l_2}\delta_{l_2l_3}\delta_{l_3l_4}\delta_{x,y}\,g_{l_1}.
  \label{eq:multiband_interaction}
\end{align}

We define the Fourier transform
\begin{align}
  &\hat{c}_{x,1s}=\frac{1}{\sqrt{N}}\sum_{k}\hat{c}_{k,1s}e^{ikxa},\\
  &\hat{c}_{x,2s}=\frac{1}{\sqrt{N}}\sum_{k}\hat{c}_{k,2s}e^{ik(x+1/2)a},
\end{align}
where $N$ denotes the number of unit cells and $a$ is the lattice constant.
The interaction term can be expressed as
\begin{align}
  \hat{H}_{\mathrm{int}}=-\frac{1}{N}\sum_{k,p,q,l}g_l\hat{c}_{k+q,l\uparrow}^\dagger\hat{c}_{-k+q,l\downarrow}^\dagger\hat{c}_{-p+q,l\downarrow}\hat{c}_{p+q,l\uparrow}
\end{align}
in momentum space.
We apply the mean-field approximation in the Cooper channel and define the gap function as
\begin{align}
  \Delta_l=-\frac{g_l}{N}\sum_{p}\langle\hat{c}_{-p+q,l\downarrow}\hat{c}_{p+q,l\uparrow}\rangle.
\end{align}
Thus, the mean-field Hamiltonian is expressed as 
\begin{align}
  \hat{H}_{\mathrm{MF}}&=\sum_k\hat{\psi}_k^\dagger\begin{pmatrix}
    H_N(k)&\Delta\\
    \Delta&-H_N(-k)^T
  \end{pmatrix}\hat{\psi}_k,
\end{align}
where
\begin{align}
  H_N(k)&=(t_2\cos ka-\mu)\sigma_0+t\cos(ka/2)\sigma_1\nonumber\\
  &\qquad-\delta t\sin(ka/2)\sigma_2+m\sigma_3,
\end{align}
and $\Delta=\mathrm{diag}\{\Delta_1,\Delta_2\}$.
In this model, previous studies have investigated the linear optical response of collective modes~\cite{Kamatani2022} and the linear and second-order optical responses without vertex corrections~\cite{Xu2019}.
For these specific examples, we apply the generalized CFOP method.
We investigate the effects of vertex corrections on the nonlinear optical response

Since our derivation of the Ward identity in Secs.~\ref{sec:general_theory} and \ref{sec:application} was for continuum models, let us check the Ward identity for lattice models
using the Rice-Mele model in Eq.~\eqref{eq:model} as an example.
The bare one-photon vertex of this model is given by
\begin{align}
  &\gamma(k,q)\nonumber\\
  &=\frac{a}{2}\tau_0\otimes\bigg[t_2\bigg(\sin(ka+\frac{qa}{4})+\sin(ka+\frac{3qa}{4})\bigg)\sigma_0\nonumber\\
  &\qquad +t\sin(\frac{ka}{2}+\frac{qa}{4})\sigma_1+\delta t\cos(\frac{ka}{2}+\frac{qa}{4})\sigma_2\bigg],
\end{align}
which satisfies
\begin{align}
  \frac{4}{a}\sin(\frac{qa}{4})\gamma(k,q)+q_0\tau_3=G_0^{-1}(k+q)\tau_3-\tau_3G_0^{-1}(k).
\end{align}
This equation corresponds to the Ward identity~\eqref{eq:wi_bare} for the bare one-photon vertex of continuum models.
The Ward identities for the correction parts in Eq.~\eqref{eq:wi_correction} are also satisfied.
This is because the proof in the continuum model presented in Eq.~\eqref{eq:wi_verify} remains valid even if the spatial derivative $\partial_z$ is replaced by the lattice difference operator.

\subsubsection{Calculations of the full photon vertices based on the CFOP method}
Let us introduce the gauge field and consider the electromagnetic responses.
Substituting Eq.~\eqref{eq:multiband_interaction} into Eq.~\eqref{eq:su2_gapeq}, the gap function is given by
\begin{align}
  &[\Delta_{A}(x,y)]_{ll'}=\delta_{ll'}\Delta_{l,A}(x,y),\\
  &\Delta_{l,A}(x,y)=-V_l(x-y)\mathrm{Tr}\big[P_{ll} G_A(x,y)\big],
\end{align}
where $V_l(x-y)=g_l\delta(\bm{x-y})\delta(x_0-y_0-0^-)$.
Let us separate the gap function into the real part and the imaginary part as $\Delta_{l,A}(x,y)=\Delta_{1,l,A}(x,y)-i\Delta_{2,l,A}(x,y)$, 
where each part is determined by the gap equation
\begin{align}
  \Delta_{i,l,A}(x,y)=-\frac{V_l(x-y)}{2}\mathrm{Tr}\big[\tau_i\otimes E_{ll}G_A(x,y)\big].\label{eq:gapeq_multi}
\end{align}
The self-energy can be rewritten as
\begin{align}
  \Sigma_A(x,y)=\sum_{l=1,2}\sum_{i=1,2}\Delta_{i,l,A}(x,y)\tau_i\otimes E_{ll}.
\end{align}
By definition~[Eq.~\eqref{eq:def_correction}], the correction part of the full one-photon vertex is given by
\begin{align}
  \Lambda^\mu(x,y,w)&=-\left.\fdv{\Sigma_A(x,y)}{A_\mu(w)}\right|_{A=0}\nonumber\\
  &=-\sum_{i,l}\left.\fdv{\Delta_{i,l,A}(x,y)}{A_\mu(w)}\right|_{A=0}\tau_i\otimes E_{ll}.
\end{align}
Plugging the gap equation [Eq.~\eqref{eq:gapeq_multi}] into this, we obtain the following equation that each component should satisfy:
\begin{align}
  &-\left.\fdv{\Delta_{i,l,A}(x,y)}{A_\mu(w)}\right|_{A=0}\nonumber\\
  &=\frac{V_l(x-y)}{2}\left.\fdv{A_\mu(w)}\mathrm{Tr}\big[\tau_i\otimes E_{ll}G_A(x,y)\big]\right|_{A=0}\nonumber\\
  &=-\frac{V_l(x-y)}{2}\mathrm{Tr}\big[\tau_i\otimes E_{ll}G(x,\overline{x}')\nonumber\\
  &\qquad\times\big(\gamma^\mu(\overline{x}',\overline{y}',w)+\Lambda^\mu(\overline{x}',\overline{y}',w)\big)G(\overline{y}',y)\big].
\end{align}
This is the integral equation of the correction parts $\Lambda^\mu$.
This can be solved if we assume a solution of the form
\begin{align}
  -\left.\fdv{\Delta_{i,l,A}(x,y)}{A_\mu(w)}\right|_{A=0}=\Lambda_{il}^\mu(x,w)\delta(x-y).
\end{align}
Performing the Fourier transform, the integral equation reduces to the easily solvable matrix equation 
\begin{align}
  \sum_{j,l'}\bigg(\frac{2}{g_l}\delta_{il,jl'}-Q_{il,jl'}(q)\bigg)\Lambda_{jl'}^\mu(q)=Q_{il}^\mu(q),
  \label{eq:cfop}
\end{align}
where
\begin{align}
  &Q_{il}^\mu(q)=-\mathrm{Tr}\big[\tau_i\otimes E_{ll}G(\overline{p}+q)\gamma^\mu(p,q)G(\overline{p})\big],\\
  &Q_{il,jl'}(q)=-\mathrm{Tr}\big[\tau_i\otimes E_{ll}G(\overline{p}+q)\tau_j\otimes E_{l'l'}G(\overline{p})\big]
\end{align}
are correlation functions related to fluctuations of order parameters.
By solving this, we can obtain the gauge-invariant electromagnetic responses.
The optical responses can be obtained by the limit $\bm{q}\to 0$.

Similarly, the correction part of the full two-photon vertex can be obtained by calculating
\begin{align}
  \Lambda^{\mu\nu}(x,y,w_1,w_2)=\sum_{i,l}\frac{\delta^2\Delta_{i,l,A}(x,y)}{\delta A_\mu(w_1)\delta A_\nu(w_2)}\tau_i\otimes E_{ll}.
\end{align}
Detailed calculations are discussed in Appendix~\ref{ap:cfop_2nd}.

\begin{figure}[t]
  \centering
  \includegraphics[width=1.0\linewidth]{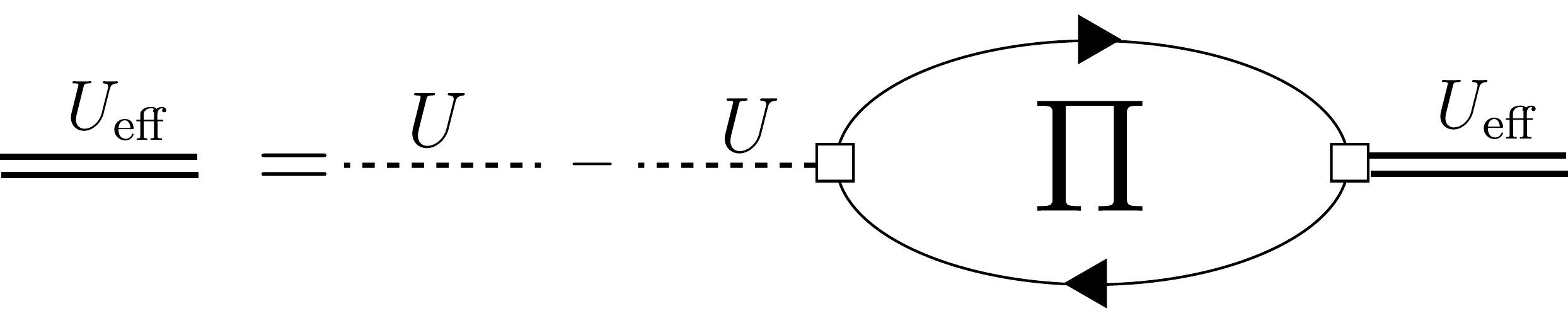}
  \caption{Diagrammatic representation of the effective interaction $U_{\mathrm{eff}}$ within the RPA.}
  \label{fig:rpa}
\end{figure}

Let us make a comment on the relation between our work and the previous studies of the collective mode in superconductors~\cite{Kamatani2022,Nagashima2024,Nagashima2024a}.
The vertex corrections in our formulation turn out to be equivalent to the random phase approximation (RPA) at the linear response level. 
To see this, note that the matrix equation~\eqref{eq:cfop} for $\Lambda^\mu$ can be rewritten as
\begin{align}
  (1+U\Pi)\vec{\Lambda}^\mu=U\vec{Q}^\mu,
\end{align}
where $U=\mathrm{diag}\{g_1,g_2,g_1,g_2\}$ is the interaction matrix,
\begin{align}
  [\Pi]_{il,jl'}=-\frac{1}{2}Q_{il,jl'}(q)
\end{align}
is the matrix corresponding to the bubble diagram, $\vec{\Lambda^\mu}=(\Lambda_{11}^\mu,\Lambda_{12}^\mu,\Lambda_{21}^\mu,\Lambda_{22}^\mu)^T$ and $\vec{Q}^\mu=(Q_{11}^\mu,Q_{12}^\mu,Q_{21}^\mu,Q_{22}^\mu)^T/2$.
Since the effective interaction within the RPA~[Fig.~\ref{fig:rpa}] is given by
\begin{align}
  U_{\mathrm{eff}}=\frac{U}{1+U\Pi},
\end{align}
the correction parts of the full one-photon vertices are
\begin{align}
  \vec{\Lambda}^\mu=U_{\mathrm{eff}}\vec{Q}^\mu.
\end{align}
Therefore, the generalized CFOP method is equivalent to the RPA in this case.
A significant advantage of our framework is that it easily extends to nonlinear responses and clearly demonstrates their gauge invariance.
Our approach emphasizes the perspective that collective excitations in superconductors restore the gauge invariance.

\subsubsection{Collective mode excitations}
First, let us see responses of collective mode.
In multi-band superconductors, there can occur a collective excitation known as the Leggett mode, corresponding to the fluctuations of the phase difference between the two order parameters~\cite{Leggett1966}.
Here, we calculate the linear and second-order optical responses based on the generalized CFOP method.

We set the parameters as $t=0.2$, $\mu=-0.3$, $\delta t=0.1$, $m=0.3$, $t_2=1.0$, $\Delta_1=0.2$, $\Delta_2=0.175$, $\eta=1\times 10^{-2}$, and $a=1$ so that our calculation reproduce the result of the previous study~\cite{Kamatani2022}.
Instead of solving the gap equation with given coupling constant $g_l$, we set the magnitude of the gap $\Delta_l$ and determined the corresponding coupling constant $g_l$ through numerical calculations, yielding $g_1=g_2\simeq 0.990$.

\begin{figure}[t]
  \centering
  \includegraphics[width=1\linewidth]{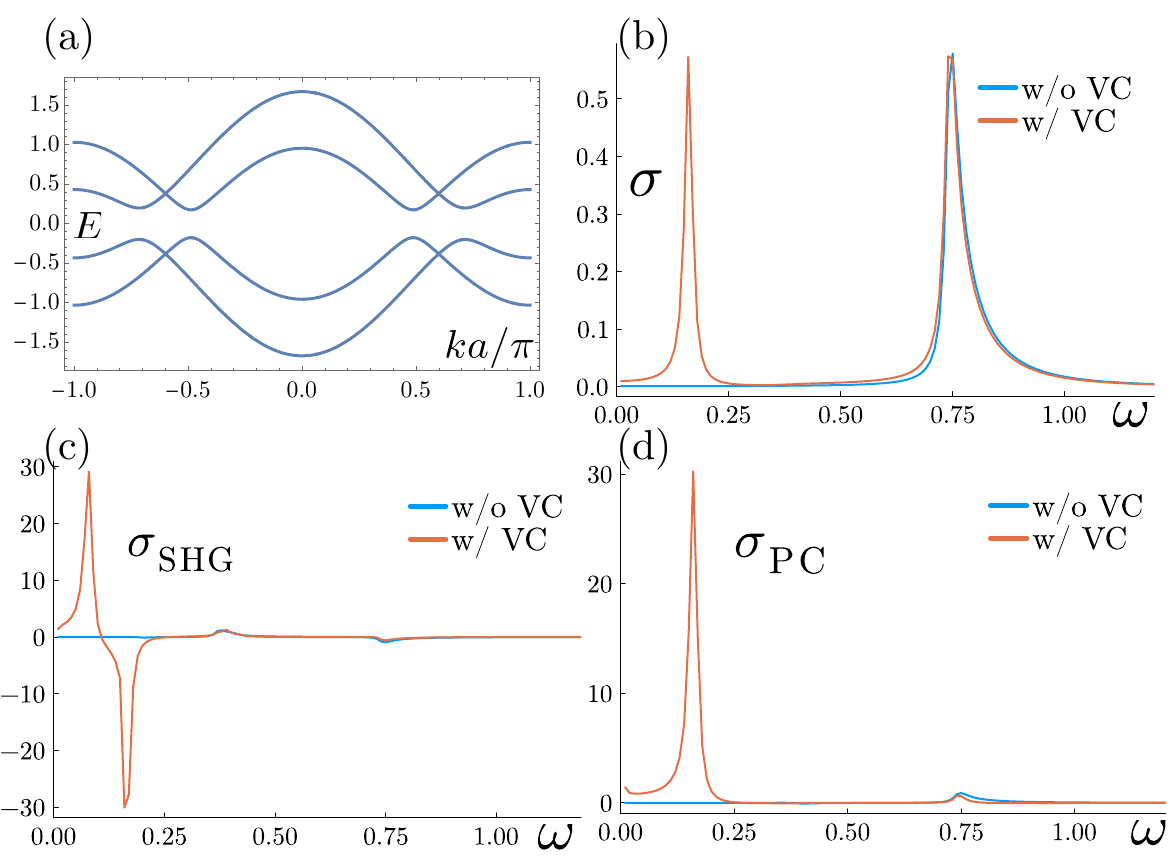}
\caption{Linear and second-order optical responses with and without vertex corrections in a multiband superconductor. 
(a) Band structure of the model. 
(b) Real part of the linear optical conductivity $\sigma^{xx}(\omega)$. Blue and orange lines indicate results without and with vertex corrections, respectively. 
(c) Real part of the second-harmonic generation response $\sigma^{xxx}(\omega, \omega)$. 
(d) Real part of the photocurrent response $\sigma^{xxx}(\omega, -\omega)$.}
  \label{fig:leggett}  
\end{figure}
The band dispersion of the BdG Hamiltonian under this parameter setting is shown in Fig.~\ref{fig:leggett}(a), and the calculated linear optical response is presented in Fig.~\ref{fig:leggett}(b).
It is evident that while the quasiparticle excitation peak around $\omega\simeq0.7$ remains nearly unchanged, a new excitation peak emerges near $\omega\simeq0.2$. 
This newly observed excitation peak corresponds to the Leggett mode, successfully reproducing the result of the previous study~\cite{Kamatani2022}.

While previous study~\cite{Kamatani2022} focused solely on linear responses, the generalized CFOP method allows for the straightforward calculation of nonlinear responses with vertex corrections.
The results of the second-order optical response are shown in Fig.~\ref{fig:leggett}(d) and Fig.~\ref{fig:leggett}(e).
Similar to the linear optical response, sharp new excitation peaks emerge around $\omega\simeq 0.2$ and $2\omega\simeq 0.2$.

\subsubsection{The suppression of the nonlinear optical conductivity}

Next, we consider the situation discussed in Ref.~\cite{Xu2019}, which analyzed systems with spatial inversion symmetry in the normal conducting phase, where this symmetry is broken by the superconducting gap, leading to finite linear and second-order optical responses. 
However, their analysis neglected many-body effects and lacked a gauge-invariant treatment. 
Here, we demonstrate how significant differences emerge when these effects are properly taken into account using our generalized CFOP method.

\begin{figure}[t]
  \centering
  \includegraphics[width=1\linewidth]{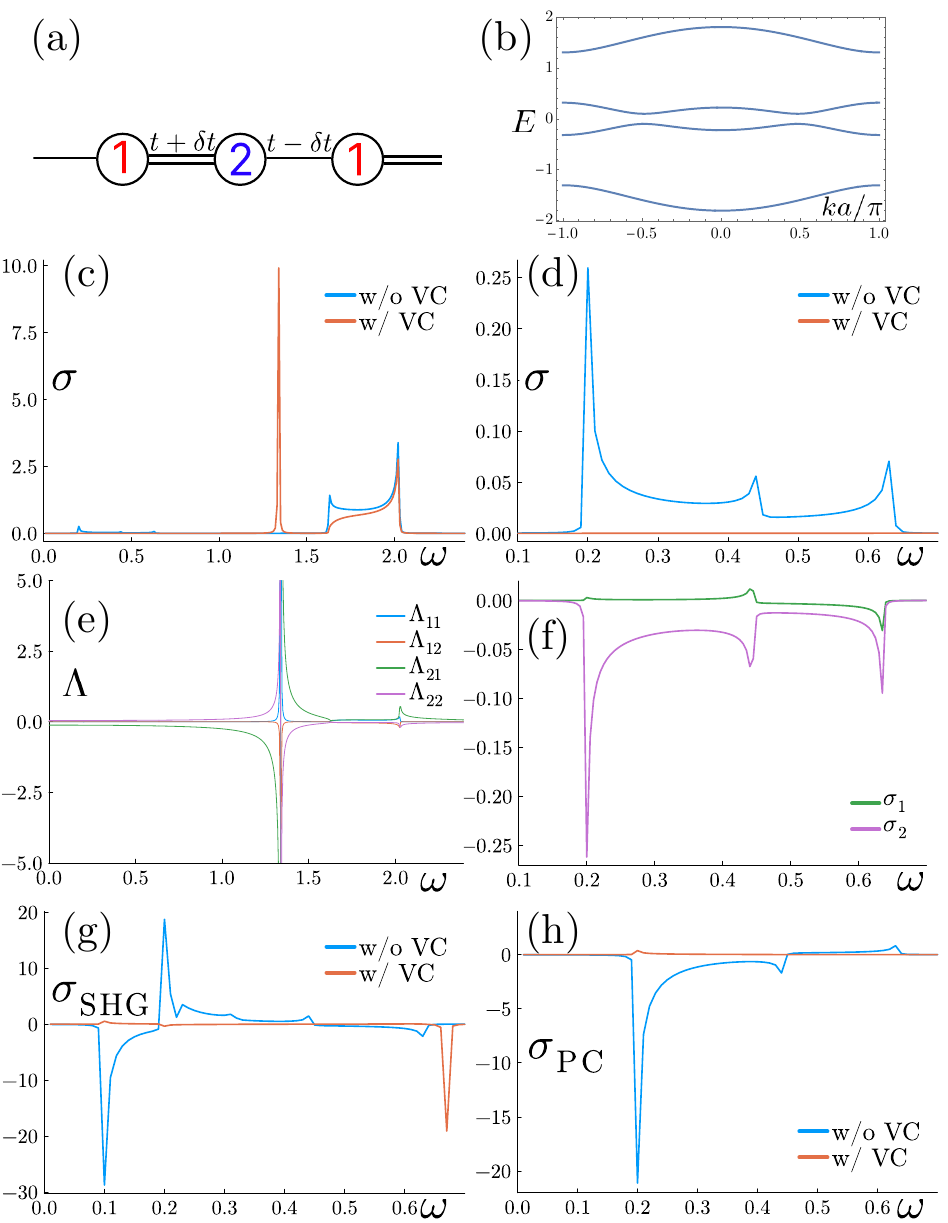}
  \caption{
  Linear and second-order optical responses with and without vertex corrections in a multiband superconductor.
(a) Schematic illustration of the model with parameters $t_2 = 0$ and $m = 0$.
(b) Band structure.
(c) Real part of the linear optical conductivity $\sigma^{xx}(\omega)$. Blue and orange lines indicate results without and with vertex corrections, respectively.
(d) Low-energy behavior of the real part of $\sigma^{xx}(\omega)$ near the superconducting gap.
(e) Real part of the correction part $\Lambda_{in}^x(\omega)$.
(f) Contribution to the optical conductivities arising from the the real (green) and the imaginary (purple) parts of fluctuations of order parameters.
(g) Real part of the second-harmonic generation $\sigma^{xxx}(\omega, \omega)$.
(h) Real part of the photocurrent generation $\sigma^{xxx}(\omega, -\omega)$.
}
  \label{fig:xu}
\end{figure}

We set the model parameters to $t=1$, $\mu=0.8$, $\delta t=0.5$, $m=0$, $t_2=0$, $\Delta_1=0.15$, $\Delta_2=0.05$, $\eta=1\times10^{-3}$, and $a=2$ to reproduce the bare calculation results of Ref.~\cite{Xu2019}.
The calculated linear optical response is shown in Fig.~\ref{fig:xu}(c). 
Even in this case, a new excitation peak appears near $\omega \simeq 1.3$ due to vertex corrections. 

Since the prior study~\cite{Xu2019} focused on low-energy quasiparticle excitations near the superconducting gap, we examine the impact of vertex corrections at this energy scale. 
The linear response in the low-energy region is shown in Fig.~\ref{fig:xu}(d).
Calculations without vertex corrections successfully reproduce the results of Ref.~\cite{Xu2019}. 
However, the inclusion of vertex corrections leads to substantial differences. 
The low-energy linear optical response is strongly suppressed and nearly vanishes.
We presented the frequency dependence of the correction part obtained by solving Eq.~\eqref{eq:cfop} in Fig.~\ref{fig:xu}(e).
There is a strong peak near $\omega\simeq 1.3$.
Comparing  Figs.~\ref{fig:xu}(c) and \ref{fig:xu}(e), we can attribute the strong peak in optical conductivities to the many-body effects.
To clarify the suppression mechanism, we decompose the gauge-invariant linear optical conductivity as $\sigma(\omega)=\sigma_{\mathrm{bare}}+\sigma_1(\omega)+\sigma_2(\omega)$ where $\sigma_1(\omega)$ and $\sigma_2(\omega)$ represent the contributions from the fluctuations of real and imaginary parts of the order parameters, respectively.
Each component is presented in Figs.~\ref{fig:xu}(d) and \ref{fig:xu}(f).
We find that the bare conductivity $\sigma_{\mathrm{bare}}(\omega)$ (blue line in Fig.~\ref{fig:xu}(d)) is strongly canceled by the contributions from $\sigma_2(\omega)$ (purple line in Fig.~\ref{fig:xu}(f)), highlighting the dominant role of fluctuations in the imaginary component.

Vertex corrections also have a significant impact on second-order optical responses.
The second harmonic generation $\sigma^{xxx}(\omega,\omega)$ and the photocurrent generation $\sigma^{xxx}(\omega,-\omega)$ are shown in Figs.~\ref{fig:xu}(g) and \ref{fig:xu}(h), respectively.
In both cases, the magnitude of the second-order optical conductivity $\sigma^{xxx}(\omega_1,\omega_2)$ is significantly reduced, and its sign may even change.

\begin{figure}[t]
  \centering
  \includegraphics[width=1\linewidth]{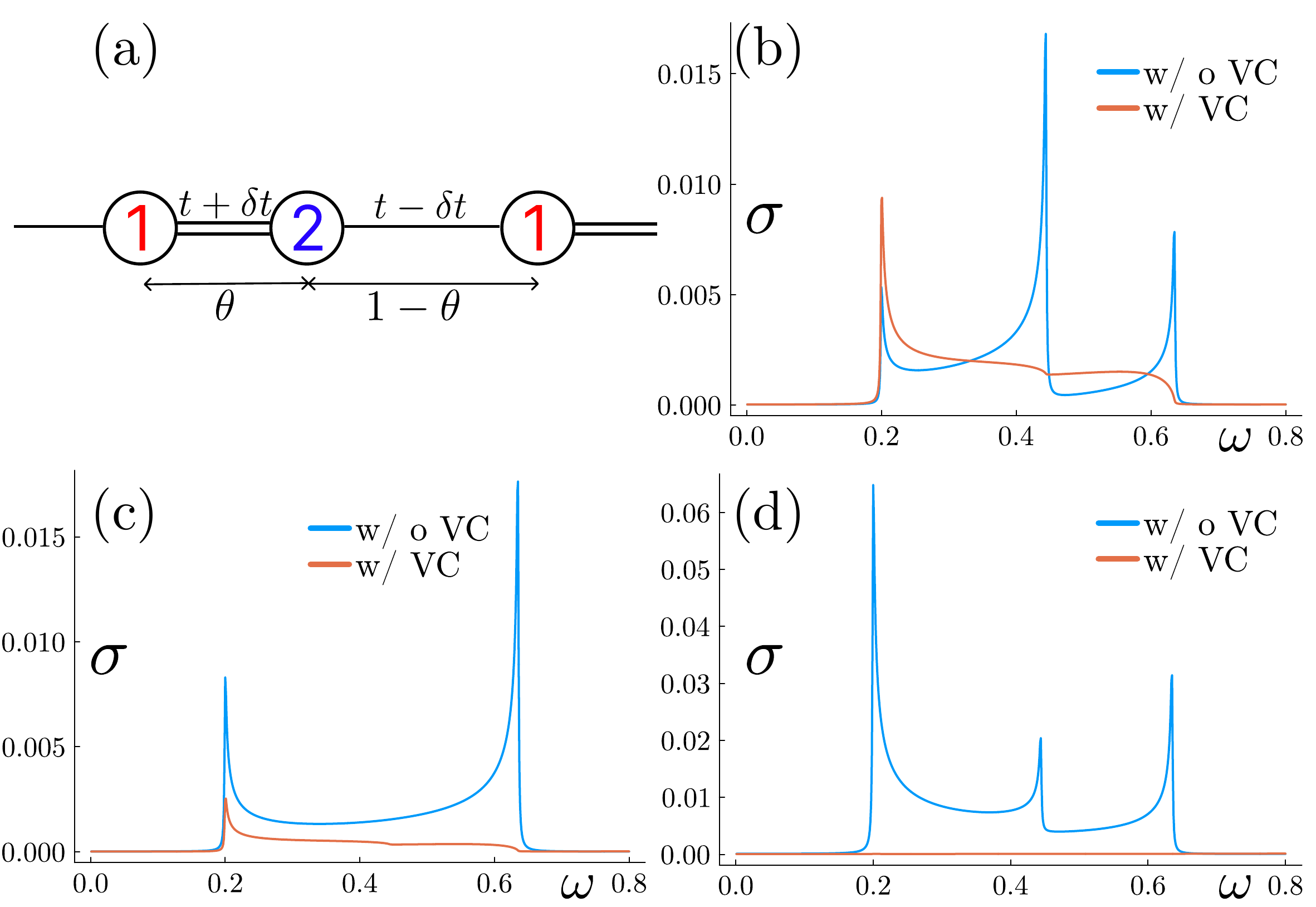}
  \caption{
(a) Schematic illustration of the Hamiltonian defined in Eq.~\eqref{eq:ssh_theta}. 
(b)--(d) Real part of the linear optical conductivity for (b) $\theta = 0$, (c) $\theta = 1/4$, and (d) $\theta = 1/2$.
}
  \label{fig:ssh_theta}
\end{figure}

To examine whether this strong suppression is a generic feature, we consider a modified model. In the previous setup, the distance between sublattices $l=1$ and $l=2$ was fixed at $a/2$. We now introduce a parameter $\theta$ that controls this distance, leading to the following normal-state Hamiltonian:
\begin{align}
  &H_N(k,\theta)\nonumber\\
  &=\frac{1}{2}\big[t\big(\cos(k\theta)+\cos(k-k\theta)\big)\nonumber\\
  &\qquad+\delta t\big(\cos(k\theta)-\cos(k-k\theta)\big)\big]\sigma_1\nonumber\\
  &-\frac{1}{2}\big[t\big(\sin(k\theta)-\sin(k-k\theta)\big)\nonumber\\
  &\qquad+\delta t\big(\sin(k\theta)+\sin(k-k\theta)\big)\big]\sigma_2-\mu\sigma_0,
  \label{eq:ssh_theta}
\end{align}
where we set $a=1$ for simplicity. This geometry is illustrated in Fig.~\ref{fig:ssh_theta}(a). Note that $\theta = 1/2$ corresponds to the original configuration.

Figures~\ref{fig:ssh_theta}(b)--(d) show the optical conductivities for different values of $\theta$. While the optical conductivity near the gap is almost entirely suppressed by vertex corrections at $\theta = 1/2$, such suppression is not observed for other values of $\theta$. This suggests that the strong suppression at $\theta = 1/2$ is likely an accidental feature of that particular parameter choice rather than a generic property. Nevertheless, our results consistently demonstrate that vertex corrections significantly affect the optical response across different parameter regimes.

\subsection{Optical responses in $d$-wave superconductors}
\subsubsection{Model}
Our formulation can be applied to the anisotropic pairing.
As an example, let us consider spin-singlet $d$-wave superconductors on a square lattice.
The normal-state Hamiltonian is defined by
\begin{align}
  \hat{H}_N=\sum_{\bm{k}s}\hat{c}_{\bm{k}s}^\dagger\epsilon(\bm{k})\hat{c}_{\bm{k}s}
\end{align}
where $\epsilon(\bm{k})=t(2-\cos k_x-\cos k_y)-\mu$.
The microscopic interaction in real space is given by
\begin{align}
  V^{s_1s_2s_3s_4}(\bm{x-y})=\frac{1}{2}V(\bm{x-y})(i\sigma_2)_{s_1s_2}(i\sigma_2)_{s_3s_4}^\dagger,
  \label{eq:interaction_dwave}
\end{align}
where $V(\bm{x-y})=g\sum_{\mu=x,y}(\delta_{\bm{x,y+e_\mu}}+\delta_{\bm{x,y-e_\mu}})$.
In momentum space, this interaction can be expressed as
\begin{align}
  \hat{H}_{\mathrm{int}}=-\frac{g}{N}\sum_{\bm{k,p,q}}\phi_d(\bm{k})\phi_d(\bm{p})\hat{c}_{\bm{k+q}\uparrow}^\dagger\hat{c}_{\bm{-k+q}\downarrow}^\dagger\hat{c}_{\bm{-p+q}\downarrow}\hat{c}_{\bm{p+q}\uparrow},
  \label{eq:dwave_interaction_kspace}
\end{align}
where $\phi_d(\bm{k})=\cos k_x-\cos k_y$ is the $d$-wave form factor.
We note that this expression relies on a specific property of the square lattice.
On the square lattice, the following operator identity holds~\cite{Paramekanti2000}:
\begin{align}
  \sum_{\bm{k,p,q}}&\big[\phi_d(\bm{k})\phi_d(\bm{p})-\phi_s(\bm{k})\phi_s(\bm{p})\big]\nonumber\\
  &\times \hat{c}_{\bm{k+q}\uparrow}^\dagger\hat{c}_{\bm{-k+q}\downarrow}^\dagger\hat{c}_{\bm{-p+q}\downarrow}\hat{c}_{\bm{p+q}\uparrow}\equiv 0,
\end{align}
where $\phi_s(\bm{k})=\cos k_x+\cos k_y$ is the extended $s$-wave form factor.
This identity leads to an ambiguity in the momentum space representation of the interaction.
In this work, we adopt the form given in Eq.~\eqref{eq:dwave_interaction_kspace} for the realization of the $d$-wave gap and the comparison with the previous study~\cite{Huang2023}.

Employing the mean-field approximation in the Cooper channel, one defines the magnitude of the gap as
\begin{align}
  \Delta_d=-\frac{g}{N}\sum_{\bm{p}}\langle\hat{c}_{-\bm{p}\downarrow}\hat{c}_{\bm{p}\uparrow}\rangle\phi_d(\bm{p})\label{eq:gap_gBCS},
\end{align}
the interaction term reduces to
\begin{align}
  \hat{H}_{\mathrm{int}}\to \sum_{\bm{k}}(\Delta_d\phi_d(\bm{k})\hat{c}_{\bm{k}\uparrow}^\dagger\hat{c}_{\bm{-k}\downarrow}^\dagger +h.c.)
\end{align}
which describes $d$-wave pairing in the superconducting state.

Since the optical transitions between the particle-hole pairs are prohibited in the presence of time-reversal symmetry and spatial inversion symmetry~\cite{Ahn2021}, we consider the situation where Cooper pairs have finite momentum and break inversion symmetry.
In this case, as already presented in Sec.~\ref{sec:finite_momentum}, the magnitude of the gap is defined as
\begin{align}
  \Delta_d=-\frac{g}{N}\sum_{\bm{p}}\phi_d(\bm{p})\langle\hat{c}_{\bm{-p+Q}\downarrow}\hat{c}_{\bm{p+Q}\uparrow}\rangle,
\end{align}
where $\bm{2}Q$ is the momentum of the Cooper pairs.
The mean-field Hamiltonian is given by
\begin{align}
  \hat{H}_{MF}=\sum_{\bm{k}}(\hat{\psi}_{\bm{k}}^{\bm{Q}})^\dagger\begin{pmatrix}
    \epsilon(\bm{k+Q})& \Delta_d\phi_d(\bm{k})\\
    \Delta_d\phi_d(\bm{k}) & -\epsilon(\bm{k-Q})
  \end{pmatrix}\hat{\psi}_{\bm{k}}^{\bm{Q}}.
\end{align}
The previous study~\cite{Huang2023} investigated the optical responses in this model with vertex corrections, although its theoretical treatment is different.
In the subsequent sections, we compare the generalized CFOP method with that of the previous work~\cite{Huang2023}.

\subsubsection{The full photon vertices}
Now, let us calculate the full vertices based on the generalized CFOP method.
The microscopic interaction in Eq.~\eqref{eq:interaction_dwave} leads to the the self-energy of superconductors~[Eqs.~\eqref{eq:su2_se} and~\eqref{eq:su2_gapeq}]:
\begin{align}
  \Sigma_A(x,y)=\begin{pmatrix}
    &\Delta_A(x,y)\\
    \Delta_A^\dagger(x,y)&
  \end{pmatrix},\\
  \Delta_{A}(x,y)=-V(x-y)\mathrm{Tr}\big[PG_A(x,y)\big],
\end{align}
where $V(x-y)=V(\bm{x-y})\delta(x_0-y_0-0^-)$.
If we separate the real part and imaginary part of the gap function as $\Delta_A(x,y)=\Delta_{1,A}(x,y)-i\Delta_{2,A}(x,y)$,
each part satisfies
\begin{align}
  \Delta_{i,A}(x,y)=-\frac{1}{2}V(x-y)\mathrm{Tr}\big[\tau_iG_A(x,y)\big],
  \label{eq:gapeq_dwave}
\end{align}
and the self-energy can be expressed as
\begin{align}
  \Sigma_A(x,y)=\Delta_{1,A}(x,y)\tau_1+\Delta_{2,A}(x,y)\tau_2.
  \label{eq:se_dwave_cfop}
\end{align}
The correction part $\Lambda^\mu$ of the full one-photon vertex is given by the functional derivatives of the self-energy as in Eq.~\eqref{eq:def_correction}:
\begin{align}
  \Lambda^\mu(x,y,z)=\sum_{i=1,2}\Lambda_i^\mu(x,y,z)\tau_i,
  \label{eq:correction_multi}
\end{align}
where
\begin{align}
  \Lambda_i^\mu(x,y,z)=-\left.\fdv{\Delta_{i,A}(x,y)}{A_\mu(z)}\right|_{A=0}.
\end{align}
The functional derivative of the gap equation in Eq.~\eqref{eq:gapeq_dwave} and the Fourier transform lead to
\begin{align}
  \Lambda_i^\mu(k,q)&=-V(k-p)\mathrm{Tr}\big[\tau_iG(\overline{p}+q)\nonumber\\
  &\times\big(\gamma^\mu(\overline{p},q)+\sum_{j=1}^2\tau_j\Lambda_j^\mu(\overline{p},q)\big)G(\overline{p})\big],\label{eq:cfop_dwave}
\end{align}
where $V(k-p)$ is the Fourier transform of $V(x-y)$.

To solve this integral equation, let us expand $V(k-p)$ as
\begin{align}
  V(k-p)=2g\sum_{n=1}^4f_n(\bm{k})f_n(\bm{p}),
\end{align}
where we defined $f_1(\bm{k})=\cos k_x,f_2(\bm{k})=\cos k_y, f_3(\bm{k})=\sin k_x, f_4(\bm{k})=\sin k_y.$
Similarly, we expand the solution to the integral equation as
\begin{align}
  \Lambda_i^\mu(k,q)&=\sum_{n=1}^4f_n(\bm{k})\Lambda_{in}^\mu(q).
  \label{eq:correction_cfop}
\end{align}
If each component $\Lambda_{i,n}^\mu(q)$ satisfies
\begin{align}
  \Lambda_{in}^\mu(q)&=-gf_n(\overline{\bm{p}})\mathrm{Tr}\big[\tau_iG(\overline{p}+q)&\nonumber\\
  &\times\big(\gamma^\mu(\overline{p},q)+\sum_{j,m}\tau_jf_m(\overline{p})\big)G(\overline{p})\big]\Lambda_{jm}^\mu(q),\label{eq:cfop_mat}
\end{align}
it will be the solution to Eq.~\eqref{eq:cfop_dwave}.
Therefore, we need to solve the matrix equation given by
\begin{align}
  \sum_{j,m}\bigg(\frac{1}{g}\delta_{in,jm}-Q_{in,jm}(q)\bigg)\Lambda_{jm}^\mu(q)=Q_{in}^\mu(q),
\end{align}
where
\begin{align}
  Q_{in,jm}(q)=-f_n(\overline{p})f_m(\overline{p})\mathrm{Tr}\big[\tau_iG(\overline{p}+q)\tau_jG(\overline{p})\big],\\
  Q_{in}^\mu(q)=-f_n(\overline{p})\mathrm{Tr}\big[\tau_iG(\overline{p}+q)\gamma^\mu(\overline{p},q)G(\overline{p})\big].
\end{align}

Before proceeding to numerical calculations, let us review the construction of the full photon vertex in the previous study~\cite{Huang2023}.
In the previous study, the Fock approximation of the self-energy
\begin{align}
  \Sigma(k)=V(\bm{k,\overline{p}})\tau_3 G(\overline{p})\tau_3
  \label{eq:fock_gBCS}
\end{align}
is employed and the full photon vertex is given as the solution to the Bethe-Salpeter equation
\begin{align}
  \Gamma^\nu(k,q)=\gamma^\nu(k,q)+V(\bm{k,\overline{p}})\tau_3G(\overline{p}+q)\Gamma^\nu(\overline{p},q)G(\overline{p})\tau_3\label{eq:bse_gBCS}
\end{align}
where $V(\bm{k,p})=\phi_d(\bm{k})\phi_d(\bm{p})$.
However, this approach does not necessarily satisfy the Ward identity strictly.
When checking the Ward identity for the correction part~[Eq.~\eqref{eq:wi_correction}], one obtains
\begin{align}
  &\Lambda^\nu(k,q)q_\nu\nonumber\\
  &=V(\bm{k,\overline{p}})\tau_3G(\overline{p}+q)\Gamma^\nu(\overline{p},q)q_\nu G(\overline{p})\tau_3\nonumber\\
  &=V(\bm{k,\overline{p}})\tau_3G(\overline{p}+q)\nonumber\\
  &\qquad\times\big(G^{-1}(\overline{p}+q)\tau_3-\tau_3G^{-1}(\overline{p})\big)G(\overline{p})\tau_3\nonumber\\
  &=V(\bm{k,\overline{p}})G(\overline{p})\tau_3-V(\bm{k,\overline{p}})\tau_3G(\overline{p}+q)\nonumber\\
  &=\tau_3\Sigma(k)-V(\bm{k,\overline{p}})\tau_3G(\overline{p}+q),
\end{align}
which deviates from the expected form.

Therefore, in such a formulation, gauge-invariant responses might not be obtained.
In the generalized CFOP method, such issues do not arise, and we can obtain full photon vertices that manifestly satisfy the Ward identities.

We compare the correction parts obtained by two different methods.
The correction part derived by the generalized CFOP method is given by Eqs.~\eqref{eq:correction_multi} and~\eqref{eq:correction_cfop}. 
In the case of the Bethe-Salpeter equation, the correction part of the full one-photon vertex is expanded in the Pauli basis as 
\begin{align}
  \Lambda^\mu(k,q)=\sum_{i=0}^3\Lambda_i^\mu(k,q)\tau_i,
\end{align}
and the $k$-dependence is assumed to be
\begin{align}
  \Lambda_i^\mu(k,q)=\phi_d(\bm{k})\Lambda_i^\mu(q).
  \label{eq:bse_solution}
\end{align}
These two methods for constructing the full vertex differ in whether the $\tau_0,\tau_3$ components are present in the correction part.
This difference stems from the form of the self-energies in Eqs.~\eqref{eq:se_dwave_cfop} and~\eqref{eq:fock_gBCS}.
The $k$-dependence of $\Lambda_i^\mu(k,q)$ is also different. 
The solutions to the Bethe-Salpeter equation always respect $d$-wave symmetry $\phi_d(\bm{k})$, whereas the solutions to Eq.~\eqref{eq:cfop_dwave} do not.
We will demonstrate these differences using numerical calculations in the next section.

\subsubsection{Numerical calculations}
\begin{figure}[t]
  \centering
  \includegraphics[width=1\linewidth]{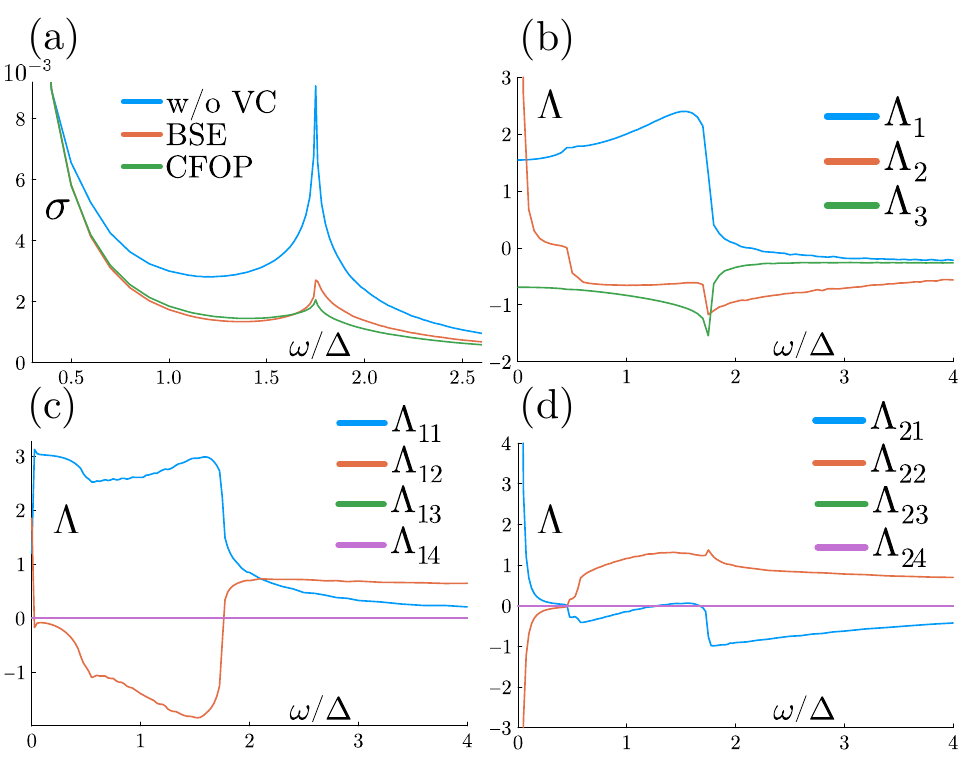}
  \caption{
(a) Real part of the optical conductivity $\sigma^{xx}(\omega)$ with and without vertex corrections. 
Blue, orange, and green lines correspond to calculations without vertex corrections, with vertex corrections based on the Bethe--Salpeter equation, and with the full vertex derived in this study, respectively. 
(b) Real part of the solution $\Lambda_i^x(\omega)$ to the Bethe--Salpeter equation. 
The $\sigma_0$ component $\Lambda_0^\mu$ is not shown as it is identically zero. 
(c) Real part of $\Lambda_{1n}^x(\omega)$. 
(d) Real part of $\Lambda_{2n}^x(\omega)$.}
  \label{fig:dwave_VC}
\end{figure}

We perform numerical calculations to investigate how the results obtained using the generalized CFOP method differ from those of a previous study~\cite{Huang2023}. 
To this end, we introduce a finite center-of-mass momentum of Cooper pairs along the $x$-direction, $\bm{Q} = Q_x \bm{e}_x$, and set the parameters as $t = 1.0 \times 10^2$, $\mu = 9.0 \times 10^1$, $\Delta_d = 2.3 \times 10^1$, $\eta = 3 \times 10^{-2}$, and $Q_x = 0.07$. 
These parameters yield a coupling constant of $g \simeq 1.8 \times 10^2$, chosen to reproduce the results of Ref.~\cite{Huang2023}. 
The corresponding optical response results are summarized in Fig.~\ref{fig:dwave_VC}.

Figure~\ref{fig:dwave_VC}(a) compares the optical conductivity obtained by different methods. 
While the results based on the generalized CFOP method show only minor quantitative deviations from those obtained using the Bethe-Salpeter equation, the differences are nonetheless noticeable.

To better understand these differences, we analyze the correction part of the full one-photon vertex in detail. 
First, we examine each Pauli matrix component $\Lambda_i^\mu$.
The correction term derived from the generalized CFOP method contains only the off-diagonal components $\tau_1$ and $\tau_2$. 
In contrast, the numerical solution to the Bethe-Salpeter equation includes three finite components: $\tau_1$, $\tau_2$, and $\tau_3$. 
The $\tau_0$ component is analytically found to vanish identically.

Next, we investigate the momentum dependence of the correction term $\Lambda^\mu(k, q)$. 
The solution to the Bethe--Salpeter equation strictly respects the $d$-wave symmetry, as given in Eq.~\eqref{eq:bse_solution}. 
However, the corrections obtained via the generalized CFOP method do not necessarily preserve this symmetry. 
The $d$-wave symmetry is satisfied when the following conditions hold:
\begin{align}
  \Lambda_{i1}^\mu(q)&=-\Lambda_{i2}^\mu(q),\label{eq:cosine}\\
  \Lambda_{i3}^\mu(q)&=\Lambda_{i4}^\mu(q)=0\label{eq:sine}.
\end{align}
As shown in Figs.~\ref{fig:dwave_VC}(c) and \ref{fig:dwave_VC}(d), these conditions are clearly violated. 
This indicates that the correction terms do not exhibit $d$-wave symmetry, which reflects the explicit breaking of $C_4$ rotational symmetry due to the finite momentum $Q_x$ carried by the Cooper pairs.
\begin{figure}[t]
  \centering
  \includegraphics[width=1\linewidth]{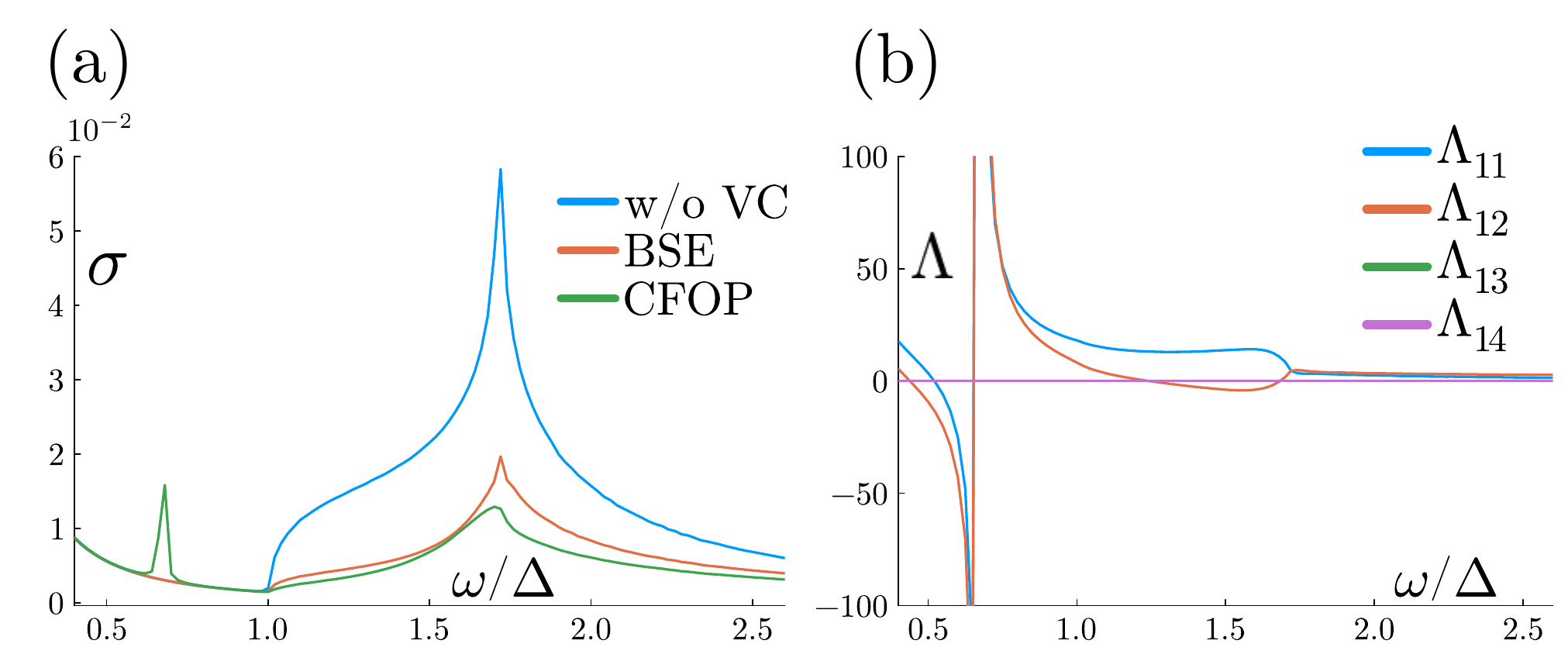}
  \caption{Optical responses for $Q_x=0.2$.
  (a) Real part of the linear optical conductivity $\sigma^{xx}(\omega)$.
  (b) Real part of the correction part $\Lambda_{1i}(\omega)$.}
  \label{fig:dwave_q02}
\end{figure}

We also investigate the dependence of optical responses on $Q_x$. 
So far, we have used $Q_x = 0.07$, consistent with Ref.~\cite{Huang2023}, where this value was chosen to be experimentally realistic. 
To explore the generality of our findings, we perform calculations with a larger value $Q_x = 0.2$ while keeping all other parameters fixed. 
The calculated optical conductivities are shown in Fig.~\ref{fig:dwave_q02}(a).
We observe substantial differences in the optical conductivities between the results obtained from the Bethe-Salpeter equation and those from the generalized CFOP method.
More interestingly, a new excitation peak emerges at $\omega \simeq 0.6\Delta$.
To investigate the feature of this new peak, we show the correction part $\Lambda_{1i}^x(\omega)$ in Fig.~\ref{fig:dwave_q02}(b).
There also exists a peak at $\omega \simeq 0.6\Delta$.
As previously discussed, the correction term obtained within the generalized CFOP framework does not necessarily obey $d$-wave symmetry. 
The correction terms with extended $s$-wave symmetry satisfy the following conditions:
\begin{align}
  &\Lambda_{i1}^\mu(q)=\Lambda_{i2}^\mu(q),\\
  &\Lambda_{i3}^\mu(q)=\Lambda_{i4}^\mu(q)=0.
\end{align}
We observe that the frequency dependence of the correction parts near the peak is more consistent with extended $s$-wave symmetry than with $d$-wave symmetry.
This behavior is strongly different from the case for $Q_x=0.02$ where the exact $d$-wave symmetry is not realized but the frequency dependence resembles it (see Fig.~\ref{fig:dwave_VC}(c)).
Although the extended $s$-wave and $d$-wave cannot be mixed in a $C_4$ symmetric system due to their different parity, in our setting, the supercurrent flow breaks $C_4$ symmetry and mixes two pairings.
Our method can include such effects and results in the new excitation peak which is not observed when employing the Bethe-Salpeter equation.
These results indicate that the discrepancies between our method and that of the previous studies become more pronounced with increasing $Q_x$.

\section{Conclusion}
\label{sec:conclusion}

In this paper, we presented a gauge-invariant formulation of electromagnetic responses in superconductors. 
We summarized the Feynman rules for calculating electromagnetic response kernels in interacting systems and proved their gauge invariance using Ward identities. 
We also developed a systematic method for constructing full photon vertices that explicitly satisfy the Ward identities. 
Using this framework, we numerically investigated the effects of vertex corrections on optical responses and found that they lead to significant qualitative and quantitative changes. 
In particular, for unconventional superconductors, our full photon vertices differ from those in previous studies, highlighting the importance of a fully gauge-invariant treatment of optical responses in superconductors.

In this work, we assumed that the electron-electron interactions in the microscopic Hamiltonian are unaffected by the gauge field. 
As a result, interaction terms such as pair hopping are beyond the scope of the present formulation. 
Extending the theory to include such interactions remains an important direction for future work.

\begin{acknowledgments}
  We thank Y.~Yanase, A.~Daido, H.~Tanaka, T.~Hayata, and T.~Morimoto for useful discussions.
  The work of S.W. is supported by World-leading Innovative Graduate Study Program for Materials Research, Information, and Technology~(MERIT-WINGS) of the University of Tokyo.
  The work of H.W. is supported by JSPS KAKENHI Grant No.~JP24K00541.
\end{acknowledgments}

\bibliography{bcs_library}

\begin{thebibliography}{59}%
\makeatletter
\providecommand \@ifxundefined [1]{%
 \@ifx{#1\undefined}
}%
\providecommand \@ifnum [1]{%
 \ifnum #1\expandafter \@firstoftwo
 \else \expandafter \@secondoftwo
 \fi
}%
\providecommand \@ifx [1]{%
 \ifx #1\expandafter \@firstoftwo
 \else \expandafter \@secondoftwo
 \fi
}%
\providecommand \natexlab [1]{#1}%
\providecommand \enquote  [1]{``#1''}%
\providecommand \bibnamefont  [1]{#1}%
\providecommand \bibfnamefont [1]{#1}%
\providecommand \citenamefont [1]{#1}%
\providecommand \href@noop [0]{\@secondoftwo}%
\providecommand \href [0]{\begingroup \@sanitize@url \@href}%
\providecommand \@href[1]{\@@startlink{#1}\@@href}%
\providecommand \@@href[1]{\endgroup#1\@@endlink}%
\providecommand \@sanitize@url [0]{\catcode `\\12\catcode `\$12\catcode
  `\&12\catcode `\#12\catcode `\^12\catcode `\_12\catcode `\%12\relax}%
\providecommand \@@startlink[1]{}%
\providecommand \@@endlink[0]{}%
\providecommand \url  [0]{\begingroup\@sanitize@url \@url }%
\providecommand \@url [1]{\endgroup\@href {#1}{\urlprefix }}%
\providecommand \urlprefix  [0]{URL }%
\providecommand \Eprint [0]{\href }%
\providecommand \doibase [0]{https://doi.org/}%
\providecommand \selectlanguage [0]{\@gobble}%
\providecommand \bibinfo  [0]{\@secondoftwo}%
\providecommand \bibfield  [0]{\@secondoftwo}%
\providecommand \translation [1]{[#1]}%
\providecommand \BibitemOpen [0]{}%
\providecommand \bibitemStop [0]{}%
\providecommand \bibitemNoStop [0]{.\EOS\space}%
\providecommand \EOS [0]{\spacefactor3000\relax}%
\providecommand \BibitemShut  [1]{\csname bibitem#1\endcsname}%
\let\auto@bib@innerbib\@empty
\bibitem [{\citenamefont {Schrieffer}(1999)}]{Schrieffer1999}%
  \BibitemOpen
  \bibfield  {author} {\bibinfo {author} {\bibfnamefont {J.~R. J.~R.}\
  \bibnamefont {Schrieffer}},\ }\href@noop {} {\emph {\bibinfo {title} {Theory
  of superconductivity}}},\ \bibinfo {edition} {rev. printing}\ ed.,\ Advanced
  book classics\ (\bibinfo  {publisher} {Perseus Books},\ \bibinfo {year}
  {1999})\BibitemShut {NoStop}%
\bibitem [{\citenamefont {Bardeen}\ \emph {et~al.}(1957)\citenamefont
  {Bardeen}, \citenamefont {Cooper},\ and\ \citenamefont
  {Schrieffer}}]{Bardeen1957}%
  \BibitemOpen
  \bibfield  {author} {\bibinfo {author} {\bibfnamefont {J.}~\bibnamefont
  {Bardeen}}, \bibinfo {author} {\bibfnamefont {L.~N.}\ \bibnamefont
  {Cooper}},\ and\ \bibinfo {author} {\bibfnamefont {J.~R.}\ \bibnamefont
  {Schrieffer}},\ }\bibfield  {title} {\bibinfo {title} {Theory of
  superconductivity},\ }\href {https://doi.org/10.1103/PhysRev.108.1175}
  {\bibfield  {journal} {\bibinfo  {journal} {Phys. Rev.}\ }\textbf {\bibinfo
  {volume} {108}},\ \bibinfo {pages} {1175} (\bibinfo {year}
  {1957})}\BibitemShut {NoStop}%
\bibitem [{\citenamefont {Sipe}\ and\ \citenamefont
  {Shkrebtii}(2000)}]{Sipe2000}%
  \BibitemOpen
  \bibfield  {author} {\bibinfo {author} {\bibfnamefont {J.~E.}\ \bibnamefont
  {Sipe}}\ and\ \bibinfo {author} {\bibfnamefont {A.~I.}\ \bibnamefont
  {Shkrebtii}},\ }\bibfield  {title} {\bibinfo {title} {Second-order optical
  response in semiconductors},\ }\href
  {https://doi.org/10.1103/PhysRevB.61.5337} {\bibfield  {journal} {\bibinfo
  {journal} {Phys. Rev. B}\ }\textbf {\bibinfo {volume} {61}},\ \bibinfo
  {pages} {5337} (\bibinfo {year} {2000})}\BibitemShut {NoStop}%
\bibitem [{\citenamefont {Morimoto}\ and\ \citenamefont
  {Nagaosa}(2016)}]{Morimoto2016}%
  \BibitemOpen
  \bibfield  {author} {\bibinfo {author} {\bibfnamefont {T.}~\bibnamefont
  {Morimoto}}\ and\ \bibinfo {author} {\bibfnamefont {N.}~\bibnamefont
  {Nagaosa}},\ }\bibfield  {title} {\bibinfo {title} {Topological nature of
  nonlinear optical effects in solids},\ }\href
  {https://doi.org/10.1126/sciadv.1501524} {\bibfield  {journal} {\bibinfo
  {journal} {Science Advances}\ }\textbf {\bibinfo {volume} {2}},\ \bibinfo
  {pages} {e1501524} (\bibinfo {year} {2016})}\BibitemShut {NoStop}%
\bibitem [{\citenamefont {Orenstein}\ \emph {et~al.}(2021)\citenamefont
  {Orenstein}, \citenamefont {Moore}, \citenamefont {Morimoto}, \citenamefont
  {Torchinsky}, \citenamefont {Harter},\ and\ \citenamefont
  {Hsieh}}]{Orenstein2021}%
  \BibitemOpen
  \bibfield  {author} {\bibinfo {author} {\bibfnamefont {J.}~\bibnamefont
  {Orenstein}}, \bibinfo {author} {\bibfnamefont {J.}~\bibnamefont {Moore}},
  \bibinfo {author} {\bibfnamefont {T.}~\bibnamefont {Morimoto}}, \bibinfo
  {author} {\bibfnamefont {D.}~\bibnamefont {Torchinsky}}, \bibinfo {author}
  {\bibfnamefont {J.}~\bibnamefont {Harter}},\ and\ \bibinfo {author}
  {\bibfnamefont {D.}~\bibnamefont {Hsieh}},\ }\bibfield  {title} {\bibinfo
  {title} {Topology and symmetry of quantum materials via nonlinear optical
  responses},\ }\href
  {https://doi.org/https://doi.org/10.1146/annurev-conmatphys-031218-013712}
  {\bibfield  {journal} {\bibinfo  {journal} {Annual Review of Condensed Matter
  Physics}\ }\textbf {\bibinfo {volume} {12}},\ \bibinfo {pages} {247}
  (\bibinfo {year} {2021})}\BibitemShut {NoStop}%
\bibitem [{\citenamefont {Wakatsuki}\ and\ \citenamefont
  {Nagaosa}(2018)}]{Wakatsuki2018}%
  \BibitemOpen
  \bibfield  {author} {\bibinfo {author} {\bibfnamefont {R.}~\bibnamefont
  {Wakatsuki}}\ and\ \bibinfo {author} {\bibfnamefont {N.}~\bibnamefont
  {Nagaosa}},\ }\bibfield  {title} {\bibinfo {title} {Nonreciprocal current in
  noncentrosymmetric rashba superconductors},\ }\href
  {https://doi.org/10.1103/PhysRevLett.121.026601} {\bibfield  {journal}
  {\bibinfo  {journal} {Phys. Rev. Lett.}\ }\textbf {\bibinfo {volume} {121}},\
  \bibinfo {pages} {026601} (\bibinfo {year} {2018})}\BibitemShut {NoStop}%
\bibitem [{\citenamefont {Xu}\ \emph {et~al.}(2019)\citenamefont {Xu},
  \citenamefont {Morimoto},\ and\ \citenamefont {Moore}}]{Xu2019}%
  \BibitemOpen
  \bibfield  {author} {\bibinfo {author} {\bibfnamefont {T.}~\bibnamefont
  {Xu}}, \bibinfo {author} {\bibfnamefont {T.}~\bibnamefont {Morimoto}},\ and\
  \bibinfo {author} {\bibfnamefont {J.~E.}\ \bibnamefont {Moore}},\ }\bibfield
  {title} {\bibinfo {title} {Nonlinear optical effects in
  inversion-symmetry-breaking superconductors},\ }\href
  {https://doi.org/10.1103/PhysRevB.100.220501} {\bibfield  {journal} {\bibinfo
   {journal} {Phys. Rev. B}\ }\textbf {\bibinfo {volume} {100}},\ \bibinfo
  {pages} {220501} (\bibinfo {year} {2019})}\BibitemShut {NoStop}%
\bibitem [{\citenamefont {Daido}\ \emph {et~al.}(2022)\citenamefont {Daido},
  \citenamefont {Ikeda},\ and\ \citenamefont {Yanase}}]{Daido2022}%
  \BibitemOpen
  \bibfield  {author} {\bibinfo {author} {\bibfnamefont {A.}~\bibnamefont
  {Daido}}, \bibinfo {author} {\bibfnamefont {Y.}~\bibnamefont {Ikeda}},\ and\
  \bibinfo {author} {\bibfnamefont {Y.}~\bibnamefont {Yanase}},\ }\bibfield
  {title} {\bibinfo {title} {Intrinsic superconducting diode effect},\ }\href
  {https://doi.org/10.1103/PhysRevLett.128.037001} {\bibfield  {journal}
  {\bibinfo  {journal} {Phys. Rev. Lett.}\ }\textbf {\bibinfo {volume} {128}},\
  \bibinfo {pages} {037001} (\bibinfo {year} {2022})}\BibitemShut {NoStop}%
\bibitem [{\citenamefont {Yuan}\ and\ \citenamefont {Fu}(2022)}]{Yuan2022}%
  \BibitemOpen
  \bibfield  {author} {\bibinfo {author} {\bibfnamefont {N.~F.~Q.}\
  \bibnamefont {Yuan}}\ and\ \bibinfo {author} {\bibfnamefont {L.}~\bibnamefont
  {Fu}},\ }\bibfield  {title} {\bibinfo {title} {Supercurrent diode effect and
  finite-momentum superconductors},\ }\href
  {https://doi.org/10.1073/pnas.2119548119} {\bibfield  {journal} {\bibinfo
  {journal} {Proceedings of the National Academy of Sciences}\ }\textbf
  {\bibinfo {volume} {119}},\ \bibinfo {pages} {e2119548119} (\bibinfo {year}
  {2022})}\BibitemShut {NoStop}%
\bibitem [{\citenamefont {Watanabe}\ \emph {et~al.}(2022)\citenamefont
  {Watanabe}, \citenamefont {Daido},\ and\ \citenamefont
  {Yanase}}]{Watanabe2022}%
  \BibitemOpen
  \bibfield  {author} {\bibinfo {author} {\bibfnamefont {H.}~\bibnamefont
  {Watanabe}}, \bibinfo {author} {\bibfnamefont {A.}~\bibnamefont {Daido}},\
  and\ \bibinfo {author} {\bibfnamefont {Y.}~\bibnamefont {Yanase}},\
  }\bibfield  {title} {\bibinfo {title} {Nonreciprocal optical response in
  parity-breaking superconductors},\ }\href
  {https://doi.org/10.1103/PhysRevB.105.024308} {\bibfield  {journal} {\bibinfo
   {journal} {Phys. Rev. B}\ }\textbf {\bibinfo {volume} {105}},\ \bibinfo
  {pages} {024308} (\bibinfo {year} {2022})}\BibitemShut {NoStop}%
\bibitem [{\citenamefont {Tanaka}\ \emph {et~al.}(2023)\citenamefont {Tanaka},
  \citenamefont {Watanabe},\ and\ \citenamefont {Yanase}}]{Tanaka2023}%
  \BibitemOpen
  \bibfield  {author} {\bibinfo {author} {\bibfnamefont {H.}~\bibnamefont
  {Tanaka}}, \bibinfo {author} {\bibfnamefont {H.}~\bibnamefont {Watanabe}},\
  and\ \bibinfo {author} {\bibfnamefont {Y.}~\bibnamefont {Yanase}},\
  }\bibfield  {title} {\bibinfo {title} {Nonlinear optical responses in
  noncentrosymmetric superconductors},\ }\href
  {https://doi.org/10.1103/PhysRevB.107.024513} {\bibfield  {journal} {\bibinfo
   {journal} {Phys. Rev. B}\ }\textbf {\bibinfo {volume} {107}},\ \bibinfo
  {pages} {024513} (\bibinfo {year} {2023})}\BibitemShut {NoStop}%
\bibitem [{\citenamefont {Tanaka}\ \emph {et~al.}(2024)\citenamefont {Tanaka},
  \citenamefont {Watanabe},\ and\ \citenamefont {Yanase}}]{Tanaka2024}%
  \BibitemOpen
  \bibfield  {author} {\bibinfo {author} {\bibfnamefont {H.}~\bibnamefont
  {Tanaka}}, \bibinfo {author} {\bibfnamefont {H.}~\bibnamefont {Watanabe}},\
  and\ \bibinfo {author} {\bibfnamefont {Y.}~\bibnamefont {Yanase}},\
  }\bibfield  {title} {\bibinfo {title} {Nonlinear optical response in
  superconductors in magnetic field: Quantum geometry and topological
  superconductivity},\ }\href {https://doi.org/10.1103/PhysRevB.110.014520}
  {\bibfield  {journal} {\bibinfo  {journal} {Phys. Rev. B}\ }\textbf {\bibinfo
  {volume} {110}},\ \bibinfo {pages} {014520} (\bibinfo {year}
  {2024})}\BibitemShut {NoStop}%
\bibitem [{\citenamefont {Buckingham}(1957)}]{Buckingham1957}%
  \BibitemOpen
  \bibfield  {author} {\bibinfo {author} {\bibfnamefont {M.~J.}\ \bibnamefont
  {Buckingham}},\ }\bibfield  {title} {\bibinfo {title} {A note on the energy
  gap model of superconductivity},\ }\href {https://doi.org/10.1007/BF02856067}
  {\bibfield  {journal} {\bibinfo  {journal} {Il Nuovo Cimento (1955-1965)}\
  }\textbf {\bibinfo {volume} {5}},\ \bibinfo {pages} {1763} (\bibinfo {year}
  {1957})}\BibitemShut {NoStop}%
\bibitem [{\citenamefont {Schafroth}(1958)}]{Schafroth1958}%
  \BibitemOpen
  \bibfield  {author} {\bibinfo {author} {\bibfnamefont {M.~R.}\ \bibnamefont
  {Schafroth}},\ }\bibfield  {title} {\bibinfo {title} {Remarks on the meissner
  effect},\ }\href {https://doi.org/10.1103/PhysRev.111.72} {\bibfield
  {journal} {\bibinfo  {journal} {Phys. Rev.}\ }\textbf {\bibinfo {volume}
  {111}},\ \bibinfo {pages} {72} (\bibinfo {year} {1958})}\BibitemShut
  {NoStop}%
\bibitem [{\citenamefont {Rickayzen}(1958)}]{Rickayzen1958}%
  \BibitemOpen
  \bibfield  {author} {\bibinfo {author} {\bibfnamefont {G.}~\bibnamefont
  {Rickayzen}},\ }\bibfield  {title} {\bibinfo {title} {Meissner effect and
  gauge invariance},\ }\href {https://doi.org/10.1103/PhysRev.111.817}
  {\bibfield  {journal} {\bibinfo  {journal} {Phys. Rev.}\ }\textbf {\bibinfo
  {volume} {111}},\ \bibinfo {pages} {817} (\bibinfo {year}
  {1958})}\BibitemShut {NoStop}%
\bibitem [{\citenamefont {Wentzel}(1958)}]{Wentzel1958}%
  \BibitemOpen
  \bibfield  {author} {\bibinfo {author} {\bibfnamefont {G.}~\bibnamefont
  {Wentzel}},\ }\bibfield  {title} {\bibinfo {title} {Meissner effect},\ }\href
  {https://doi.org/10.1103/PhysRev.111.1488} {\bibfield  {journal} {\bibinfo
  {journal} {Phys. Rev.}\ }\textbf {\bibinfo {volume} {111}},\ \bibinfo {pages}
  {1488} (\bibinfo {year} {1958})}\BibitemShut {NoStop}%
\bibitem [{\citenamefont {Pines}\ and\ \citenamefont
  {Schrieffer}(1958)}]{Pines1958}%
  \BibitemOpen
  \bibfield  {author} {\bibinfo {author} {\bibfnamefont {D.}~\bibnamefont
  {Pines}}\ and\ \bibinfo {author} {\bibfnamefont {J.~R.}\ \bibnamefont
  {Schrieffer}},\ }\bibfield  {title} {\bibinfo {title} {Gauge invariance in
  the theory of superconductivity},\ }\href
  {https://doi.org/10.1007/BF02859833} {\bibfield  {journal} {\bibinfo
  {journal} {Il Nuovo Cimento (1955-1965)}\ }\textbf {\bibinfo {volume} {10}},\
  \bibinfo {pages} {496} (\bibinfo {year} {1958})}\BibitemShut {NoStop}%
\bibitem [{\citenamefont {Anderson}(1958)}]{Anderson1958}%
  \BibitemOpen
  \bibfield  {author} {\bibinfo {author} {\bibfnamefont {P.~W.}\ \bibnamefont
  {Anderson}},\ }\bibfield  {title} {\bibinfo {title} {Random-phase
  approximation in the theory of superconductivity},\ }\href
  {https://doi.org/10.1103/PhysRev.112.1900} {\bibfield  {journal} {\bibinfo
  {journal} {Phys. Rev.}\ }\textbf {\bibinfo {volume} {112}},\ \bibinfo {pages}
  {1900} (\bibinfo {year} {1958})}\BibitemShut {NoStop}%
\bibitem [{\citenamefont {Bogoliubov}\ \emph {et~al.}(1958)\citenamefont
  {Bogoliubov}, \citenamefont {Tolmachev},\ and\ \citenamefont
  {Shirkov}}]{Bogoliubov1958}%
  \BibitemOpen
  \bibfield  {author} {\bibinfo {author} {\bibfnamefont {N.~N.}\ \bibnamefont
  {Bogoliubov}}, \bibinfo {author} {\bibfnamefont {V.~V.}\ \bibnamefont
  {Tolmachev}},\ and\ \bibinfo {author} {\bibfnamefont {D.~V.}\ \bibnamefont
  {Shirkov}},\ }\href@noop {} {\emph {\bibinfo {title} {A new method in the
  theory of superconductivity}}}\ (\bibinfo  {publisher} {Academy of Sciences
  of USSR Press, Moskow},\ \bibinfo {year} {1958})\BibitemShut {NoStop}%
\bibitem [{\citenamefont {Yosida}(1959)}]{Yosida1959}%
  \BibitemOpen
  \bibfield  {author} {\bibinfo {author} {\bibfnamefont {K.}~\bibnamefont
  {Yosida}},\ }\bibfield  {title} {\bibinfo {title} {Collective excitations in
  superconductors},\ }\href {https://doi.org/10.1143/PTP.21.731} {\bibfield
  {journal} {\bibinfo  {journal} {Progress of Theoretical Physics}\ }\textbf
  {\bibinfo {volume} {21}},\ \bibinfo {pages} {731} (\bibinfo {year}
  {1959})}\BibitemShut {NoStop}%
\bibitem [{\citenamefont {Rickayzen}(1959{\natexlab{a}})}]{Rickayzen1959a}%
  \BibitemOpen
  \bibfield  {author} {\bibinfo {author} {\bibfnamefont {G.}~\bibnamefont
  {Rickayzen}},\ }\bibfield  {title} {\bibinfo {title} {Collective excitations
  and the meissner effect},\ }\href {https://doi.org/10.1103/PhysRevLett.2.90}
  {\bibfield  {journal} {\bibinfo  {journal} {Phys. Rev. Lett.}\ }\textbf
  {\bibinfo {volume} {2}},\ \bibinfo {pages} {90} (\bibinfo {year}
  {1959}{\natexlab{a}})}\BibitemShut {NoStop}%
\bibitem [{\citenamefont {Rickayzen}(1959{\natexlab{b}})}]{Rickayzen1959}%
  \BibitemOpen
  \bibfield  {author} {\bibinfo {author} {\bibfnamefont {G.}~\bibnamefont
  {Rickayzen}},\ }\bibfield  {title} {\bibinfo {title} {Collective excitations
  in the theory of superconductivity},\ }\href
  {https://doi.org/10.1103/PhysRev.115.795} {\bibfield  {journal} {\bibinfo
  {journal} {Phys. Rev.}\ }\textbf {\bibinfo {volume} {115}},\ \bibinfo {pages}
  {795} (\bibinfo {year} {1959}{\natexlab{b}})}\BibitemShut {NoStop}%
\bibitem [{\citenamefont {Nambu}(1960)}]{Nambu1960}%
  \BibitemOpen
  \bibfield  {author} {\bibinfo {author} {\bibfnamefont {Y.}~\bibnamefont
  {Nambu}},\ }\bibfield  {title} {\bibinfo {title} {Quasi-particles and gauge
  invariance in the theory of superconductivity},\ }\href
  {https://doi.org/10.1103/PhysRev.117.648} {\bibfield  {journal} {\bibinfo
  {journal} {Phys. Rev.}\ }\textbf {\bibinfo {volume} {117}},\ \bibinfo {pages}
  {648} (\bibinfo {year} {1960})}\BibitemShut {NoStop}%
\bibitem [{\citenamefont {Martin}\ and\ \citenamefont
  {Schwinger}(1959)}]{Martin1959}%
  \BibitemOpen
  \bibfield  {author} {\bibinfo {author} {\bibfnamefont {P.~C.}\ \bibnamefont
  {Martin}}\ and\ \bibinfo {author} {\bibfnamefont {J.}~\bibnamefont
  {Schwinger}},\ }\bibfield  {title} {\bibinfo {title} {Theory of many-particle
  systems. i},\ }\href {https://doi.org/10.1103/PhysRev.115.1342} {\bibfield
  {journal} {\bibinfo  {journal} {Phys. Rev.}\ }\textbf {\bibinfo {volume}
  {115}},\ \bibinfo {pages} {1342} (\bibinfo {year} {1959})}\BibitemShut
  {NoStop}%
\bibitem [{\citenamefont {Kadanoff}\ and\ \citenamefont
  {Martin}(1961)}]{Kadanoff1961}%
  \BibitemOpen
  \bibfield  {author} {\bibinfo {author} {\bibfnamefont {L.~P.}\ \bibnamefont
  {Kadanoff}}\ and\ \bibinfo {author} {\bibfnamefont {P.~C.}\ \bibnamefont
  {Martin}},\ }\bibfield  {title} {\bibinfo {title} {Theory of many-particle
  systems. ii. superconductivity},\ }\href
  {https://doi.org/10.1103/PhysRev.124.670} {\bibfield  {journal} {\bibinfo
  {journal} {Phys. Rev.}\ }\textbf {\bibinfo {volume} {124}},\ \bibinfo {pages}
  {670} (\bibinfo {year} {1961})}\BibitemShut {NoStop}%
\bibitem [{\citenamefont {Baym}\ and\ \citenamefont
  {Kadanoff}(1961)}]{Baym1961}%
  \BibitemOpen
  \bibfield  {author} {\bibinfo {author} {\bibfnamefont {G.}~\bibnamefont
  {Baym}}\ and\ \bibinfo {author} {\bibfnamefont {L.~P.}\ \bibnamefont
  {Kadanoff}},\ }\bibfield  {title} {\bibinfo {title} {Conservation laws and
  correlation functions},\ }\href {https://doi.org/10.1103/PhysRev.124.287}
  {\bibfield  {journal} {\bibinfo  {journal} {Phys. Rev.}\ }\textbf {\bibinfo
  {volume} {124}},\ \bibinfo {pages} {287} (\bibinfo {year}
  {1961})}\BibitemShut {NoStop}%
\bibitem [{\citenamefont {Baym}(1962)}]{Baym1962}%
  \BibitemOpen
  \bibfield  {author} {\bibinfo {author} {\bibfnamefont {G.}~\bibnamefont
  {Baym}},\ }\bibfield  {title} {\bibinfo {title} {Self-consistent
  approximations in many-body systems},\ }\href
  {https://doi.org/10.1103/PhysRev.127.1391} {\bibfield  {journal} {\bibinfo
  {journal} {Phys. Rev.}\ }\textbf {\bibinfo {volume} {127}},\ \bibinfo {pages}
  {1391} (\bibinfo {year} {1962})}\BibitemShut {NoStop}%
\bibitem [{\citenamefont {Ambegaokar}\ and\ \citenamefont
  {Kadanoff}(1961)}]{Ambegaokar1961}%
  \BibitemOpen
  \bibfield  {author} {\bibinfo {author} {\bibfnamefont {V.}~\bibnamefont
  {Ambegaokar}}\ and\ \bibinfo {author} {\bibfnamefont {L.~P.}\ \bibnamefont
  {Kadanoff}},\ }\bibfield  {title} {\bibinfo {title} {Electromagnetic
  properties of superconductors},\ }\href {https://doi.org/10.1007/BF02787879}
  {\bibfield  {journal} {\bibinfo  {journal} {Il Nuovo Cimento (1955-1965)}\
  }\textbf {\bibinfo {volume} {22}},\ \bibinfo {pages} {914} (\bibinfo {year}
  {1961})}\BibitemShut {NoStop}%
\bibitem [{\citenamefont {Kulik}\ \emph {et~al.}(1981)\citenamefont {Kulik},
  \citenamefont {Entin-Wohlman},\ and\ \citenamefont {Orbach}}]{Kulik1981}%
  \BibitemOpen
  \bibfield  {author} {\bibinfo {author} {\bibfnamefont {I.~O.}\ \bibnamefont
  {Kulik}}, \bibinfo {author} {\bibfnamefont {O.}~\bibnamefont
  {Entin-Wohlman}},\ and\ \bibinfo {author} {\bibfnamefont {R.}~\bibnamefont
  {Orbach}},\ }\bibfield  {title} {\bibinfo {title} {Pair susceptibility and
  mode propagation in superconductors: A microscopic approach},\ }\href
  {https://doi.org/10.1007/BF00115617} {\bibfield  {journal} {\bibinfo
  {journal} {Journal of Low Temperature Physics}\ }\textbf {\bibinfo {volume}
  {43}},\ \bibinfo {pages} {591} (\bibinfo {year} {1981})}\BibitemShut
  {NoStop}%
\bibitem [{\citenamefont {Zha}\ \emph {et~al.}(1995)\citenamefont {Zha},
  \citenamefont {Levin},\ and\ \citenamefont {Liu}}]{Zha1995}%
  \BibitemOpen
  \bibfield  {author} {\bibinfo {author} {\bibfnamefont {Y.}~\bibnamefont
  {Zha}}, \bibinfo {author} {\bibfnamefont {K.}~\bibnamefont {Levin}},\ and\
  \bibinfo {author} {\bibfnamefont {D.~Z.}\ \bibnamefont {Liu}},\ }\bibfield
  {title} {\bibinfo {title} {Collective modes and implications for c-axis
  optical experiments in layered cuprates},\ }\href
  {https://doi.org/10.1103/PhysRevB.51.6602} {\bibfield  {journal} {\bibinfo
  {journal} {Phys. Rev. B}\ }\textbf {\bibinfo {volume} {51}},\ \bibinfo
  {pages} {6602} (\bibinfo {year} {1995})}\BibitemShut {NoStop}%
\bibitem [{\citenamefont {Guo}\ \emph {et~al.}(2013)\citenamefont {Guo},
  \citenamefont {Chien},\ and\ \citenamefont {He}}]{Guo2013}%
  \BibitemOpen
  \bibfield  {author} {\bibinfo {author} {\bibfnamefont {H.}~\bibnamefont
  {Guo}}, \bibinfo {author} {\bibfnamefont {C.-C.}\ \bibnamefont {Chien}},\
  and\ \bibinfo {author} {\bibfnamefont {Y.}~\bibnamefont {He}},\ }\bibfield
  {title} {\bibinfo {title} {Theories of linear response in bcs superfluids and
  how they meet fundamental constraints},\ }\href
  {https://doi.org/10.1007/s10909-012-0853-7} {\bibfield  {journal} {\bibinfo
  {journal} {Journal of Low Temperature Physics}\ }\textbf {\bibinfo {volume}
  {172}},\ \bibinfo {pages} {5} (\bibinfo {year} {2013})}\BibitemShut {NoStop}%
\bibitem [{\citenamefont {Lutchyn}\ \emph {et~al.}(2008)\citenamefont
  {Lutchyn}, \citenamefont {Nagornykh},\ and\ \citenamefont
  {Yakovenko}}]{Lutchyn2008}%
  \BibitemOpen
  \bibfield  {author} {\bibinfo {author} {\bibfnamefont {R.~M.}\ \bibnamefont
  {Lutchyn}}, \bibinfo {author} {\bibfnamefont {P.}~\bibnamefont {Nagornykh}},\
  and\ \bibinfo {author} {\bibfnamefont {V.~M.}\ \bibnamefont {Yakovenko}},\
  }\bibfield  {title} {\bibinfo {title} {Gauge-invariant electromagnetic
  response of a chiral ${p}_{x}+i{p}_{y}$ superconductor},\ }\href
  {https://doi.org/10.1103/PhysRevB.77.144516} {\bibfield  {journal} {\bibinfo
  {journal} {Phys. Rev. B}\ }\textbf {\bibinfo {volume} {77}},\ \bibinfo
  {pages} {144516} (\bibinfo {year} {2008})}\BibitemShut {NoStop}%
\bibitem [{\citenamefont {Anderson}\ \emph {et~al.}(2016)\citenamefont
  {Anderson}, \citenamefont {Boyack}, \citenamefont {Wu},\ and\ \citenamefont
  {Levin}}]{Anderson2016}%
  \BibitemOpen
  \bibfield  {author} {\bibinfo {author} {\bibfnamefont {B.~M.}\ \bibnamefont
  {Anderson}}, \bibinfo {author} {\bibfnamefont {R.}~\bibnamefont {Boyack}},
  \bibinfo {author} {\bibfnamefont {C.-T.}\ \bibnamefont {Wu}},\ and\ \bibinfo
  {author} {\bibfnamefont {K.}~\bibnamefont {Levin}},\ }\bibfield  {title}
  {\bibinfo {title} {Going beyond the bcs level in the superfluid path
  integral: A consistent treatment of electrodynamics and thermodynamics},\
  }\href {https://doi.org/10.1103/PhysRevB.93.180504} {\bibfield  {journal}
  {\bibinfo  {journal} {Phys. Rev. B}\ }\textbf {\bibinfo {volume} {93}},\
  \bibinfo {pages} {180504} (\bibinfo {year} {2016})}\BibitemShut {NoStop}%
\bibitem [{\citenamefont {Boyack}\ \emph {et~al.}(2016)\citenamefont {Boyack},
  \citenamefont {Anderson}, \citenamefont {Wu},\ and\ \citenamefont
  {Levin}}]{Boyack2016}%
  \BibitemOpen
  \bibfield  {author} {\bibinfo {author} {\bibfnamefont {R.}~\bibnamefont
  {Boyack}}, \bibinfo {author} {\bibfnamefont {B.~M.}\ \bibnamefont
  {Anderson}}, \bibinfo {author} {\bibfnamefont {C.-T.}\ \bibnamefont {Wu}},\
  and\ \bibinfo {author} {\bibfnamefont {K.}~\bibnamefont {Levin}},\ }\bibfield
   {title} {\bibinfo {title} {Gauge-invariant theories of linear response for
  strongly correlated superconductors},\ }\href
  {https://doi.org/10.1103/PhysRevB.94.094508} {\bibfield  {journal} {\bibinfo
  {journal} {Phys. Rev. B}\ }\textbf {\bibinfo {volume} {94}},\ \bibinfo
  {pages} {094508} (\bibinfo {year} {2016})}\BibitemShut {NoStop}%
\bibitem [{\citenamefont {Dai}\ and\ \citenamefont {Lee}(2017)}]{Dai2017}%
  \BibitemOpen
  \bibfield  {author} {\bibinfo {author} {\bibfnamefont {Z.}~\bibnamefont
  {Dai}}\ and\ \bibinfo {author} {\bibfnamefont {P.~A.}\ \bibnamefont {Lee}},\
  }\bibfield  {title} {\bibinfo {title} {Optical conductivity from pair density
  waves},\ }\href {https://doi.org/10.1103/PhysRevB.95.014506} {\bibfield
  {journal} {\bibinfo  {journal} {Phys. Rev. B}\ }\textbf {\bibinfo {volume}
  {95}},\ \bibinfo {pages} {014506} (\bibinfo {year} {2017})}\BibitemShut
  {NoStop}%
\bibitem [{\citenamefont {Boyack}(2018)}]{Boyack2018}%
  \BibitemOpen
  \bibfield  {author} {\bibinfo {author} {\bibfnamefont {R.}~\bibnamefont
  {Boyack}},\ }\bibfield  {title} {\bibinfo {title} {Restoring gauge invariance
  in conventional fluctuation corrections to a superconductor},\ }\href
  {https://doi.org/10.1103/PhysRevB.98.184504} {\bibfield  {journal} {\bibinfo
  {journal} {Phys. Rev. B}\ }\textbf {\bibinfo {volume} {98}},\ \bibinfo
  {pages} {184504} (\bibinfo {year} {2018})}\BibitemShut {NoStop}%
\bibitem [{\citenamefont {Papaj}\ and\ \citenamefont
  {Moore}(2022)}]{Papaj2022}%
  \BibitemOpen
  \bibfield  {author} {\bibinfo {author} {\bibfnamefont {M.}~\bibnamefont
  {Papaj}}\ and\ \bibinfo {author} {\bibfnamefont {J.~E.}\ \bibnamefont
  {Moore}},\ }\bibfield  {title} {\bibinfo {title} {Current-enabled optical
  conductivity of superconductors},\ }\href
  {https://doi.org/10.1103/PhysRevB.106.L220504} {\bibfield  {journal}
  {\bibinfo  {journal} {Phys. Rev. B}\ }\textbf {\bibinfo {volume} {106}},\
  \bibinfo {pages} {L220504} (\bibinfo {year} {2022})}\BibitemShut {NoStop}%
\bibitem [{\citenamefont {Oh}\ and\ \citenamefont {Watanabe}(2024)}]{Oh2024}%
  \BibitemOpen
  \bibfield  {author} {\bibinfo {author} {\bibfnamefont {C.-g.}\ \bibnamefont
  {Oh}}\ and\ \bibinfo {author} {\bibfnamefont {H.}~\bibnamefont {Watanabe}},\
  }\bibfield  {title} {\bibinfo {title} {Revisiting electromagnetic response of
  superconductors in mean-field approximation},\ }\href
  {https://doi.org/10.1103/PhysRevResearch.6.013058} {\bibfield  {journal}
  {\bibinfo  {journal} {Phys. Rev. Res.}\ }\textbf {\bibinfo {volume} {6}},\
  \bibinfo {pages} {013058} (\bibinfo {year} {2024})}\BibitemShut {NoStop}%
\bibitem [{\citenamefont {Wang}\ \emph {et~al.}(2025)\citenamefont {Wang},
  \citenamefont {Boyack},\ and\ \citenamefont {Levin}}]{Wang2025}%
  \BibitemOpen
  \bibfield  {author} {\bibinfo {author} {\bibfnamefont {K.}~\bibnamefont
  {Wang}}, \bibinfo {author} {\bibfnamefont {R.}~\bibnamefont {Boyack}},\ and\
  \bibinfo {author} {\bibfnamefont {K.}~\bibnamefont {Levin}},\ }\bibfield
  {title} {\bibinfo {title} {Higgs amplitude mode in optical conductivity in
  the presence of a supercurrent: Gauge-invariant formulation with disorder},\
  }\bibfield  {journal} {\bibinfo  {journal} {Physical Review B}\ }\textbf
  {\bibinfo {volume} {111}},\ \href
  {https://doi.org/10.1103/physrevb.111.144512} {10.1103/physrevb.111.144512}
  (\bibinfo {year} {2025})\BibitemShut {NoStop}%
\bibitem [{\citenamefont {Yu}\ and\ \citenamefont
  {Wu}(2017{\natexlab{a}})}]{Yu2017}%
  \BibitemOpen
  \bibfield  {author} {\bibinfo {author} {\bibfnamefont {T.}~\bibnamefont
  {Yu}}\ and\ \bibinfo {author} {\bibfnamefont {M.~W.}\ \bibnamefont {Wu}},\
  }\bibfield  {title} {\bibinfo {title} {Gauge-invariant theory of
  quasiparticle and condensate dynamics in response to terahertz optical pulses
  in superconducting semiconductor quantum wells. i. $s$-wave superconductivity
  in the weak spin-orbit coupling limit},\ }\href
  {https://doi.org/10.1103/PhysRevB.96.155311} {\bibfield  {journal} {\bibinfo
  {journal} {Phys. Rev. B}\ }\textbf {\bibinfo {volume} {96}},\ \bibinfo
  {pages} {155311} (\bibinfo {year} {2017}{\natexlab{a}})}\BibitemShut
  {NoStop}%
\bibitem [{\citenamefont {Yu}\ and\ \citenamefont
  {Wu}(2017{\natexlab{b}})}]{Yu2017a}%
  \BibitemOpen
  \bibfield  {author} {\bibinfo {author} {\bibfnamefont {T.}~\bibnamefont
  {Yu}}\ and\ \bibinfo {author} {\bibfnamefont {M.~W.}\ \bibnamefont {Wu}},\
  }\bibfield  {title} {\bibinfo {title} {Gauge-invariant theory of
  quasiparticle and condensate dynamics in response to terahertz optical pulses
  in superconducting semiconductor quantum wells. ii. ($s+p$)-wave
  superconductivity in the strong spin-orbit coupling limit},\ }\href
  {https://doi.org/10.1103/PhysRevB.96.155312} {\bibfield  {journal} {\bibinfo
  {journal} {Phys. Rev. B}\ }\textbf {\bibinfo {volume} {96}},\ \bibinfo
  {pages} {155312} (\bibinfo {year} {2017}{\natexlab{b}})}\BibitemShut
  {NoStop}%
\bibitem [{\citenamefont {Yang}\ and\ \citenamefont {Wu}(2018)}]{Yang2018gi}%
  \BibitemOpen
  \bibfield  {author} {\bibinfo {author} {\bibfnamefont {F.}~\bibnamefont
  {Yang}}\ and\ \bibinfo {author} {\bibfnamefont {M.~W.}\ \bibnamefont {Wu}},\
  }\bibfield  {title} {\bibinfo {title} {Gauge-invariant microscopic kinetic
  theory of superconductivity in response to electromagnetic fields},\ }\href
  {https://doi.org/10.1103/PhysRevB.98.094507} {\bibfield  {journal} {\bibinfo
  {journal} {Phys. Rev. B}\ }\textbf {\bibinfo {volume} {98}},\ \bibinfo
  {pages} {094507} (\bibinfo {year} {2018})}\BibitemShut {NoStop}%
\bibitem [{\citenamefont {Yang}\ and\ \citenamefont {Wu}(2019)}]{Yang2019gi}%
  \BibitemOpen
  \bibfield  {author} {\bibinfo {author} {\bibfnamefont {F.}~\bibnamefont
  {Yang}}\ and\ \bibinfo {author} {\bibfnamefont {M.~W.}\ \bibnamefont {Wu}},\
  }\bibfield  {title} {\bibinfo {title} {Gauge-invariant microscopic kinetic
  theory of superconductivity: Application to the optical response of
  nambu-goldstone and higgs modes},\ }\href
  {https://doi.org/10.1103/PhysRevB.100.104513} {\bibfield  {journal} {\bibinfo
   {journal} {Phys. Rev. B}\ }\textbf {\bibinfo {volume} {100}},\ \bibinfo
  {pages} {104513} (\bibinfo {year} {2019})}\BibitemShut {NoStop}%
\bibitem [{\citenamefont {Yang}\ and\ \citenamefont
  {Wu}(2020{\natexlab{a}})}]{Yang2020}%
  \BibitemOpen
  \bibfield  {author} {\bibinfo {author} {\bibfnamefont {F.}~\bibnamefont
  {Yang}}\ and\ \bibinfo {author} {\bibfnamefont {M.~W.}\ \bibnamefont {Wu}},\
  }\bibfield  {title} {\bibinfo {title} {Theory of higgs modes in $d$-wave
  superconductors},\ }\href {https://doi.org/10.1103/PhysRevB.102.014511}
  {\bibfield  {journal} {\bibinfo  {journal} {Phys. Rev. B}\ }\textbf {\bibinfo
  {volume} {102}},\ \bibinfo {pages} {014511} (\bibinfo {year}
  {2020}{\natexlab{a}})}\BibitemShut {NoStop}%
\bibitem [{\citenamefont {Yang}\ and\ \citenamefont
  {Wu}(2020{\natexlab{b}})}]{Yang2020a}%
  \BibitemOpen
  \bibfield  {author} {\bibinfo {author} {\bibfnamefont {F.}~\bibnamefont
  {Yang}}\ and\ \bibinfo {author} {\bibfnamefont {M.~W.}\ \bibnamefont {Wu}},\
  }\bibfield  {title} {\bibinfo {title} {Influence of scattering on the optical
  response of superconductors},\ }\href
  {https://doi.org/10.1103/PhysRevB.102.144508} {\bibfield  {journal} {\bibinfo
   {journal} {Phys. Rev. B}\ }\textbf {\bibinfo {volume} {102}},\ \bibinfo
  {pages} {144508} (\bibinfo {year} {2020}{\natexlab{b}})}\BibitemShut
  {NoStop}%
\bibitem [{\citenamefont {Tanaka}\ and\ \citenamefont
  {Yanase}(2025)}]{Tanaka2025}%
  \BibitemOpen
  \bibfield  {author} {\bibinfo {author} {\bibfnamefont {H.}~\bibnamefont
  {Tanaka}}\ and\ \bibinfo {author} {\bibfnamefont {Y.}~\bibnamefont
  {Yanase}},\ }\href {https://arxiv.org/abs/2502.15373} {\bibinfo {title}
  {Vertex correction for the linear and nonlinear optical responses in
  superconductors: multiband effect and topological superconductivity}}
  (\bibinfo {year} {2025}),\ \Eprint {https://arxiv.org/abs/2502.15373}
  {arXiv:2502.15373 [cond-mat.supr-con]} \BibitemShut {NoStop}%
\bibitem [{\citenamefont {Huang}\ and\ \citenamefont {Wang}(2023)}]{Huang2023}%
  \BibitemOpen
  \bibfield  {author} {\bibinfo {author} {\bibfnamefont {L.}~\bibnamefont
  {Huang}}\ and\ \bibinfo {author} {\bibfnamefont {J.}~\bibnamefont {Wang}},\
  }\bibfield  {title} {\bibinfo {title} {Second-order optical response of
  superconductors induced by supercurrent injection},\ }\href
  {https://doi.org/10.1103/PhysRevB.108.224516} {\bibfield  {journal} {\bibinfo
   {journal} {Phys. Rev. B}\ }\textbf {\bibinfo {volume} {108}},\ \bibinfo
  {pages} {224516} (\bibinfo {year} {2023})}\BibitemShut {NoStop}%
\bibitem [{\citenamefont {Watanabe}\ and\ \citenamefont
  {Watanabe}(2024)}]{SW2024L}%
  \BibitemOpen
  \bibfield  {author} {\bibinfo {author} {\bibfnamefont {S.}~\bibnamefont
  {Watanabe}}\ and\ \bibinfo {author} {\bibfnamefont {H.}~\bibnamefont
  {Watanabe}},\ }\href {https://arxiv.org/abs/2410.18679} {\bibinfo {title} {A
  gauge-invariant formulation of optical responses in superconductors}}
  (\bibinfo {year} {2024}),\ \Eprint {https://arxiv.org/abs/2410.18679}
  {arXiv:2410.18679 [cond-mat.supr-con]} \BibitemShut {NoStop}%
\bibitem [{\citenamefont {Ward}(1950)}]{Ward1950}%
  \BibitemOpen
  \bibfield  {author} {\bibinfo {author} {\bibfnamefont {J.~C.}\ \bibnamefont
  {Ward}},\ }\bibfield  {title} {\bibinfo {title} {An identity in quantum
  electrodynamics},\ }\href {https://doi.org/10.1103/PhysRev.78.182} {\bibfield
   {journal} {\bibinfo  {journal} {Phys. Rev.}\ }\textbf {\bibinfo {volume}
  {78}},\ \bibinfo {pages} {182} (\bibinfo {year} {1950})}\BibitemShut
  {NoStop}%
\bibitem [{\citenamefont {Takahashi}(1957)}]{Takahashi1957}%
  \BibitemOpen
  \bibfield  {author} {\bibinfo {author} {\bibfnamefont {Y.}~\bibnamefont
  {Takahashi}},\ }\bibfield  {title} {\bibinfo {title} {On the generalized ward
  identity},\ }\href {https://doi.org/10.1007/BF02832514} {\bibfield  {journal}
  {\bibinfo  {journal} {Il Nuovo Cimento (1955-1965)}\ }\textbf {\bibinfo
  {volume} {6}},\ \bibinfo {pages} {371} (\bibinfo {year} {1957})}\BibitemShut
  {NoStop}%
\bibitem [{\citenamefont {Rostami}\ \emph {et~al.}(2021)\citenamefont
  {Rostami}, \citenamefont {Katsnelson}, \citenamefont {Vignale},\ and\
  \citenamefont {Polini}}]{Rostami2021}%
  \BibitemOpen
  \bibfield  {author} {\bibinfo {author} {\bibfnamefont {H.}~\bibnamefont
  {Rostami}}, \bibinfo {author} {\bibfnamefont {M.~I.}\ \bibnamefont
  {Katsnelson}}, \bibinfo {author} {\bibfnamefont {G.}~\bibnamefont
  {Vignale}},\ and\ \bibinfo {author} {\bibfnamefont {M.}~\bibnamefont
  {Polini}},\ }\bibfield  {title} {\bibinfo {title} {Gauge invariance and ward
  identities in nonlinear response theory},\ }\href
  {https://doi.org/https://doi.org/10.1016/j.aop.2021.168523} {\bibfield
  {journal} {\bibinfo  {journal} {Annals of Physics}\ }\textbf {\bibinfo
  {volume} {431}},\ \bibinfo {pages} {168523} (\bibinfo {year}
  {2021})}\BibitemShut {NoStop}%
\bibitem [{\citenamefont {Parker}\ \emph {et~al.}(2019)\citenamefont {Parker},
  \citenamefont {Morimoto}, \citenamefont {Orenstein},\ and\ \citenamefont
  {Moore}}]{Parker2019}%
  \BibitemOpen
  \bibfield  {author} {\bibinfo {author} {\bibfnamefont {D.~E.}\ \bibnamefont
  {Parker}}, \bibinfo {author} {\bibfnamefont {T.}~\bibnamefont {Morimoto}},
  \bibinfo {author} {\bibfnamefont {J.}~\bibnamefont {Orenstein}},\ and\
  \bibinfo {author} {\bibfnamefont {J.~E.}\ \bibnamefont {Moore}},\ }\bibfield
  {title} {\bibinfo {title} {Diagrammatic approach to nonlinear optical
  response with application to weyl semimetals},\ }\href
  {https://doi.org/10.1103/PhysRevB.99.045121} {\bibfield  {journal} {\bibinfo
  {journal} {Phys. Rev. B}\ }\textbf {\bibinfo {volume} {99}},\ \bibinfo
  {pages} {045121} (\bibinfo {year} {2019})}\BibitemShut {NoStop}%
\bibitem [{\citenamefont {Kamatani}\ \emph {et~al.}(2022)\citenamefont
  {Kamatani}, \citenamefont {Kitamura}, \citenamefont {Tsuji}, \citenamefont
  {Shimano},\ and\ \citenamefont {Morimoto}}]{Kamatani2022}%
  \BibitemOpen
  \bibfield  {author} {\bibinfo {author} {\bibfnamefont {T.}~\bibnamefont
  {Kamatani}}, \bibinfo {author} {\bibfnamefont {S.}~\bibnamefont {Kitamura}},
  \bibinfo {author} {\bibfnamefont {N.}~\bibnamefont {Tsuji}}, \bibinfo
  {author} {\bibfnamefont {R.}~\bibnamefont {Shimano}},\ and\ \bibinfo {author}
  {\bibfnamefont {T.}~\bibnamefont {Morimoto}},\ }\bibfield  {title} {\bibinfo
  {title} {Optical response of the leggett mode in multiband superconductors in
  the linear response regime},\ }\href
  {https://doi.org/10.1103/PhysRevB.105.094520} {\bibfield  {journal} {\bibinfo
   {journal} {Phys. Rev. B}\ }\textbf {\bibinfo {volume} {105}},\ \bibinfo
  {pages} {094520} (\bibinfo {year} {2022})}\BibitemShut {NoStop}%
\bibitem [{\citenamefont {Nagashima}\ \emph
  {et~al.}(2024{\natexlab{a}})\citenamefont {Nagashima}, \citenamefont {Tian},
  \citenamefont {Haenel}, \citenamefont {Tsuji},\ and\ \citenamefont
  {Manske}}]{Nagashima2024}%
  \BibitemOpen
  \bibfield  {author} {\bibinfo {author} {\bibfnamefont {R.}~\bibnamefont
  {Nagashima}}, \bibinfo {author} {\bibfnamefont {S.}~\bibnamefont {Tian}},
  \bibinfo {author} {\bibfnamefont {R.}~\bibnamefont {Haenel}}, \bibinfo
  {author} {\bibfnamefont {N.}~\bibnamefont {Tsuji}},\ and\ \bibinfo {author}
  {\bibfnamefont {D.}~\bibnamefont {Manske}},\ }\bibfield  {title} {\bibinfo
  {title} {Classification of lifshitz invariant in multiband superconductors:
  An application to leggett modes in the linear response regime in kagome
  lattice models},\ }\href {https://doi.org/10.1103/PhysRevResearch.6.013120}
  {\bibfield  {journal} {\bibinfo  {journal} {Phys. Rev. Res.}\ }\textbf
  {\bibinfo {volume} {6}},\ \bibinfo {pages} {013120} (\bibinfo {year}
  {2024}{\natexlab{a}})}\BibitemShut {NoStop}%
\bibitem [{\citenamefont {Nagashima}\ \emph
  {et~al.}(2024{\natexlab{b}})\citenamefont {Nagashima}, \citenamefont
  {Mouilleron},\ and\ \citenamefont {Tsuji}}]{Nagashima2024a}%
  \BibitemOpen
  \bibfield  {author} {\bibinfo {author} {\bibfnamefont {R.}~\bibnamefont
  {Nagashima}}, \bibinfo {author} {\bibfnamefont {T.}~\bibnamefont
  {Mouilleron}},\ and\ \bibinfo {author} {\bibfnamefont {N.}~\bibnamefont
  {Tsuji}},\ }\href {https://arxiv.org/abs/2410.18438} {\bibinfo {title}
  {Optically active higgs and leggett modes in multiband pair-density-wave
  superconductors with lifshitz invariant}} (\bibinfo {year}
  {2024}{\natexlab{b}}),\ \Eprint {https://arxiv.org/abs/2410.18438}
  {arXiv:2410.18438 [cond-mat.supr-con]} \BibitemShut {NoStop}%
\bibitem [{\citenamefont {Rice}\ and\ \citenamefont {Mele}(1982)}]{Rice1982}%
  \BibitemOpen
  \bibfield  {author} {\bibinfo {author} {\bibfnamefont {M.~J.}\ \bibnamefont
  {Rice}}\ and\ \bibinfo {author} {\bibfnamefont {E.~J.}\ \bibnamefont
  {Mele}},\ }\bibfield  {title} {\bibinfo {title} {Elementary excitations of a
  linearly conjugated diatomic polymer},\ }\href
  {https://doi.org/10.1103/PhysRevLett.49.1455} {\bibfield  {journal} {\bibinfo
   {journal} {Phys. Rev. Lett.}\ }\textbf {\bibinfo {volume} {49}},\ \bibinfo
  {pages} {1455} (\bibinfo {year} {1982})}\BibitemShut {NoStop}%
\bibitem [{\citenamefont {Leggett}(1966)}]{Leggett1966}%
  \BibitemOpen
  \bibfield  {author} {\bibinfo {author} {\bibfnamefont {A.~J.}\ \bibnamefont
  {Leggett}},\ }\bibfield  {title} {\bibinfo {title} {{Number-Phase
  Fluctuations in Two-Band Superconductors}},\ }\href
  {https://doi.org/10.1143/PTP.36.901} {\bibfield  {journal} {\bibinfo
  {journal} {Progress of Theoretical Physics}\ }\textbf {\bibinfo {volume}
  {36}},\ \bibinfo {pages} {901} (\bibinfo {year} {1966})}\BibitemShut
  {NoStop}%
\bibitem [{\citenamefont {Paramekanti}\ \emph {et~al.}(2000)\citenamefont
  {Paramekanti}, \citenamefont {Randeria}, \citenamefont {Ramakrishnan},\ and\
  \citenamefont {Mandal}}]{Paramekanti2000}%
  \BibitemOpen
  \bibfield  {author} {\bibinfo {author} {\bibfnamefont {A.}~\bibnamefont
  {Paramekanti}}, \bibinfo {author} {\bibfnamefont {M.}~\bibnamefont
  {Randeria}}, \bibinfo {author} {\bibfnamefont {T.~V.}\ \bibnamefont
  {Ramakrishnan}},\ and\ \bibinfo {author} {\bibfnamefont {S.~S.}\ \bibnamefont
  {Mandal}},\ }\bibfield  {title} {\bibinfo {title} {Effective actions and
  phase fluctuations in d-wave superconductors},\ }\href
  {https://doi.org/10.1103/PhysRevB.62.6786} {\bibfield  {journal} {\bibinfo
  {journal} {Phys. Rev. B}\ }\textbf {\bibinfo {volume} {62}},\ \bibinfo
  {pages} {6786} (\bibinfo {year} {2000})}\BibitemShut {NoStop}%
\bibitem [{\citenamefont {Ahn}\ and\ \citenamefont {Nagaosa}(2021)}]{Ahn2021}%
  \BibitemOpen
  \bibfield  {author} {\bibinfo {author} {\bibfnamefont {J.}~\bibnamefont
  {Ahn}}\ and\ \bibinfo {author} {\bibfnamefont {N.}~\bibnamefont {Nagaosa}},\
  }\bibfield  {title} {\bibinfo {title} {Theory of optical responses in clean
  multi-band superconductors},\ }\href
  {https://doi.org/10.1038/s41467-021-21905-x} {\bibfield  {journal} {\bibinfo
  {journal} {Nature Communications}\ }\textbf {\bibinfo {volume} {12}},\
  \bibinfo {pages} {1617} (\bibinfo {year} {2021})}\BibitemShut {NoStop}%
\end{thebibliography}%

\appendix
\onecolumngrid
\newpage
\section{The detailed derivation of the Ward identities}
\label{WI}
In this appendix, we present the detailed derivation of the Ward identities.
Given a classical gauge field $A_\mu$, the generating functional for correlation functions is defined by
\begin{align}
  Z[A_\mu,\bar{\eta},\eta]=\grassfint{\psi}\exp[-S[A_\mu,\bar{\psi},\psi]+\int d^4x (\bar{\eta}(x)\psi(x)+\bar{\psi}(x)\eta(x))],
\end{align}
where $\bar{\eta},\eta$ are sources and $S[A_\mu,\bar{\psi},\psi]$ is the action of electronic system.
Furthermore, assuming that the action $S[A_\mu,\bar{\psi},\psi]$ is invariant under local $U(1)$ gauge transformation 
\begin{align}
  A_\mu\to A_\mu-\partial_\mu\theta, \bar{\psi}\to\bar{\psi}e^{-i\theta\tau_3},\psi\to e^{i\theta\tau_3}\psi,
\end{align}
we have
\begin{align}
  S[A_\mu-\partial_\mu\theta,\bar{\psi}e^{-i\theta\tau_3},e^{i\theta\tau_3}\psi]=S[A_\mu,\bar{\psi},\psi].
\end{align}

We also define the functional
\begin{align}
  W[A_\mu,\bar{\eta},\eta]=-\ln Z[A_\mu,\bar{\eta},\eta],
\end{align}
and the effective action
\begin{align}
  \Gamma[A_\mu,\bar{\phi},\phi]=W[A_\mu,\bar{\eta},\eta]+\int d^4x (\bar{\eta}(x)\phi(x)+\bar{\phi}(x)\eta(x)).
\end{align}
by the Legendre transformation of the functional $W[A_\mu,\bar{\eta},\eta]$, where
\begin{align}
  \fdv{W[A_\mu,\bar{\eta},\eta]}{\bar{\eta}(x)}=-\phi(x), \quad\fdv{W[A_\mu,\bar{\eta},\eta]}{\eta(x)}=\bar{\phi}(x).
\end{align}
Relying on the general properties of the Legendre transformation, we obtain the inverse transformation
\begin{align}
  \fdv{\Gamma[A_\mu,\bar{\phi},\phi]}{\phi(x)}=-\bar{\eta}(x), \quad\fdv{\Gamma[A_\mu,\bar{\phi},\phi]}{\bar{\phi}(x)}=\eta(x).
\end{align}

From the local $U(1)$ symmetry of the action $S[A_\mu,\bar{\psi},\psi]$, we can also demonstrate the local $U(1)$ invariance of $Z[A_\mu,\bar{\eta},\eta]$ and $W[A_\mu,\bar{\eta},\eta]$.
If we perform the local $U(1)$ transformation on the gauge field and sources as
\begin{align}
  A_\mu'=A_\mu-\partial_\mu\theta, \quad\bar{\eta}'=\bar{\eta}e^{-i\theta\tau_3}, \quad\eta'=e^{i\theta\tau_3}\eta,
\end{align}
the generating functional is modified to
\begin{align}
  Z[A_\mu',\bar{\eta}',\eta']=\grassfint{\psi}\exp[-S[A_\mu-\partial_\mu\theta,\bar{\psi},\psi]+\int d^4x\left(\bar{\eta}e^{-i\theta\tau_3}\psi+\bar{\psi}e^{i\theta\tau_3}\eta\right)].
\end{align}
Transforming the fields integrated in the functional integral as
\begin{align}
  \bar{\psi}'=\bar{\psi}e^{i\theta\tau_3}, \quad\psi'=e^{-i\theta\tau_3}\psi,
\end{align}
we obtain the local $U(1)$ symmetry of the generating functional
\begin{align}
  Z[A_\mu',\bar{\eta}',\eta']&=\grassfint{\psi'}\exp[-S[A_\mu-\partial_\mu\theta,\bar{\psi}'e^{-i\theta\tau_3},e^{i\theta\tau_3}\psi']+\int d^4x \left(\bar{\eta}\psi'+\bar{\psi}'\eta\right)]\nonumber\\
    &=\grassfint{\psi'}\exp[-S[A_\mu,\bar{\psi'},\psi']+\int d^4x\left(\bar{\eta}\psi'+\bar{\psi'}\eta\right)]\nonumber\\
    &=Z[A_\mu,\bar{\eta},\eta].
\end{align}
As a result, $W[A_\mu,\bar{\eta},\eta]$ is also invariant under the local $U(1)$ transformation.

Due to the invariance under local $U(1)$ transformation, we obtain
\begin{align}
  \int d^4z\left[-\fdv{W[A_\mu,\bar{\eta},\eta]}{A_\mu(z)}\partial_{z^\mu}\theta(z)-i\bar{\eta}(z)\tau_3\fdv{W[A_\mu,\bar{\eta},\eta]}{\bar{\eta}(z)}\theta(z)-i\fdv{W[A_\mu,\bar{\eta},\eta]}{\eta(z)}\tau_3\eta(z)\theta(z)\right]=0.
\end{align}
Since this equation holds for any function $\theta$,
\begin{align}
  \partial_{z^\mu}\fdv{W[A_\mu,\bar{\eta},\eta]}{A_\mu(z)}=i\left(\bar{\eta}(z)\tau_3\fdv{W[A_\mu,\bar{\eta},\eta]}{\bar{\eta}(z)}+\fdv{W[A_\mu,\bar{\eta},\eta]}{\eta(z)}\tau_3\eta(z)\right)
\end{align}
must hold.
Rewriting this using the effective action $\Gamma[A_\mu,\bar{\phi},\phi]$, we get
\begin{align}
  \partial_{z^\mu}\fdv{\Gamma[A_\mu,\bar{\phi},\phi]}{A_\mu(z)}=i\left(\fdv{\Gamma[A_\mu,\bar{\phi},\phi]}{\phi(z)}\tau_3\phi(z)+\bar{\phi}(z)\tau_3\fdv{\Gamma[A_\mu,\bar{\phi},\phi]}{\bar{\phi}(z)}\right).
\end{align}
By further performing functional derivatives of this equation, we can derive the Ward identities of the desired order.
First, differentiating with respect to $\phi(y)$ and setting $\phi=0$, we get
\begin{align}
  \partial_{z^\mu}\left.\frac{\delta^2 \Gamma[A_\mu,\bar{\phi},\phi]}{\delta A_\mu(z)\delta\phi(y)}\right|_{\phi=0}=i\left(\bar{\phi}(z)\tau_3\left.\frac{\delta^2\Gamma[A_\mu,\bar{\phi},\phi]}{\delta\bar{\phi}(z)\delta\phi(y)}\right|_{\phi=0}-\left.\fdv{\Gamma[A_\mu,\bar{\phi},\phi]}{\phi(z)}\right|_{\phi=0}\tau_3\delta(y-z)\right).
\end{align}
Furthermore, differentiating with respect to $\bar{\phi}(x)$ and $\bar{\phi}=0$, we get
\begin{align}
  \partial_{z^\mu}\left.\frac{\delta^3\Gamma[A_\mu,\bar{\phi},\phi]}{\delta A_\mu(z)\delta\bar{\phi}(x)\delta\phi(y)}\right|_{\bar{\phi}=\phi=0}&=i\left(\delta(x-z)\tau_3\left.\frac{\delta^2\Gamma[A_\mu,\bar{\phi},\phi]}{\delta\bar{\phi}(z)\delta\phi(y)}\right|_{\bar{\phi}=\phi=0}-\left.\frac{\delta^2\Gamma[A_\mu,\bar{\phi},\phi]}{\delta\bar{\phi}(x)\phi(z)}\right|_{\bar{\phi}=\phi=0}\tau_3\delta(y-z)\right).
  \label{eq:ap_wi}
\end{align}
The functional derivatives of the effective action are related to the inverse of the Green function.
This can be shown by the equation
\begin{align}
  \tau_0\delta(x-y)=\fdv{\bar{\eta}(y)}{\bar{\eta}(x)}=\int d^4z \fdv{\bar{\phi}(z)}{\bar{\eta}(x)}\fdv{\bar{\eta}(y)}{\bar{\phi}(z)}=\int d^4z \frac{\delta^2W[A_\mu,\bar{\eta},\eta]}{\delta\bar{\eta}(x)\delta\eta(z)}\left(-\frac{\delta^2\Gamma[A_\mu,\bar{\phi},\phi]}{\delta\bar{\phi}(z)\delta\phi(y)}\right).
\end{align}
By setting, $\bar{\eta}=\eta=\bar{\phi}=\phi=0$, we obtain
\begin{align}
  \left.\frac{\delta^2\Gamma[A_\mu,\bar{\phi},\phi]}{\delta\bar{\phi}(z)\delta\phi(y)}\right|_{\bar{\phi}=\phi=0}= G_A^{-1}(z,y),
\end{align}
where
\begin{align}
   G_A(x,y)=-\frac{1}{Z_A}\grassfint{\psi}\psi(x)\bar{\psi}(y)\exp[-S[A_\mu,\bar{\psi},\psi]]=-\left.\frac{\delta^2W[A_\mu,\bar{\eta},\eta]}{\delta\bar{\eta}(x)\delta\eta(y)}\right|_{\bar{\eta}=\eta=0}
\end{align}
is the full (two-fermion) Green function of the fermion system with the gauge field.
Plugging this relation to Eq.~\eqref{eq:ap_wi}, we obtain
\begin{align}
  \partial_{z^\mu}\fdv{ G_A^{-1}(x,y)}{A_\mu(z)}=i\bigg(\delta(x-z)\tau_3 G_A^{-1}(x,y)-G_A^{-1}(x,y)\tau_3\delta(y-z)\bigg).
  \label{eq:ap_generalwi}
\end{align}
From this equation, we can derive the relations that must be satisfied by $n$-photon vertices.

For example, by setting $A=0$ in Eq.~\eqref{eq:ap_generalwi}, we obtain
\begin{align}
  \partial_{z^\mu} \Gamma^\mu(x,y,z)=i\bigg(\delta(x-z)\tau_3 G^{-1}(x,y)-G^{-1}(x,y)\tau_3\delta(y-z)\bigg),
\end{align}
where
\begin{align}
  \Gamma^\mu(x,y,z)=\left.\frac{\delta^3\Gamma[A_\mu,\bar{\phi},\phi]}{\delta A_\mu(z)\delta\bar{\phi}(x)\delta\phi(y)}\right|_{A=\bar{\phi}=\phi=0}=\left.\fdv{ G_A^{-1}(x,y)}{A_\mu(z)}\right|_{A=0}
\end{align}
is the one-photon (two-fermion) vertex and $G$ is the full Green function of the system without the gauge field.
This equation is the well-known Ward identity for the one-photon vertex.
In the momentum space, the Ward identity for the full one-photon vertex is given by
\begin{align}
  \Gamma^\mu(k,q)q_\mu=G^{-1}(k+q)\tau_3-\tau_3G^{-1}(k).
\end{align}

Furthermore, by functionally differentiating Eq.~\eqref{eq:ap_generalwi} with respect to $A_\nu(w)$ and setting $A=0$, we obtain
\begin{align}
  \partial_{z^\mu}\left.\frac{\delta^2 G_A^{-1}(x,y)}{\delta A_\nu(w)\delta A_\mu(z)}\right|_{A=0}=i\left(\delta(x-z)\tau_3\left.\fdv{ G_A^{-1}(x,y)}{A_\nu(w)}\right|_{A=0}-\left.\fdv{ G_A^{-1}(x,y)}{A_\nu(w)}\right|_{A=0}\tau_3\delta(y-z)\right).
\end{align}
Defining the two-photon vertex as
\begin{align}
  \Gamma^{\nu\mu}(x,y,w,z)=-\left.\frac{\delta^2 G_A^{-1}(x,y)}{\delta A_\nu(w)\delta A_\mu(z)}\right|_{A=0},
\end{align}
we obtain the Ward identity for the two-photon vertex:
\begin{align}
  \partial_{z^\mu} \Gamma^{\nu\mu}(x,y,w,z)=i\left(\Gamma^\nu(x,y,w)\tau_3\delta(y-z)-\delta(x-z)\tau_3\Gamma^\nu(x,y,w)\right).
\end{align}
The Ward identities for full multi-photon vertices can be obtained in the same way.
By defining the $n$-photon vertex as
\begin{align}
  \Gamma^{\alpha_1\cdots\alpha_n}(x,y,w_1,\cdots,w_n)=(-1)^{n-1}\left.\left(\prod_{i=1}^n\fdv{A_{\alpha_i}(w_i)}\right) G_A^{-1}(x,y)\right|_{A=0},
\end{align}
the Ward identity for the $n$-photon vertex is given by
\begin{align}
  \partial_{w_n^{\alpha_n}}\Gamma^{\alpha_1\cdots\alpha_n}(x,y,w_1,\cdots,w_n)&=i\Gamma^{\alpha_1\cdots\alpha_{n-1}}(x,y,w_1,\cdots,w_{n-1})\tau_3\delta(y-w_n)\nonumber\\
    &\qquad\qquad -i\delta(x-w_n)\tau_3\Gamma^{\alpha_1\cdots\alpha_{n-1}}(x,y,w_1,\cdots,w_{n-1}).
\end{align}
When expressed in the momentum space, it becomes
\begin{align}
  (q_n)_{\alpha_n}\Gamma^{\alpha_1\cdots\alpha_n}(k,q_1,\cdots,q_n)&=\tau_3\Gamma^{\alpha_1\cdots\alpha_{n-1}}(k,q_1,\cdots,q_{n-1})\nonumber\\
    &\qquad -\Gamma^{\alpha_1\cdots\alpha_{n-1}}(k+q_n,q_1,\cdots,q_n)\tau_3.
\end{align}

\section{The diagrammatical method of the full photon vertices of superconductors}
\label{ap:BSE}

In this appendix, we employ the Fock approximation to the self-energy instead of the generalized CFOP method and discuss the gauge-invariant treatment of the electromagnetic responses in this case.
The Fock approximation is given by
\begin{align}
  \Sigma_A(x,y)=V(x-y)\tau_3G_A(x,y)\tau_3.
  \label{eq:ap_se}
\end{align}

First, we show that the Fock approximation is compatible with the Ward identities.
We have to check the satisfaction of Eq.~\eqref{wi_se}.
This can be shown by
\begin{align}
  \partial_{z^\mu}\fdv{\Sigma_A(x,y)}{A_\mu(z)}&=V(x-y)\tau_3\partial_{z^\mu}\fdv{G_A(x,y)}{A_\mu(z)}\tau_3\nonumber\\
  &=-V(x-y)\tau_3G_A(x,\overline{x}')\partial_{z^\mu}\fdv{G_A^{-1}(\overline{x}',\overline{y}')}{A_\mu(z)}G_A(\overline{y}',y)\tau_3\nonumber\\
  &=-V(x-y)\tau_3G_A(x,\overline{x}')\bigg[i\delta(\overline{x}'-z)\tau_3G_A^{-1}(\overline{x}',\overline{y}')-iG_A^{-1}(\overline{x}',\overline{y}')\tau_3\delta(\overline{y}'-z)\bigg]G_A(\overline{y}',y)\tau_3\nonumber\\
  &=iV(x-y)\delta(x-z)G_A(x,y)\tau_3-iV(x-y)\delta(y-z)\tau_3G_A(x,y)\nonumber\\
  &=i\delta(x-z)\tau_3\Sigma_A(x,y)-i\Sigma_A(x,y)\tau_3\delta(y-z).
\end{align}
Therefore, it is possible to obtain gauge-invariant electromagnetic responses in BCS superconductors within the Fock approximation.

\begin{figure}
  \centering
  \includegraphics[width=0.9\linewidth]{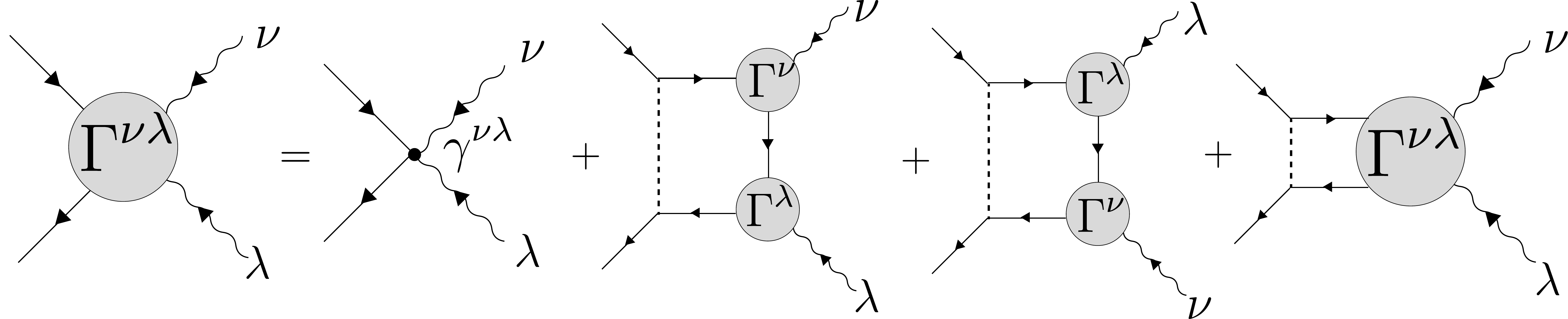}
  \caption{The diagrammatical representation of the second-order extension of the Bethe-Salpeter equation.}
  \label{fig:bse_extension}
\end{figure}

The full photon vertex is determined by definition in Eq.~\eqref{eq:def_fullvertex} and the resulting photon vertices will automatically satisfy the Ward identity.
We can easily obtain the second-order extension of the Bethe-Salepter equation~[Fig.~\ref{fig:bse_extension}]:
\begin{align}
  \Gamma^{\mu\nu}(k,q_1,q_2)&=\gamma^{\mu\nu}(k,q_1,q_2)+V(k-\overline{p})\tau_3G(\overline{p}+q)\Gamma^\nu(\overline{p}+q_1,q_2)G(\overline{p}+q_1)\Gamma^\mu(\overline{p},q_1)G(\overline{p})\tau_3\nonumber\\
    &\qquad+V(k-\overline{p})\tau_3G(\overline{p}+q)\Gamma^\mu(\overline{p}+q_2,q_1)G(\overline{p}+q_2)\Gamma^\nu(\overline{p},q_2)G(\overline{p})\tau_3\nonumber\\
    &\qquad+V(k-\overline{p})\tau_3G(\overline{p}+q)\Gamma^{\mu\nu}(\overline{p},q_1,q_2)G(\overline{p})\tau_3.
    \label{eq:bse_2}
\end{align}
In principle, by similarly calculating for higher-order cases, it is possible to derive any full photon vertex in BCS superconductors.

\begin{figure}[t]
  \centering
  \includegraphics[width=0.9\linewidth]{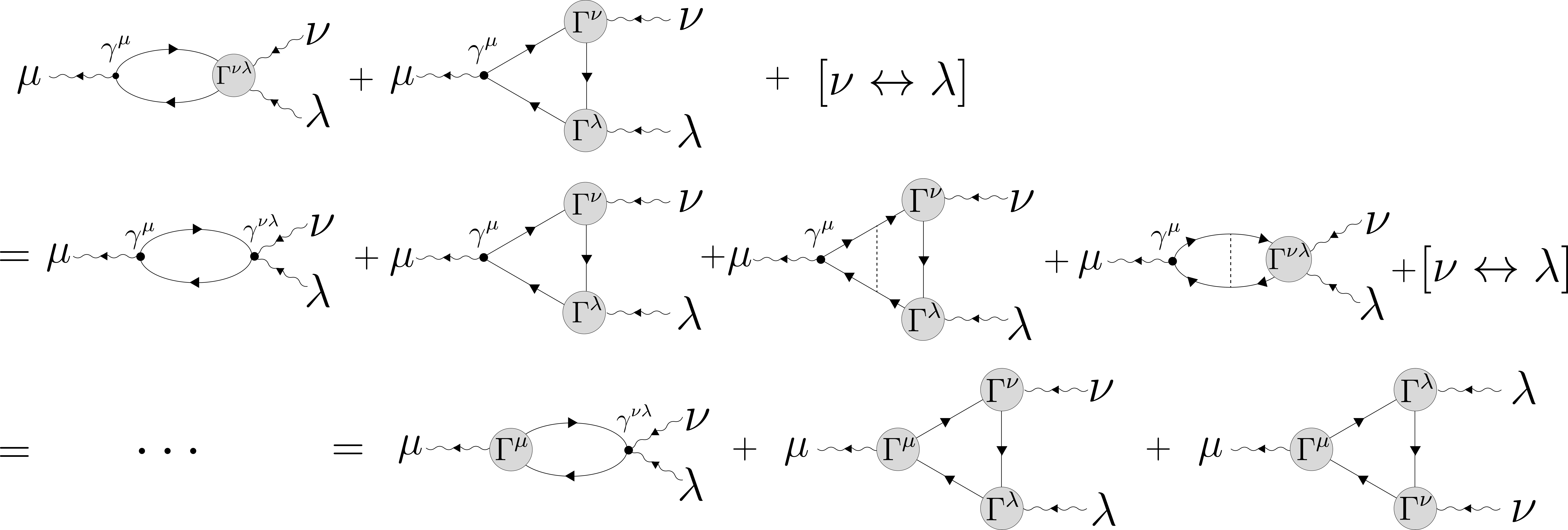}
  \caption{The diagrammatic representation of the calculation of the second-order response kernel. The last equality can be obtained by the Bethe-Salpeter equation. We can avoid explicitly solving the integral equation for the full two-photon vertex.}
  \label{fig:triangle}
\end{figure}

At higher orders, it becomes more complicated to obtain the full photon vertex by solving the integral equations.
In some cases, it is possible to avoid explicitly solving the integral equations using the diagrammatic method.
As an example, we calculate the second-order response kernel.
Let us focus on the term involving the full two-photon vertex and the triangular diagrams in the second-order response kernel in Eq.~\eqref{eq:sec_kernel_momentum}.
The corresponding diagrams are shown in the first line of Fig.~\ref{fig:triangle}.
We can sequentially substitute Eq.~\eqref{eq:bse_2} into them, and it can be shown that
\begin{align}
  &\frac{1}{2}\mathrm{Tr}\bigg[\gamma^\mu(\overline{k}+q,-q)G(\overline{k}+q)\Gamma^{\nu\lambda}(\overline{k},q_1,q_2)G(\overline{k})\bigg]\nonumber\\
  &+\mathrm{Tr}\bigg[\gamma^\mu(\overline{k}+q,-q)G(\overline{k}+q)\Gamma^\lambda(\overline{k}+q_1,q_2)G(\overline{k}+q_1)\Gamma^\nu(\overline{k},q_1)G(\overline{k})\bigg]+[(\nu,q_1)\leftrightarrow(\lambda,q_2)]\nonumber\\
  =&\frac{1}{2}\mathrm{Tr}\big[\Gamma^\mu(\overline{k}+q,-q)G(\overline{k}+q)\gamma^{\nu\lambda}(\overline{k},q_1,q_2)G(\overline{k})\big]\nonumber\\
  &+\mathrm{Tr}\bigg[\Gamma^\mu(\overline{k}+q,-q)G(\overline{k}+q)\Gamma^\lambda(\overline{k}+q_1,q_2)G(\overline{k}+q_1)\Gamma^\nu(\overline{k},q_1)G(\overline{k})\bigg]+[(\nu,q_1)\leftrightarrow(\lambda,q_2)].
\end{align}
This equation can be easily translated into Feynman diagrams presented in Fig.~\ref{fig:triangle}.
If we adopt the self-energy in Eq.~\eqref{eq:ap_se}, it is possible to calculate the second-order response without explicitly solving the integral equation for the full two-photon vertex, given the full one-photon vertex.
In the previous study~\cite{Huang2023}, the second-order response kernel is calculated in this way instead of explicitly using the full two-photon vertex.
Although this simplification of the calculation is useful, we will not adopt the Fock approximation due to the subtlety explained in the main text.

\section{The detailed calculation of the correction part of the full two-photon vertex}
\label{ap:cfop_2nd}

In this appendix, we discuss the full two-photon vertex by the generalized CFOP method.
The correction part of the full two-photon vertex is defined by
\begin{align}
  \Lambda^{\nu\lambda}(x,y,w_1,w_2)=\left.\frac{\delta^2\Sigma_A(x,y)}{\delta A_\nu(w_1)\delta A_\lambda(w_2)}\right|_{A=0}.
\end{align}
Since the self-energy of superconductors is the gap function, we have to calculate the functional derivative of $\Delta_{i,l,A}(x,y)$.
The functional derivative of the gap equation leads to
\begin{align}
  &\left.\frac{\delta^2\Delta_{i,l,A}(x,y)}{\delta A_\nu(w_1)\delta A_\lambda(w_2)}\right|_{A=0}\nonumber\\
  &=-\frac{g_l\delta(x-y)}{2}\frac{\delta^2}{\delta A_\nu(w_1)\delta A_\lambda(w_2)}\left.\mathrm{Tr}\bigg[\tau_i\otimes E_{ll}G_A(x,y)\bigg]\right|_{A=0}\nonumber\\
  &=\frac{g_l\delta(x-y)}{2}\fdv{A_\nu(w_1)}\left.\mathrm{Tr}\bigg[\tau_i\otimes E_{ll}G_A(x,\bar{z_1})\fdv{G_A^{-1}(\bar{z_1},\bar{z_2})}{A_\lambda(w_2)}G_A(\bar{z_2},y)\bigg]\right|_{A=0}\nonumber\\
  &=-\frac{g_l\delta(x-y)}{2}\mathrm{Tr}\bigg[\tau_i\otimes E_{ll}G(x,\bar{z_3})\Gamma^\nu(\bar{z_3},\bar{z_4},w_1)G(\bar{z_4},\bar{z_1})\Gamma^\lambda(\bar{z_1},\bar{z_2},w_2)G(\bar{z_2},y)\bigg]\nonumber\\
  &\qquad -\frac{g_l\delta(x-y)}{2}\mathrm{Tr}\bigg[\tau_i\otimes E_{ll}G(x,\bar{z_1})\Gamma^{\nu\lambda}(\bar{z_1},\bar{z_2},w_1,w_2)G(\bar{z_2},y)\bigg]\nonumber\\
  &\qquad -\frac{g_l\delta(x-y)}{2}\mathrm{Tr}\bigg[\tau_i\otimes E_{ll}G(x,\bar{z_1})\Gamma^\lambda(\bar{z_1},\bar{z_2},w_2)G(\bar{z_2},\bar{z_3})\Gamma^\nu(\bar{z_3},\bar{z_4},w_1)G(\bar{z_4},y)\bigg].
\end{align}
Assuming the functional derivative of the gap function takes the form
\begin{align}
  \left.\frac{\delta^2\Delta_{i,l,A}(x,y)}{\delta A_\nu(w_1)\delta A_\lambda(w_2)}\right|_{A=0}=\Lambda_{il}^{\nu\lambda}(x,w_1,w_2)\delta(x-y),
\end{align}
we obtain the matrix equation in the momentum space
\begin{align}
  &\frac{2}{g_i}\Lambda_{il}^{\nu\lambda}(q_1,q_2)+\sum_{j,l'}\mathrm{Tr}\bigg[\tau_i\otimes E_{ll}G(\bar{k}+q)\tau_j\otimes E_{l'l'}G(\bar{k})\bigg]\Lambda_{jl'}^{\nu\lambda}(q_1,q_2)\nonumber\\
  &=-\mathrm{Tr}\bigg[\tau_i\otimes E_{ll}G(\bar{k}+q)\gamma^{\nu\lambda}(\bar{k},q_1,q_2)G(\bar{k})\bigg]-\mathrm{Tr}\bigg[\tau_i\otimes E_{ll}G(\bar{k}+q)\Gamma^\lambda(\bar{k}+q_1,q_2)G(\bar{k}+q_1)\Gamma^\nu(\bar{k},q_1)G(\bar{k})\bigg]\nonumber\\
  &\qquad -\mathrm{Tr}\bigg[\tau_i\otimes E_{ll}G(\bar{k}+q)\Gamma^\nu(\bar{k}+q_2,q_1)G(\bar{k}+q_2)\Gamma^\lambda(\bar{k},q_2)G(\bar{k})\bigg].
  \label{eq:cfop_2}
\end{align}
This equation can be solved given the full one-photon vertices $\Gamma^\mu(k,q)$.

\end{document}